\documentclass[prd,preprint,superscriptaddress,nofootinbib,showpacs]{revtex4-2}

\usepackage[english]{babel}

\usepackage[utf8]{inputenc}
\usepackage{amsmath,amssymb}
\usepackage{bbold}
\usepackage{slashed}
\usepackage{subfigure}
\usepackage[colorlinks=true,
breaklinks=true,
urlcolor=blue,
citecolor=blue]{hyperref}
\usepackage[usenames,dvipsnames]{color}
\usepackage{appendix}   
\usepackage{braket,bm}
\usepackage{multirow}
\usepackage{enumitem}
\usepackage{color}
\usepackage{cancel}
\usepackage{dsfont}
\usepackage{orcidlink}
\usepackage{array}
\usepackage{indentfirst}
\usepackage{bbm}
\usepackage{epstopdf}
\usepackage{graphicx}
\usepackage{caption}
\usepackage[normalem]{ulem}
\usepackage{booktabs}

\DeclareCaptionLabelFormat{figprefix}{FIG.#2.} 
\allowdisplaybreaks          

\AtBeginDocument{}

\begin{document}

\title{Hadronic vacuum polarization and light-by-light scattering in representative infrared-improved soft-wall holographic QCD models}

\author{Jin-Yang Shen}
\email{Jinyang\_Shen@hnu.edu.cn}
\affiliation{School of Physics and Electronics, Hunan University, Changsha 410082, China}
\affiliation{Hunan Provincial Key Laboratory of High-Energy Scale Physics and Applications, Hunan University, Changsha 410082, China}

\author{Wen-Yuan Peng}
\email{S240700462@hnu.edu.cn}
\affiliation{School of Physics and Electronics, Hunan University, Changsha 410082, China}
\affiliation{Hunan Provincial Key Laboratory of High-Energy Scale Physics and Applications, Hunan University, Changsha 410082, China}

\author{Ling-Yun Dai}
\email{dailingyun@hnu.edu.cn}
\affiliation{School of Physics and Electronics, Hunan University, Changsha 410082, China}
\affiliation{Hunan Provincial Key Laboratory of High-Energy Scale Physics and Applications, Hunan University, Changsha 410082, China}

\author{Zhen Fang}
\email{zhenfang@hnu.edu.cn}
\affiliation{School of Physics and Electronics, Hunan University, Changsha 410082, China}
\affiliation{Hunan Provincial Key Laboratory of High-Energy Scale Physics and Applications, Hunan University, Changsha 410082, China}
\begin{abstract}
We investigate the hadronic contributions to the anomalous magnetic moment of the muon \(a_\mu\) in three representative infrared-improved soft-wall AdS/QCD models, focusing on the leading-order hadronic vacuum polarization (LO-HVP) and hadronic light-by-light (HLbL) contributions.  The model parameters are fixed by the ground-state \(\rho\)-meson mass, pion mass, and pion decay constant, with the low-lying excited spectrum used as an additional phenomenological constraint. We find that the LO-HVP contribution remains below dispersive determinations, mainly because the \(\rho\)-meson decay constant is underestimated. In the HLbL sector, the standard Chern-Simons term gives well-defined ground-state pseudoscalar-pole contributions but leads to divergent excited-state sums in soft-wall models. We therefore also adopt a scalar-extended Chern-Simons term motivated by the brane--antibrane construction, which provides infrared suppression and allows finite estimates for pseudoscalar excitations and axial-vector mesons. The resulting HLbL predictions are internally convergent but still quantitatively model dependent, showing that further refinement of the infrared completion and short-distance constraints is required.
\end{abstract}

\maketitle
\newpage

\section{Introduction}

The anomalous magnetic moment of the muon, $a_\mu=(g-2)_\mu/2$, constitutes a key precision observable in particle physics and has a long history rooted in early foundational studies~\cite{Schwinger:1948iu}. Over the past several decades, sustained experimental efforts have been devoted to continuously improving the measurement precision~\cite{CERN-Mainz-Daresbury:1978ccd,Muong-2:2006rrc,Muong-2:2021ojo,Muong-2:2023cdq}. In 2025, the Fermilab Muon $g-2$ Collaboration reported its final result based on the complete dataset, achieving an unprecedented precision of 127 parts per billion and establishing an experimental benchmark that is expected to stand for the foreseeable future~\cite{Muong-2:2025xyk}. Despite this remarkable achievement, the comparison between experiment and the Standard Model prediction has developed into a subtle and evolving situation, with the primary source of tension now residing in the theoretical determination of hadronic contributions~\cite{Aoyama:2020ynm,Aliberti:2025beg}.

The Standard Model prediction for the anomalous magnetic moment of the muon receives contributions from three distinct sectors: electromagnetic, electroweak, and hadronic. Among these, the hadronic sector constitutes the dominant source of theoretical uncertainty. It is conventionally decomposed into two components: the hadronic vacuum polarization (HVP) contribution~\cite{Davier:2017zfy,Hoid:2020xjs,Hoferichter:2021wyj,Benayoun:2021ody,Yi:2021ccc,Qin:2020udp,Wang:2023njt,Qin:2024ulb} and the hadronic light-by-light (HLbL) scattering contribution~\cite{Hayakawa:1997rq,Guevara:2018rhj,Raya:2019dnh,Estrada:2024cfy,Estrada:2025bty,Colangelo:2014dfa,Hoferichter:2018kwz,Danilkin:2019mhd,ExtendedTwistedMass:2022ofm,ExtendedTwistedMass:2023hin,Holz:2024lom,Zhang:2025ijd}. At present, the evaluation of these hadronic contributions relies on two complementary and increasingly sophisticated theoretical frameworks.

On the one hand, a broad class of phenomenological and theory-driven approaches is employed to constrain both the HVP and HLbL contributions. These include traditional data-driven methods based on dispersion relations, as well as effective field theory descriptions such as resonance chiral theory and vector meson dominance models. In particular, the data-driven approach exploits precise measurements of electron--positron annihilation into hadrons, together with dispersive techniques, to determine the HVP function $\Pi^{\mathrm{had}}_{\mathrm{em}}(Q^2)$~\cite{Jegerlehner:2017gek}. However, recent high-precision measurements of the $e^+e^-\to\pi^+\pi^-$ cross section have revealed tensions among different experimental datasets~\cite{CMD-3:2023alj}. These discrepancies have propagated into dispersive analyses and have thus far hindered the construction of a single, unified data-driven estimate~\cite{Aliberti:2025beg}.

On the other hand, first-principles calculations based on lattice QCD provide a systematically improvable and fully nonperturbative alternative that is independent of experimental cross-section data. Recent lattice determinations of the HVP contribution have reached sub-percent precision and can be consistently combined into a lattice average. When these results are supplemented by improved estimates of the HLbL contribution obtained through a synergistic combination of lattice QCD~\cite{Blum:2019ugy} and data-driven constraints~\cite{Colangelo:2014pva}, the resulting Standard Model prediction becomes compatible with the experimental measurement of the muon $g-2$~\cite{Borsanyi:2020mff,RBC:2023pvn,Aliberti:2025beg,Boccaletti:2024guq}. This evolving theoretical landscape underscores the importance of reconciling lattice QCD results, data-driven determinations, and effective Lagrangian approaches in order to elucidate the origin of the muon $g-2$ puzzle.

In this context, holographic QCD models inspired by the AdS/CFT correspondence~\cite{Witten:1998qj,Gubser:1998bc,Maldacena:1997re} provide a valuable framework for exploring nonperturbative hadronic dynamics~\cite{Erlich:2005qh,Karch:2006pv,DaRold:2005mxj,Jarvinen:2011qe,Arean:2012mq,Li:2013oda,FolcoCapossoli:2019imm,Li:2016smq,Chen:2020ath,Li:2012ay,Chen:2024ckb,He:2013qq,Fang:2015ytf,deTeramond:2005su,Sui:2009xe,Gherghetta:2009ac,Chelabi:2015gpc,Boschi-Filho:2005xct,Shen:2025zkj,Liu:2026cpr}. In recent years, such models have been increasingly applied to the study of the anomalous magnetic moment of the muon~\cite{Cappiello:2010uy,Leutgeb:2019zpq,Leutgeb:2019gbz,Leutgeb:2021mpu,Leutgeb:2022lqw,Mager:2025pvz,Colangelo:2024xfh,Cappiello:2021vzi,Cappiello:2019hwh,Cappiello:2025fyf}. Most existing holographic evaluations of $a_\mu$ have been performed within the standard hard-wall or original soft-wall constructions, which provide useful qualitative insights into hadronic contributions. A distinctive feature of the holographic approach is the emergence of an infinite tower of vector and axial-vector resonances, allowing for the implementation of QCD short-distance constraints on correlation functions and form factors. Nevertheless, both the hard-wall and original soft-wall models suffer from well-known quantitative shortcomings. The hard-wall model fails to reproduce linear Regge trajectories, while the original soft-wall framework typically underestimates the decay constants of low-lying vector and axial-vector mesons. Consequently, these models are unable to simultaneously describe key hadronic observables within a single consistent parametrization, leading to sizable model-dependent uncertainties in holographic predictions for $a_\mu$. 

This motivates a more modest comparative study of \(a_\mu\) within three representative infrared-improved soft-wall constructions~\cite{Cui:2013xva,Fang:2016nfj,Fang:2019lmd}. We do not claim to survey the full space of holographic QCD models. Instead, we ask how these specific backgrounds, which incorporate more realistic infrared dynamics such as a scalar vacuum expectation value with linear growth at large \(z\), perform for the vector correlator, pion transition form factor, and selected HLbL contributions. The HVP contribution is determined by the renormalized vector current two-point function, while the HLbL contribution is governed by transition form factors generated by the Chern-Simons sector. In the standard Chern-Simons setup, the ground-state pseudoscalar contribution can be computed, but the extension to the full tower of excited pseudoscalars and axial-vector mesons is not well defined in soft-wall backgrounds: the corresponding transition form factors and mode sums diverge~\cite{Leutgeb:2025jmv}. We therefore first present the standard ground-state calculation and then introduce a scalar-extended Chern-Simons term~\cite{Leutgeb:2024rfs} as an effective infrared completion. This extension renders the excited pseudoscalar and axial-vector contributions finite, but it should be viewed as a repair of the soft-wall HLbL sector rather than as a unique fundamental modification of the holographic action.

The remainder of this paper is organized as follows. In Sec.~\ref{sec2}, we introduce the three infrared-improved soft-wall models considered in this work. In Sec.~\ref{sec33}, the model parameters are fixed by reproducing the ground-state $\rho$-meson mass, the pion mass, and the pion decay constant, while simultaneously providing a reasonable description of the excited meson spectrum. In Sec.~\ref{sec3}, we compute the leading-order HVP contribution to the anomalous magnetic moment of the muon in the two-flavor case within these models. In Sec.~\ref{sec4}, we first compute the ground-state pseudoscalar-pole contribution to $a_\mu^{\mathrm{HLbL}}$ using the standard Chern-Simons term. We then discuss the limitations of this standard treatment for higher resonances and employ a scalar-extended Chern-Simons term to evaluate the contributions from pseudoscalar excited states and axial-vector mesons. Finally, Sec.~\ref{sec5} summarizes our main findings and discusses their implications.

\section{The Holographic QCD Models}\label{sec2}

\subsection{The infrared-improved soft-wall model (SW1)}

In the original soft-wall AdS/QCD model, chiral symmetry breaking cannot be realized in a fully consistent manner, since the scalar vacuum expectation value (VEV) is not compatible with the corresponding equation of motion (EOM)~\cite{Colangelo:2008us,Li:2012ay}. In order to simultaneously and coherently describe linear confinement in low-energy QCD, chiral symmetry breaking, and the resulting meson spectra, an infrared-improved soft-wall AdS/QCD model was proposed in Ref.~\cite{Cui:2013xva}. The infrared-improved soft-wall model is formulated in an AdS$_5$ spacetime with the metric ansatz
\begin{equation}
ds^2 = z^{-2} \left( \eta_{\mu\nu} \, dx^\mu dx^\nu - dz^2 \right),
\end{equation}
where $\eta_{\mu\nu} = \mathrm{diag}\{1, -1, -1, -1\}$.
The structure of the model is specified by the following five-dimensional action
\begin{equation}\label{2action}
S = \int d^5x \, \sqrt{g} \, e^{-\Phi(z)} \,
\mathrm{Tr} \left[
 |D X|^2 - m_5^2(z) |X|^2 - \lambda_X |X|^4
- \frac{1}{4 g_5^2} \left( F_L^2 + F_R^2 \right)
\right],
\end{equation}
where the covariant derivative and field strengths are given by
\[
D^M X = \partial^M X - i A_L^M X + i X A_R^M,
\qquad
F_{MN}^{L,R} = \partial_M A_N^{L,R} - \partial_N A_M^{L,R}
- i\,[A_M^{L,R}, A_N^{L,R}].
\]
The gauge fields are written as $A_{L,R}^M = A_{L,R}^{M a} t^a$, with the generators normalized according to $\mathrm{Tr}(t^a t^b) = \delta^{ab}/2$. The five-dimensional gauge coupling is fixed by matching to perturbative QCD, yielding $g_5^2 = 12\pi^2/N_c$, where $N_c$ denotes the number of colors~\cite{Erlich:2005qh}.

For the dilaton field $\Phi(z)$, the infrared (IR) behavior is required to grow quadratically in order to ensure that the mass spectrum of highly excited mesons exhibits approximately linear Regge trajectories. At the same time, since the introduction of the dilaton field explicitly breaks the original conformal symmetry, $\Phi(z)$ must approach zero in the ultraviolet (UV) limit so that conformality is recovered. Following the SW1 construction of Ref.~\cite{Cui:2013xva}, we use the dilaton profile
\begin{equation}
\Phi(z) = \mu_g^2 z^2 - \frac{\lambda_g^4 \mu_g^4 z^4}{(1 + \mu_g^2 z^2)^3}.
\end{equation}
The leading term fixes the required ultraviolet and infrared powers. The second term is subleading in both limits and deforms the intermediate-$z$ region sampled by the low-lying normalizable modes. It was introduced phenomenologically in Ref.~\cite{Cui:2013xva} to improve the ground-to-first-excitation separation and the description of the vector and other low-lying radial spectra. We retain this functional form in the present comparison.

The bulk scalar field can be decomposed as
\begin{equation}
X(x,z) = \left( \frac{\chi(z)}{2} + S(x,z) \right) e^{2 i \pi(x,z)} ,
\end{equation}
where $\pi(x,z) = \pi^a(x,z) t^a$ denotes the pseudoscalar meson field and $S(x,z)$ represents the scalar meson field. The VEV of the bulk scalar field is assumed to take the form
\begin{equation}
X_0(z)\equiv\langle X(z)\rangle = \frac{\chi(z)}{2} I_2 ,
\end{equation}
with $I_2$ the $2\times2$ identity matrix. The scalar VEV $\chi(z)$ satisfies the asymptotic behaviors
\begin{equation}
\chi(z \to 0) = m_q \zeta \, z + \frac{\sigma}{\zeta} z^3,
\qquad
\chi(z \to \infty) = v_q z ,
\end{equation}
where $m_q$ denotes the current quark mass, $\sigma$ is the chiral condensate, $\zeta$ is a normalization constant fixed to be $\zeta = \sqrt{3}/(2\pi)$ (and is taken to have this value throughout this work), and $v_q$ is a constant parameter characterizing the energy scale of dynamically generated spontaneous chiral symmetry breaking. To satisfy the above ultraviolet and infrared asymptotic conditions simultaneously, the scalar vacuum profile is parameterized as
\begin{equation}\label{chiSW1}
\chi(z) = \frac{A z + B z^3}{1 + C z^2},
\end{equation}
with $A = m_q \zeta$, $B = \sigma/\zeta + m_q \zeta C$, $C = \mu_c^2 / \zeta$, and $v_q = B / C$, where the constant parameter $\mu_c$ characterizes the QCD confinement scale.

The linear infrared behavior $\chi(z)\sim v_qz$ is motivated by the vector--axial mass-squared splitting. The transverse axial equation contains the asymptotic term $g_5^2\chi^2(z)/z^2$, which approaches a nonzero constant when $\chi\sim z$ and permits asymptotically parallel vector and axial trajectories with a finite offset. If $\chi\to\mathrm{const.}$, this term instead vanishes as $z^{-2}$, corresponding to an asymptotically vanishing splitting. Both behaviors have been considered in the soft-wall literature~\cite{Shifman:2007xn,Gherghetta:2009ac}; the SW1 construction adopts the former phenomenological pattern.

The rational form of $\chi(z)$ is a parametrization chosen to satisfy the prescribed UV and IR asymptotic behavior. In SW1, the profiles $\Phi(z)$, $\lambda_X(z)$, and $\chi(z)$ are specified first, and the coordinate-dependent scalar mass $m_5^2(z)$ is subsequently reconstructed from the scalar-background EOM. The resulting background therefore satisfies this EOM by construction. With the conventions used here,
\begin{equation}\label{chiEOMSW1}
\chi''-\left(\frac{3}{z}+\Phi'\right)\chi'
-\frac{1}{z^2}\left[m_5^2(z)\chi+\frac{\lambda_X(z)}{2}\chi^3\right]=0,
\end{equation}
\begin{equation}\label{m5SW1reconstructed}
m_5^2(z)=z^2\frac{\chi''}{\chi}
-\left(3z+z^2\Phi'\right)\frac{\chi'}{\chi}
-\frac{\lambda_X(z)}{2}\chi^2.
\end{equation}
Thus the parametrized $\chi(z)$ and the reconstructed $m_5^2(z)$ form an EOM-consistent background. The term infrared-improved refers to the complete SW1 background, designed to accommodate linear radial trajectories, chiral symmetry breaking, a nonvanishing vector--axial splitting, and the light-meson spectrum; the rational form of $\chi(z)$ is one phenomenological element of that construction.

In the action~(\ref{2action}), a quartic interaction term for the bulk scalar field is introduced. For SW1, we adopt the following parametrization introduced in Ref.~\cite{Cui:2013xva}:
\begin{equation}
\lambda_X(z)=\frac{\mu_g^2 z^2}{1+\mu_g^2 z^2}\,\lambda .
\end{equation}
A vanishing UV limit is a feature of this phenomenological ansatz, and constant bulk quartic couplings are also admissible in modified soft-wall constructions~\cite{Gherghetta:2009ac}. Ref.~\cite{Cui:2013xva} compared several $z$-dependent parametrizations approaching a constant at large $z$ and found the resonance mass spectra to be relatively insensitive to the specific functional form of $\lambda_X(z)$. 

\subsection{The improved soft-wall model with $2$ flavors (SW2)}

In the SW1 model, the scalar VEV $\chi(z)$ is parameterized to reproduce the required UV and IR asymptotic
behaviors, while the corresponding coordinate-dependent scalar mass
$m_5^2(z)$ is determined from the scalar-background EOM. The SW2 model reverses this procedure: the bulk scalar mass,
dilaton profile, and scalar potential are specified first, and
$\chi(z)$ is then obtained by numerically solving the scalar-background EOM~\cite{Fang:2016nfj}.
SW2 retains the same field content, AdS$_5$ metric, and general
form of the bulk action as SW1, but employs different scalar-sector background functions and a constant quartic coupling. Its dilaton profile is chosen in the simplest quadratic form
\begin{equation}
\Phi(z) = \mu_g^2 z^2 ,
\end{equation}
whose infrared growth ($z \to \infty$) provides a soft cutoff scale for the action and constitutes a defining feature of the soft-wall framework.

In order to achieve a dynamically consistent realization of chiral symmetry breaking, the improved soft-wall model further introduces a $z$-dependent five-dimensional mass term for the scalar field. This bulk mass approaches the conventional constant value in the ultraviolet region, consistent with the ultraviolet asymptotic analysis of the scalar EOM. It becomes increasingly negative in the infrared. In Refs.~\cite{Fang:2016nfj,Fang:2019lmd}, asymptotic analyses of the scalar-background EOM lead to a constant UV limit and a negative quadratic leading behavior in the IR. The following expression is adopted to realize both asymptotic behaviors:
\begin{equation}\label{mass2}
m_5^2(z) = -3 - \mu_c^2 z^2 .
\end{equation}
The coordinate-dependent mass in Eq.~(\ref{mass2}) becomes more negative than the AdS Breitenlohner--Freedman (BF) value at large \(z\). This does not imply an instability, because the BF bound applies to a constant-mass scalar in pure AdS, whereas the present fluctuation problem contains a nontrivial dilaton, a \(z\)-dependent mass, and a nonzero scalar background. The relevant check is the spectrum of normalizable scalar fluctuations \(S(x,z)\). After transforming the scalar fluctuation equation to Schr\"odinger form, we find a well-behaved effective potential and a discrete spectrum with positive lowest eigenvalues for the parameter sets used below. Thus no tachyonic four-dimensional scalar mode appears at the level of linear fluctuations. 

Starting from the bulk action, the EOM for the scalar VEV $\chi(z)$ can be derived as
\begin{equation}
\chi''(z)
- \left( \frac{3}{z} + \Phi'(z) \right) \chi'(z)
- \frac{1}{z^2} \left( m_5^2(z)\, \chi(z) + \frac{\lambda}{2} \chi^3(z) \right) = 0 ,
\end{equation}
and the asymptotic behaviors of $\chi(z)$ in this model take the same form as those of the SW1 model.

\subsection{The improved soft-wall model with $2+1$ flavors (SW3)}

In Ref.~\cite{Fang:2019lmd}, a $2+1$-flavor construction was developed within the same improved soft-wall class as SW2, enabling the calculation of octet meson spectra and their associated decay constants. We emphasize that SW3 retains coupled light--strange scalar backgrounds, a different dilaton profile, a determinant interaction, and an independently calibrated parameter set. It therefore does not reduce numerically to the calibrated SW2 background when an observable is restricted to the light channel. This construction makes it possible to investigate the chiral phase transition both at finite baryon chemical potential and for different quark masses in the $2+1$-flavor case. The action of the improved soft-wall model with $2+1$ flavors can be written as
\begin{equation}
S = \int d^5x \sqrt{g}\, e^{-\Phi(z)} \left[ \mathrm{Tr} \left( |DX|^2 - m_5^2(z)|X|^2 - \lambda |X|^4 - \frac{1}{4 g_5^2} (F_L^2 + F_R^2) \right) - \gamma \, \mathrm{Re}\{\det X\} \right],
\end{equation}
where the last term corresponds to the 't~Hooft determinant introduced to ensure a correct realization of the chiral transition in the $2+1$-flavor case. In order to satisfy the required boundary behavior, the dilaton field is parametrized as
\begin{equation}
\Phi(z) = \mu_g^2 z^2 \left( 1 - e^{-\frac{1}{4} \mu_g^2 z^2} \right).
\end{equation}
The second term of the dilaton was introduced to improve the $\rho$-meson radial spectrum~\cite{Fang:2019lmd}.

For $2+1$ flavors, the VEV of the bulk scalar field takes the form
\begin{equation}
X_0(z)\equiv\langle X(z)\rangle = \frac{1}{\sqrt{2}}
\begin{pmatrix}
\chi_u(z) & 0 & 0 \\
0 & \chi_d(z) & 0 \\
0 & 0 & \chi_s(z)
\end{pmatrix},
\end{equation}
with $\chi_u = \chi_d$. Substituting this scalar VEV into the action, the EOMs for $\chi_u$ and $\chi_s$ are obtained as
\begin{align}
\chi_u'' - \left( \frac{3}{z} + \Phi' \right) \chi_u' - \frac{1}{z^2} \left( m_5^2 \chi_u + \lambda \chi_u^3 + \frac{\gamma}{2\sqrt{2}} \chi_u \chi_s \right) &= 0, \\
\chi_s'' - \left( \frac{3}{z} + \Phi' \right) \chi_s' - \frac{1}{z^2} \left( m_5^2 \chi_s + \lambda \chi_s^3 + \frac{\gamma}{2\sqrt{2}} \chi_u^2 \right) &= 0,
\end{align}
with the boundary conditions $\chi_{u,s}'(0) = \frac{1}{\sqrt{2}} m_{u,s} \zeta$ and $\chi_{u,s}(z \to \infty) \sim z$. The factor $1/\sqrt{2}$ belongs to the matrix-valued VEV $X_0$, not to the UV source coefficient of $\chi_{u,s}$.

\section{The meson spectra and decay constants}\label{sec33}

The parameters of the improved soft-wall models are determined by the constrained phenomenological calibration described below. The ground-state $\rho$ mass, neutral-pion mass, and pion decay constant are imposed as inputs, while the low-lying radial spectra guide the remaining parameter choices.

The vector meson masses are obtained by solving the EOM,
\begin{align}
\partial_z \left( \frac{e^{-\Phi}}{z} \partial_z V_\rho(z) \right) + m_{\rho}^2 \frac{e^{-\Phi}}{z} V_\rho(z) = 0,
\end{align}
subject to the boundary conditions $V_\rho(0)=0$ and $V_\rho'(\infty)=0$. The decay constant of the $\rho$ meson is given by
\begin{equation}
F_\rho = \frac{1}{g_5}\left.\frac{e^{-\Phi}\partial_z V_\rho(z)}{z} \right|_{z\to0}
 ,
\end{equation}
where $V_\rho(z)$ denotes the normalized ground-state wave function satisfying
$\int dz\, e^{-\Phi} V_\rho^2/z = 1$.

To determine the masses of the $\pi$ and axial-vector meson $a_1$, we derive the equations of motion for the axial-vector and pseudoscalar sectors from the bulk action. In the two-flavor improved soft-wall model, the equations take the form
\begin{align}
\partial_z \left( \frac{e^{-\Phi}}{z} \partial_z A_n \right) + \frac{e^{- \Phi}}{z} m_{a_1}^2 A_n - g_5^2 \frac{e^{ - \Phi}}{z^3} \chi^2(z) A_n &= 0, \\
\partial_z \left( \frac{e^{-\Phi}}{z} \partial_z \varphi \right)
+ g_5^2 \frac{\chi^2(z)e^{-\Phi}}{z^3} (\pi - \varphi) &= 0, \\
m_\pi^2 \partial_z \varphi
- g_5^2 \frac{\chi^2(z)}{z^2} \partial_z \pi &= 0 ,
\end{align}
with boundary conditions $A_n(0)=\partial_z A_n(\infty)=0$ and
$\varphi^a(0) =\partial_z \varphi^a(\infty) =\pi^a(0) =0$.
The wave functions $\varphi(z)$ of the pion and $A_n(z)$ of the axial-vector meson are normalized according to
\begin{align}
\label{nor1}
\int_0^\infty dz\, \frac{e^{-\Phi}}{g_5^2}
\left[
\frac{(\partial_z \varphi)^2}{z}
+ \frac{g_5^2 \chi^2 (\pi-\varphi)^2}{z^3}
\right] = 1,\\
\label{Anor}
\int_0^\infty dz\, \frac{e^{-\Phi}}{g_5^2}
\frac{A_n
(z)^2}{z}= 1.
\end{align}
For the $2+1$-flavor soft-wall model, the above equations remain valid after replacing
$\chi(z)$ by $\sqrt{2}\,\chi_u(z)$. Since the derivation of these equations has been discussed extensively in the literature, we do not repeat it here and refer the reader to Refs.~\cite{Fang:2016nfj,Fang:2019lmd} for details. The $a_1$ and pion decay constants are then computed from their normalized wave functions as
\begin{align}
f_{a_1}
&= \frac{1}{g_5^2}\frac{1}{m_{a_1}}
\left.
\frac{e^{-\Phi}\partial_z A_n(z)}{z}
\right|_{z\to0},\\
f_\pi
&= \frac{1}{g_5^2}
\left.
\frac{e^{-\Phi}\partial_z \varphi(z)}{z}
\right|_{z\to0}.
\end{align}

Our QCD conventions for the vector and axial-vector current matrix elements are
\begin{align}
\langle0|J_{V_\mu}^a(0)|\rho_n^b(p,\lambda)\rangle
&=\delta^{ab}F_{\rho_n}\,\epsilon_\mu^{(\lambda)},\\
\langle0|J_{A_\mu}^a(0)|a_{1,n}^b(p,\lambda)\rangle
&=\delta^{ab}f_{a_{1,n}}m_{a_{1,n}}\,\epsilon_\mu^{(\lambda)}.
\end{align}
This leads to the explicit factor $1/m_{a_{1}}$ in the formula for $f_{a_1}$. For the pion, the axial-current matrix element is proportional to $f_\pi p_\mu$, and therefore no analogous factor $1/m_\pi$ appears. The powers of $g_5$ appearing in the axial-vector and pseudoscalar decay constants differ from that in the vector-meson expression above because
different normalization conventions are adopted for the corresponding normalizable bulk modes. 

To solve the above EOMs, we employ two independent numerical approaches implemented in \textit{Mathematica}. The first is the conventional shooting method, which transforms the boundary-value problem into an initial-value problem. The second combines a spectral method with the \texttt{Eigensystem} routine and is numerically more stable; both methods give consistent results. The calibration is performed without a statistically weighted global $\chi^2$. First, the inputs \(m_{\rho,0}=775~\mathrm{MeV}\), \(m_{\pi^0,0}=135~\mathrm{MeV}\), and \(f_\pi=92.4~\mathrm{MeV}\) are imposed in all three models. The remaining freedom is adjusted phenomenologically to obtain an acceptable simultaneous description of the displayed low-lying radial \(\rho\), \(\pi\), and \(a_1\) spectra; no experimental-error weights are assigned to uncertain resonance identifications. The quantities \(F_\rho\) and \(f_{a_1}\) are predictions rather than fitting inputs. With this prescription the parameter sets are
\begin{itemize}
\item SW1: $\mu_g=473~\mathrm{MeV}$, $\lambda_g=1.7$, $\mu_c=382.5~\mathrm{MeV}$,
$\sigma^{1/3}=293~\mathrm{MeV}$, $m_q=3.218~\mathrm{MeV}$;
\item SW2: $\mu_g=387.5~\mathrm{MeV}$, $\lambda=65.72$, $\mu_c=1489~\mathrm{MeV}$,
$m_q=3.265~\mathrm{MeV}$;
\item SW3: $\mu_g=480~\mathrm{MeV}$, $\lambda=110$, $\mu_c=1191~\mathrm{MeV}$,
$m_u=3.38~\mathrm{MeV}$, $m_s=98~\mathrm{MeV}$, and $\gamma=-25.25$.
\end{itemize}

For SW1, the dilaton parameters $\mu_g$ and $\lambda_g$ are determined jointly from the vector sector. The parameter $\mu_g$ mainly sets the overall mass scale and the average radial slope, while $\lambda_g$ controls the intermediate-$z$ deformation and thus strongly affects the low-lying level spacings. We fix the ground-state $\rho$-meson mass to its physical value and use the displayed excited $\rho$ masses to constrain these two parameters. In the chiral sector, $m_q$, $\sigma$, and $\mu_c$ are varied simultaneously to reproduce $m_{\pi^0}$ and $f_\pi$ while providing a satisfactory description of the displayed low-lying $\pi$ and $a_1$ spectra. For SW2, the purely quadratic dilaton renders the vector equation independent of the scalar background and yields $m_{\rho,n}^2=4\mu_g^2(n+1)$. The original choice $\mu_g=440~\mathrm{MeV}$ therefore gives $m_{\rho,0}=880~\mathrm{MeV}$. In the present calibration, fixing the ground-state vector mass to its physical value gives $\mu_g=m_\rho/2=387.5~\mathrm{MeV}$. The parameters $m_q$, $\mu_c$, and $\lambda$ are then adjusted to reproduce the neutral-pion mass and $f_\pi$ while retaining a satisfactory pseudoscalar spectrum. The SW3 parameters differ from those used in Ref.~\cite{Fang:2019lmd} because the present calculation simultaneously reproduces the neutral-pion mass and $f_\pi$ and imposes the infrared-suppression condition in Eq.~\eqref{IRsuppressionCondition}. This additional condition restricts the allowed SW3 parameter sets without compromising the phenomenological description of the low-lying spectrum.

Table~\ref{table1} summarizes the $\rho$, $a_1$, and pion mass spectra obtained in the three models, with asterisks denoting quantities used as input. Overall, the improved soft-wall models agree well with the experimental data, except for the excited vector-meson spectrum in SW2, which deviates from experiment once the ground-state mass is fixed. Table~\ref{table1} also presents selected vector-meson decay constants for several radial excitations, which are used below to calculate the vector-current two-point correlation function. However, all three models predict ground-state $\rho$-meson decay constants that are systematically lower than the experimental value, $F_{\rho}^{1/2}\simeq 348(1)~\text{MeV}$~\cite{Leutgeb:2022cvg}. Each model yields the ground-state pion decay constant $f_\pi=92.4~\text{MeV}$. The corresponding ground-state $a_1$-meson decay constants are $f_{a_1}^{\mathrm{SW1}}=133~\text{MeV}$, $f_{a_1}^{\mathrm{SW2}}=153~\text{MeV}$, and $f_{a_1}^{\mathrm{SW3}}=120~\text{MeV}$.

\begin{table}[htbp]
  \centering
  \resizebox{0.95\textwidth}{!}{\begin{tabular}{c c c c c c c c c}
    \hline\hline
    Model &  & 0 & 1 & 2 & 3 & 4 & 5 & ~~6 \\
    \hline
    \multirow{3}{*}{Exp. [MeV]}
    & $m_\rho$  & $775.26\pm0.25$  & $1465\pm25$ & 1720$\pm$20 & 1909$\pm$17 & 2150$\pm$40 & - & ~~ - \\
    & $m_\pi$  & 135  & 1300$\pm$100 & 1812$\pm$12 & 2070$\pm$35 & 2360$\pm$25 & - &~~ - \\
    & $m_{a_1}$  & 1230$\pm$40  & $1655 \pm 16$ & $1930^{+30}_{-70}$ & $2096 \pm 17$ & $2270^{+55}_{-40}$ & - &~~ - \\
    \cline{2-9}
    \multirow{4}{*}{SW1}
    & $m_\rho$  & $775^\star$  & 1354 & 1667 & 1919 & 2140 & 2339 & ~~ 2523 \\
    & $F_\rho^{1/2}$ & 329.4 & 365.5 & 405 & 439.4 & 468.3 & 492.7 & ~~ 513.8 \\
    & $m_\pi$  & $135^{\star}$  & 1465 & 1812 & 2067 & 2285 & 2481 & ~~ 2660 \\
    & $m_{a_1}$  & 1097  & 1625 & 1912 & 2144 & 2349 & 2536 & ~~ 2709 \\
    \cline{2-9}
    \multirow{4}{*}{SW2}
    & $m_\rho$  & $775^\star$  & 1096 & 1342 & 1550 & 1733 & 1898 & ~~ 2050 \\
    & $F_\rho^{1/2}$ & 260  & 309.2 & 342.2 & 367.7 & 388.8 & 406.9 & ~~ 422.9 \\
    & $m_\pi$  &  $135^\star$  & 1649 & 1890 & 2061 & 2212 & 2351 & ~~ 2480 \\
    & $m_{a_1}$  & 1459  & 1774 & 1952 & 2109 & 2252 & 2385& ~~ 2511 \\
    \cline{2-9}
    \multirow{4}{*}{SW3}
    & $m_\rho$  & $775^\star$  & 1335 & 1714 & 1986 & 2201 & 2393 & ~~ 2572 \\
    & $F_\rho^{1/2}$ & 308  & 405.5 & 448.5 & 465 & 482.6 & 503.4 & ~~ 523 \\
    & $m_\pi$  &  $135^\star$  & 1476 & 1854 & 2122 & 2329 & 2512 & ~~ 2684 \\
    & $m_{a_1}$  & 1168  & 1583 & 1915 & 2163 & 2362 & 2542& ~~ 2712 \\
    \hline\hline
  \end{tabular}}
  \caption{The \(\rho\)-meson, pion, and \(a_1\)-meson mass spectra, together with the corresponding \(\rho\)-meson decay constants \(F_\rho^{1/2}\), calculated in units of MeV within different holographic QCD models. Entries marked by an asterisk are used as input. Experimental values are taken from the suggested quark-model assignments for observed light mesons \cite{ParticleDataGroup:2024cfk}.}
  \label{table1}
\end{table}

\section{The LO-HVP contribution of $a_\mu$}\label{sec3}

To set the stage, we introduce the HVP function, which plays a central role in the leading-order HVP (LO-HVP) contribution to the muon $(g-2)_\mu$, following Refs.~\cite{Blum:2002ii,Aubin:2006xv}:
\begin{equation}\label{MI1}
a_\mu^{\text{LO-HVP}}
= 4\pi^2 \left(\frac{\alpha}{\pi}\right)^2
\int_0^{\infty} dQ^2 \, f(Q^2)\, \Pi^{\text{had}}_{\text{em}}(Q^2),
\end{equation}
where $Q^2=-q^2$ denotes the spacelike momentum squared, and
\begin{equation}
f(Q^2) = \frac{m_\mu^2 Q^2 Z^3 \left(1 - Q^2 Z\right)}{1 + m_\mu^2 Q^2 Z^2},
\qquad
Z = -\frac{Q^2 - \sqrt{Q^4 + 4 m_\mu^2 Q^2}}{2 m_\mu^2 Q^2}.
\end{equation}
Here $\Pi^{\text{had}}_{\text{em}}(Q^2)$ denotes the hadronic component of the electromagnetic vacuum polarization function, defined through the two-point correlator of electromagnetic currents evaluated at spacelike momentum $Q^2$, and renormalized such that $\Pi^{\text{had}}_{\text{em}}(0)=0$. In holographic QCD, it can be extracted from the renormalized vector-current correlator; in the flavor-symmetric case,
\begin{equation}
\Pi^{\text{had}}_{\text{em}}(-q^2)
= 2\, \mathrm{Tr}\!\left(Q_{\text{em}}^2\right)\,
  \Pi_V^{\text{ren}}(-q^2),
\end{equation}
where $Q_{\text{em}} = \mathrm{diag}\!\left(\frac{2}{3},-\frac{1}{3},-\frac{1}{3},\ldots\right)$ is the quark charge matrix. Consequently, for the two-flavor and three-flavor cases, one obtains \cite{Hong:2009jv}
\begin{equation}\label{PiV}
\Pi^{\text{had}}_{\text{em}(N_f=2)}(-q^2)
= \frac{10}{9}\,\Pi_V^{\text{ren}}(-q^2),
\qquad
\Pi^{\text{had}}_{\text{em}(N_f=3)}(-q^2)
= \frac{4}{3}\,\Pi_V^{\text{ren}}(-q^2).
\end{equation}
In this work, we restrict our analysis to the $N_f = 2$ case.

The vector current two-point function $\Pi_V(Q^2)$ is defined as
\begin{equation}
i \int d^4x \, e^{iqx}
\langle 0 | T \{ J_V^{a\mu}(x) J_V^{b\nu}(0) \} | 0 \rangle
= \delta^{ab} \left( q^2 \eta^{\mu\nu} - q^\mu q^\nu \right)
\Pi_V(Q^2).
\end{equation}
 For a general dilaton background, the vector current two-point function is obtained by evaluating the on-shell vector action at the UV boundary and adding the standard holographic counterterm that subtracts the logarithmic divergence. We define the renormalized correlator as
\begin{equation}\label{2pf1}
\Pi_V^{\text{ren}}(-q^2)
= -\frac{1}{g_5^2}
\left[\frac{e^{-\Phi}\partial_z V(q,z)}{q^2 z}-\ln(\mu z)+C\right]_{z=\epsilon},
\end{equation}
where the limit $\epsilon \to 0$ is implied, $\mu$ is the renormalization scale, and $C$ denotes a finite subtraction constant associated with the renormalization scheme, corresponding to the non-uniqueness of the finite counterterms in holographic renormalization. The constant is fixed by the subtraction condition \(\Pi_V^{\mathrm{ren}}(0)=0\) used in the HVP integral. $V(q,z)$ denotes the vector bulk-to-boundary propagator, which is determined by solving the EOM for the transverse component of the gauge field,
\begin{equation}\label{Vqz}
\partial_z \left( \frac{e^{-\Phi}}{z} \partial_z V_\mu^a(q,z) \right)
+ \frac{q^2 e^{-\Phi}}{z} V_\mu^a(q,z)
= 0,
\end{equation}
subject to the boundary conditions $V(q,0)=1$ and $\partial_z V(q,\infty)=0$. Equations~(\ref{2pf1}) and (\ref{Vqz}) are not special to SW2; they apply to the vector sector of all three backgrounds. The special feature of SW2 is only that its quadratic dilaton makes the bulk-to-boundary solution analytic. Since the vector sector of this model coincides with that of the original soft-wall model, an analytic solution for $V(q,z)$ can be obtained \cite{Grigoryan:2007my},
\begin{equation}
V(q,z)
= \Gamma\!\left(1 - \frac{q^2}{4\mu_g^2}\right)
U\!\left(-\frac{q^2}{4\mu_g^2},\,0,\,\mu_g^2 z^2\right),
\end{equation}
where $\Gamma$ is the Gamma function and $U$ denotes the Tricomi confluent hypergeometric function. Substituting this expression into Eq.~(\ref{2pf1}), expanding in the small parameter $\epsilon$, and performing the renormalization using $\psi(1)=-\gamma$, one obtains
\begin{equation}
\Pi_V^{\text{ren}}(Q^2)
= \frac{1}{2g_5^2}
\left[
\psi\!\left(\frac{Q^2}{4\mu_g^2}+1\right)
+ \gamma
\right],
\end{equation}
where $\psi(z)=\frac{d}{dz}\ln\Gamma(z)$ is the digamma function.

For the SW1 and SW3 models, the increased functional complexity prevents us from deriving a closed-form analytic expression. One may solve Eq.~(\ref{Vqz}) numerically for each momentum; in practice we use the equivalent low-energy spectral representation, which is numerically more stable for the HVP kernel-dominated region and makes direct use of the vector masses and decay constants listed in Table~\ref{table1}. Therefore, following Refs.~\cite{Hong:2009jv,Leutgeb:2022cvg}, we evaluate the renormalized vector correlator, with the condition $\Pi_V^{\text{ren}}(0)=0$, using
\begin{equation}\label{PiV2}
\Pi_V^{\text{ren}}(-q^2)
= \sum_{n=1}^{l}
\frac{q^2 F_{\rho_n}^2}
{(q^2 - m_{\rho_n}^2)\, m_{\rho_n}^4}
+ \mathcal{O}\!\left(\frac{q^2}{m_{\rho_{l+1}}^2}\right),
\end{equation}
where \(m_{\rho_n}\) denotes the mass of the \(n\)-th vector meson and \(F_{\rho_n}^{1/2}\) the corresponding decay constant. This representation is a low-energy approximation to the renormalized correlator. It does not reproduce the perturbative logarithm of \(\Pi_V^{\mathrm{ren}}(Q^2)\) at asymptotically large \(Q^2\), and we therefore use it only in the momentum region relevant for the LO-HVP kernel, which is strongly weighted toward low \(Q^2\). We have tested the approximation in the hard-wall and SW2 models, where analytic correlators are available. With a truncation \(l=100\), the resulting \(a_\mu^{\mathrm{LO-HVP}}\) differs from the analytic SW2 result by less than \(1\%\)\footnote{For SW2, \(a_\mu\times10^{10}=274.6\) from the \(l=100\) resonance sum, compared with \(276.4\) from the analytic renormalized correlator, a relative difference of \(0.66\%\). For SW1, increasing the cutoff from \(l=80\) to \(90\) and \(100\) gives \(482.2\), \(482.4\), and \(482.5\), respectively.}. We therefore set \(l=100\) throughout the HVP calculation.

\begin{table}[htbp]
  \centering
  \begin{tabular}{ccccccc}
    \hline\hline
     Model & SW1 & SW2 & SW3 & HW & SW & TC (fit 1|2) \\
    \hline
     $a_{\mu(N_f=2)}^{\text{LO-HVP}}\times10^{10}$   & 482.5  &  276.4  & 404.0 & 476.9 & 276.4 & 442.3|403.6 \\
    \hline\hline
  \end{tabular}
  \caption{The LO-HVP contribution to $a_\mu$ for $N_f=2$ obtained from different holographic QCD models. The results for the hard-wall (HW), original soft-wall (SW), and tachyon condensation (TC) models are taken from Ref.~\cite{Leutgeb:2022cvg}.}
  \label{table2}
\end{table}

\begin{figure}[htbp]
    \centering   \includegraphics[width=0.6\linewidth]{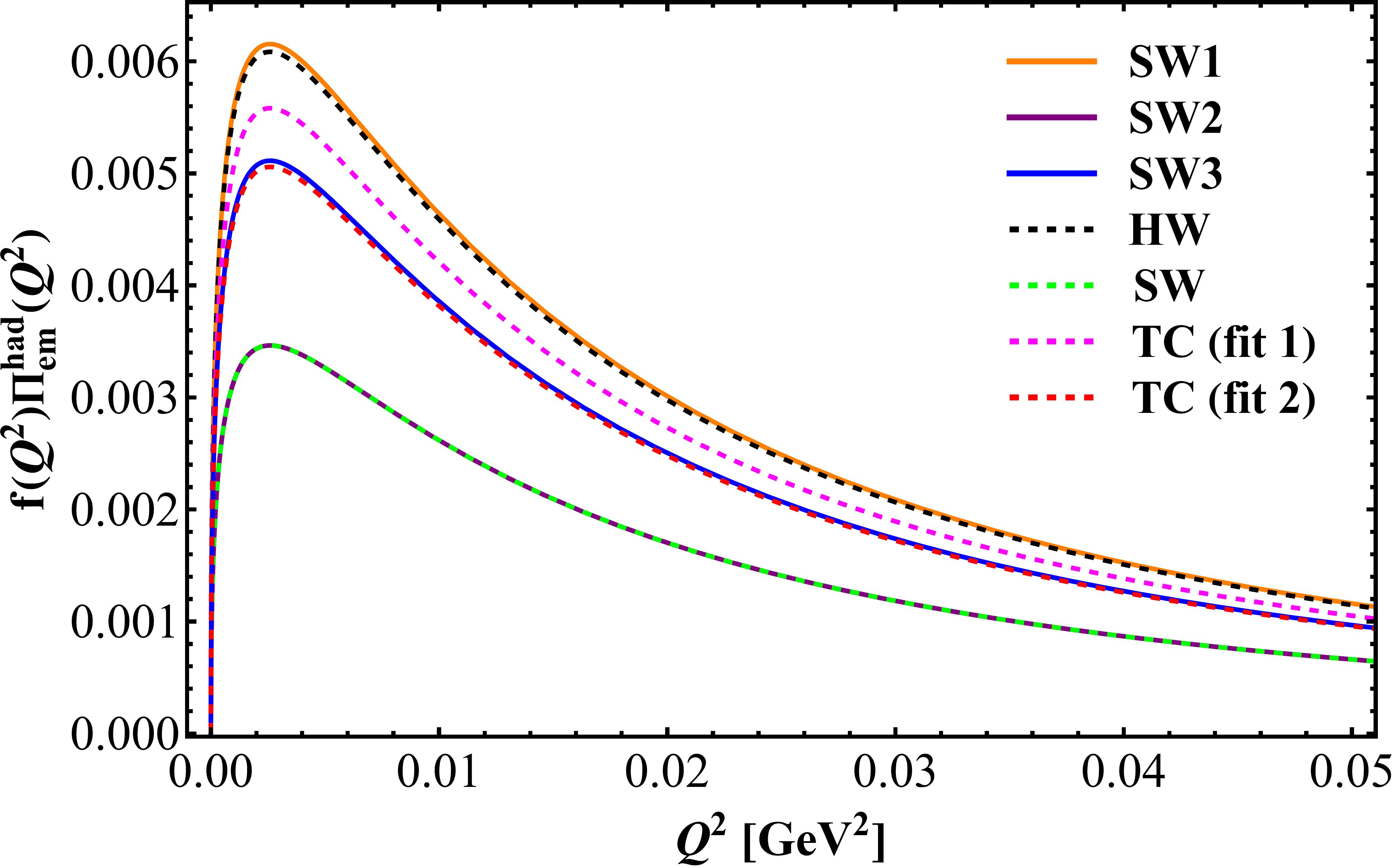}
    \caption{The integrand $f(Q^2)\Pi^{\text{had}}_{\text{em}}(Q^2)$ (in GeV$^{-2}$) as a function of $Q^2$ computed in different models: SW1 (orange solid line), SW2 (purple solid line), SW3 (blue solid line), the hard-wall model (black dashed line), the original soft-wall model (green dashed line), and the tachyon condensation model (magenta and red dashed lines).}
    \label{fig:2}
\end{figure}

Substituting the vector-current correlator into Eq.~(\ref{MI1}), we compute the $N_f=2$ LO-HVP contribution to the anomalous magnetic moment of the muon in the various improved soft-wall models, with the results summarized in Table~\ref{table2}. For comparison, we also quote the corresponding results obtained in the hard-wall model~\cite{Erlich:2005qh,DaRold:2005mxj}, the original soft-wall model~\cite{Karch:2006pv}, and the tachyon-condensation model~\cite{Casero:2007ae,Iatrakis:2010zf,Iatrakis:2010jb}. Figure~\ref{fig:2} further shows the integrand of Eq.~(\ref{MI1}) as a function of $Q^2$. SW1 yields the largest contribution, slightly exceeding the hard-wall result, whereas the SW3 prediction lies in an intermediate range and nearly overlaps with that of the TC model (fit 2). The SW2 result coincides with that of the original soft-wall model and is the smallest among the models considered.


The smaller SW2 result originates from its vector/dilaton sector rather than from its scalar sector. Within the leading-order treatment adopted here, the LO-HVP contribution probes the vector-current correlator determined by the transverse vector equation, which depends only on the dilaton profile $\Phi(z)$ and has no direct dependence on $\chi(z)$, $m_5^2(z)$, or $\lambda_X(z)$. The principal scalar-sector improvement of SW2 therefore does not enter the present HVP calculation directly. Because SW2 employs a purely quadratic dilaton, it has the same vector background as the original soft-wall model. Once the physical ground-state $\rho$-meson mass is fixed, the entire radial trajectory is determined, yielding a radial slope smaller than that favored by the excited-state spectrum. The resulting lower excited vector-meson masses are accompanied by an underestimated ground-state pole residue $F_\rho$. Since the muon HVP kernel strongly weights the low-Euclidean-momentum region, this reduced low-energy pole strength directly suppresses the LO-HVP contribution. By contrast, SW1 and SW3 retain additional intermediate-$z$ structure in their dilaton profiles, introduced in their source models to improve the low-lying vector spectrum.


It has been argued that the two-flavor holographic QCD predictions are most appropriately compared with the contributions from the two- and three-pion states as well as the $\pi^0\gamma$ intermediate state~\cite{Leutgeb:2022cvg}. The corresponding dispersive determinations yield $(558 \pm 4)\times 10^{-10}$~\cite{Donoghue_Golowich_Holstein_2014} and $(555 \pm 3)\times 10^{-10}$~\cite{Davier:2019can}, while resonance chiral theory gives $(559.18 \pm 2.75)\times 10^{-10}$~\cite{Qin:2020udp,Wang:2023njt}. We find that all holographic QCD models exhibit sizable deviations from the dispersive and resonance chiral theory results for the LO-HVP contribution. Even in the most favorable case, the holographic prediction reaches only about $87\%$ of the dispersive value. This discrepancy may be attributed to the fact that the vector-meson decay constants predicted by these models remain systematically lower than their experimental counterparts \cite{Leutgeb:2022cvg}. To this end, we adjust the parameters of the SW1 model by setting $\lambda_g = 1.8$ and $\mu_g = 503.5~\text{MeV}$. At the expense of the agreement between the mass spectrum of higher vector meson excited states and experimental data, this adjustment yields a ground-state mass $m_{\rho} = 775.2~\text{MeV}$ and a decay constant $F_{\rho}^{1/2} = 347.2~\text{MeV}$, both of which are consistent with the experimental values. Recalculating the LO-HVP contribution with these adjusted parameters, we obtain $a_{\mu}^{\text{LO-HVP}} = 573.5 \times 10^{-10}$, which is slightly higher than the corresponding estimates from dispersion relations and resonance chiral theory.

\section{HADRONIC LIGHT-BY-LIGHT
CONTRIBUTION}\label{sec4}

\subsection{Pion transition form factors from the standard Chern-Simons term}

\begin{figure}[htbp]
    \centering   \includegraphics[width=0.5\linewidth]{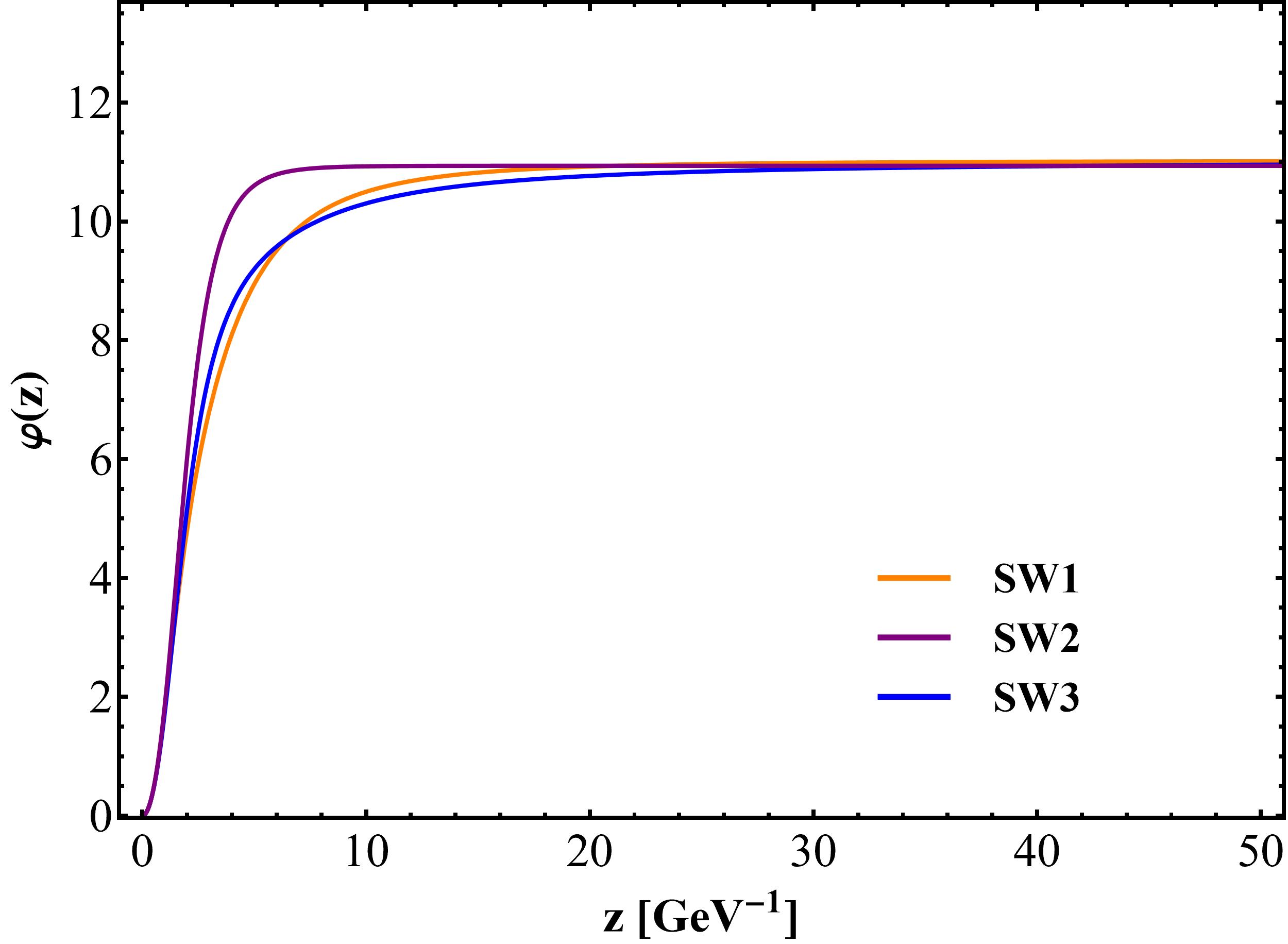}
    \caption{Normalized ground-state wave functions $\varphi(z)$ of $\pi_0$ in different holographic models.}
    \label{fig:22}
\end{figure}

To evaluate the pseudoscalar-meson contribution to the HLbL scattering part of the anomalous magnetic moment of the muon, we first consider the transition form factors (TFFs) of the neutral pion into two photons, defined by
\begin{equation}
i \int d^4 x \, e^{i q \cdot x}
\langle 0 | T \{ j_\mu(x) j_\nu(0) \} | \pi^0(p) \rangle
= \epsilon_{\mu\nu\rho\sigma} q^\rho p^\sigma
F_{\pi^0\gamma^\ast\gamma^\ast}(q^2,(p-q)^2),
\end{equation}
where $j_\mu(x)$ denotes the hadronic electromagnetic current. The holographic computation of $F_{\pi^0\gamma^\ast\gamma^\ast}$ leads to the standard expression \cite{Grigoryan:2007wn,Kwee:2007dd,Grigoryan:2008up,Zuo:2011sk,Brodsky:2011xx},
\begin{equation}
F(Q_1^2,Q_2^2)
= -\frac{N_c}{12\pi^2}\, K(Q_1^2,Q_2^2).
\end{equation}
The function $K(Q_1^2,Q_2^2)$ is derived from the standard Chern-Simons action, which is required to incorporate the chiral anomaly. The standard Chern-Simons action is given by
\begin{align}
S_{CS}=S_{CS}^L-S_{CS}^R,
\end{align}
where
\begin{align}
S_{CS}^{L{(R)}}=\frac{N_c}{24\pi^2}\int\text{Tr}\left[\mathcal{AF}^2-\frac{i}{2}\mathcal{A}^3\mathcal{F}-\frac{1}{10}\mathcal{A}^5\right]^{L(R)}.
\end{align}
Evaluated from this action, $K(Q_1^2,Q_2^2)$ can be expressed as an integral over the product of the bulk-to-boundary propagators for the two photons and the pion wave function:
\begin{equation}
K(Q_1^2,Q_2^2)
= -\int_0^\infty V(Q_1,z)\, V(Q_2,z)\, \partial_z \varphi(z)\, dz.
\end{equation}
Here, $V(q,z)$ is determined by solving Eq.~(\ref{Vqz}), and $\varphi(z)$ denotes the normalized pion wave function according to Eq.~(\ref{nor1}). Fig.~\ref{fig:22} shows the normalized ground-state wave functions of $\pi^0$ calculated from different models. It can be observed that the values of  wave functions exhibit certain discrepancies at small radial coordinate $z$, whereas as $z$ increases, these solutions approach a common constant value. Numerically, this asymptotic value is comparable to $1/f_\pi$, but does not coincide with it exactly. This deviation reflects the effect of explicit chiral symmetry breaking at finite quark mass, which modifies the normalization of the wave function.

For the evaluation of $F(Q_1^2,Q_2^2)$, we follow the convention of Ref.~\cite{Colangelo:2023een}, noting that the expression given in Ref.~\cite{Leutgeb:2019zpq} differs slightly. This discrepancy originates from different normalization conventions for the pion wave function. If we adopt the following normalization condition for the SW1 and SW2 models\footnote{For the SW3 case, one simply replaces $\chi$ with $\sqrt{2}\chi_u$.}:
\begin{equation}
\int_0^\infty dz \,\frac{e^{-\Phi}}{g_5^2 f_\pi^2}
\left(\frac{(\partial_z \varphi)^2}{z}
+ \frac{g_5^2 \chi^2 (\pi - \varphi)^2}{z^3}
\right) = 1,
\end{equation}
the wave function approaches unity at large $z$. With this normalization, one may equivalently follow the procedure of Ref.~\cite{Leutgeb:2019zpq} to compute $F(Q_1^2,Q_2^2)$.

\begin{figure}[htbp]
    \centering   \includegraphics[width=0.5\linewidth]{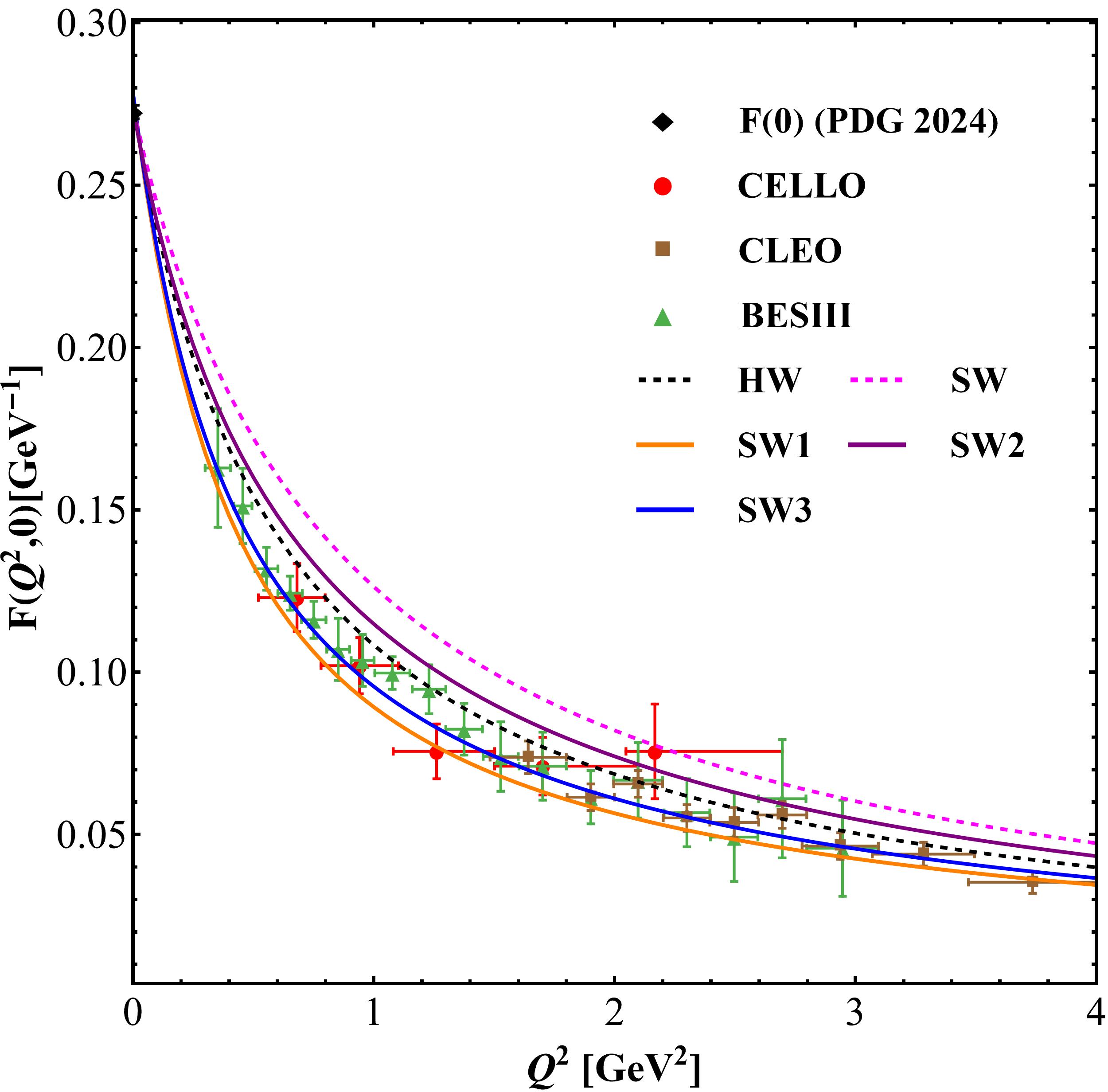}
    \caption{Comparison of the $Q^2$ dependence of $F(Q^2,0)$ computed in various holographic models with experimental data from CELLO, CLEO, and BESIII~\cite{Danilkin:2019mhd}.}
    \label{fig:3}
\end{figure}

\begin{figure}[htbp]
    \centering   \includegraphics[width=0.5\linewidth]{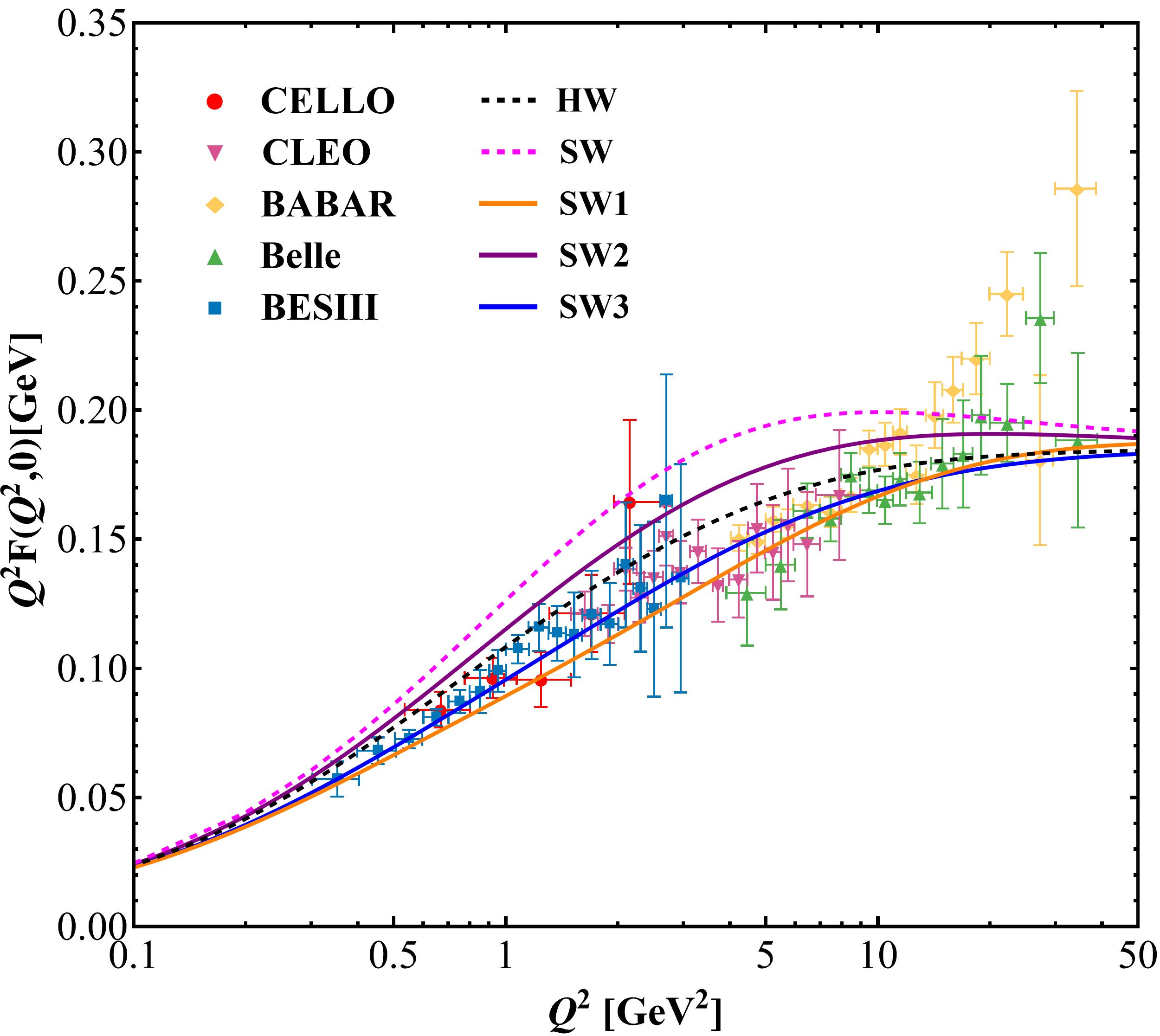}
    \caption{The singly-virtual form factor $Q^2F(Q^2,0)$ calculated in the holographic models, compared with the experimental data \cite{CELLO:1990klc,CLEO:1997fho,BaBar:2009rrj,Belle:2012wwz,Redmer:2018uew}.}
    \label{fig:4}
\end{figure}

In Fig.~\ref{fig:3}, we compare the spacelike $\pi^{0}$ TFFs obtained from various holographic models with the experimental data compiled in Ref.~\cite{Danilkin:2019mhd}, focusing on the momentum range $0 \le Q^{2} \le 4~\mathrm{GeV}^{2}$; for reference, we also display the results from the hard-wall model and the standard soft-wall model~\cite{Colangelo:2023een}. At $Q^{2}=0$, the three models give $F(0,0) =0.2782~\mathrm{GeV}^{-1}$ for SW1, $F(0,0)=0.277~\mathrm{GeV}^{-1}$ for SW2, and $F(0,0)=0.2786~\mathrm{GeV}^{-1}$ for SW3. In the $Q^{2}$ range displayed in the figure, all three improved soft-wall models show good agreement with the experimental data, with the SW2 curve lying slightly above that of the hard-wall model, whereas the SW1 and SW3 results are somewhat lower.
Fig.~\ref{fig:4} shows the $Q^{2}$ dependence of the singly-virtual TFFs computed in different holographic models over a broad kinematic range. As can be seen, the form factors obtained from the three improved soft-wall models are consistent with the available experimental data. At large $Q^{2}$, the asymptotic Brodsky--Lepage behavior is correctly reproduced, with the quantity $Q^{2}F(Q^{2},0)$ approaching the limit $2f_{\pi}$~\cite{Lepage:1980fj}. Furthermore, no significant rise is observed for $Q^{2} > 10~\mathrm{GeV}^{2}$, in contrast to the trend reported by the BaBar Collaboration.

\begin{figure}[htbp]
    \centering   \includegraphics[width=0.5\linewidth]{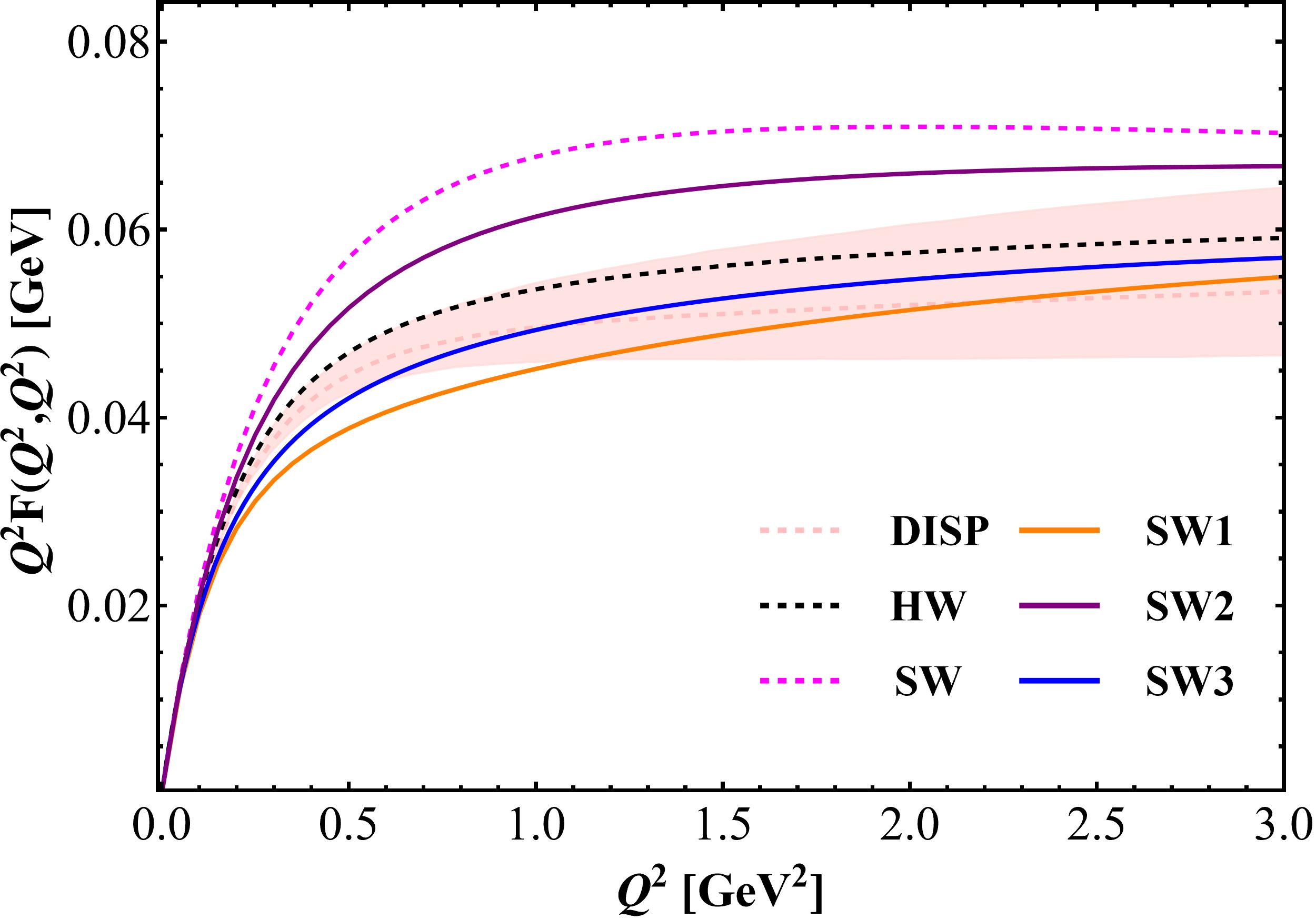}
    \caption{The $Q^{2}$ dependence of the diagonal form factor $Q^2F(Q^2,Q^2)$ in the low-$Q^{2}$ region, where the light-red error band denotes the result obtained from the dispersive approach~\cite{Hoferichter:2018kwz}.}
    \label{fig:5}
\end{figure}

In the momentum-symmetric case $Q_{1}^{2}=Q_{2}^{2}$, the diagonal $\pi^{0}$ TFFs computed from different holographic models are shown in Fig.~\ref{fig:5}, together with the results and uncertainties obtained from the dispersive approach. The SW1 and SW3 models lie closest to the dispersive determination and remain slightly below the prediction of the hard-wall model, whereas the SW2 model yields comparatively larger values than both the other holographic models and the dispersive result.

\begin{figure}[htbp]
    \centering   \includegraphics[width=0.4\linewidth]{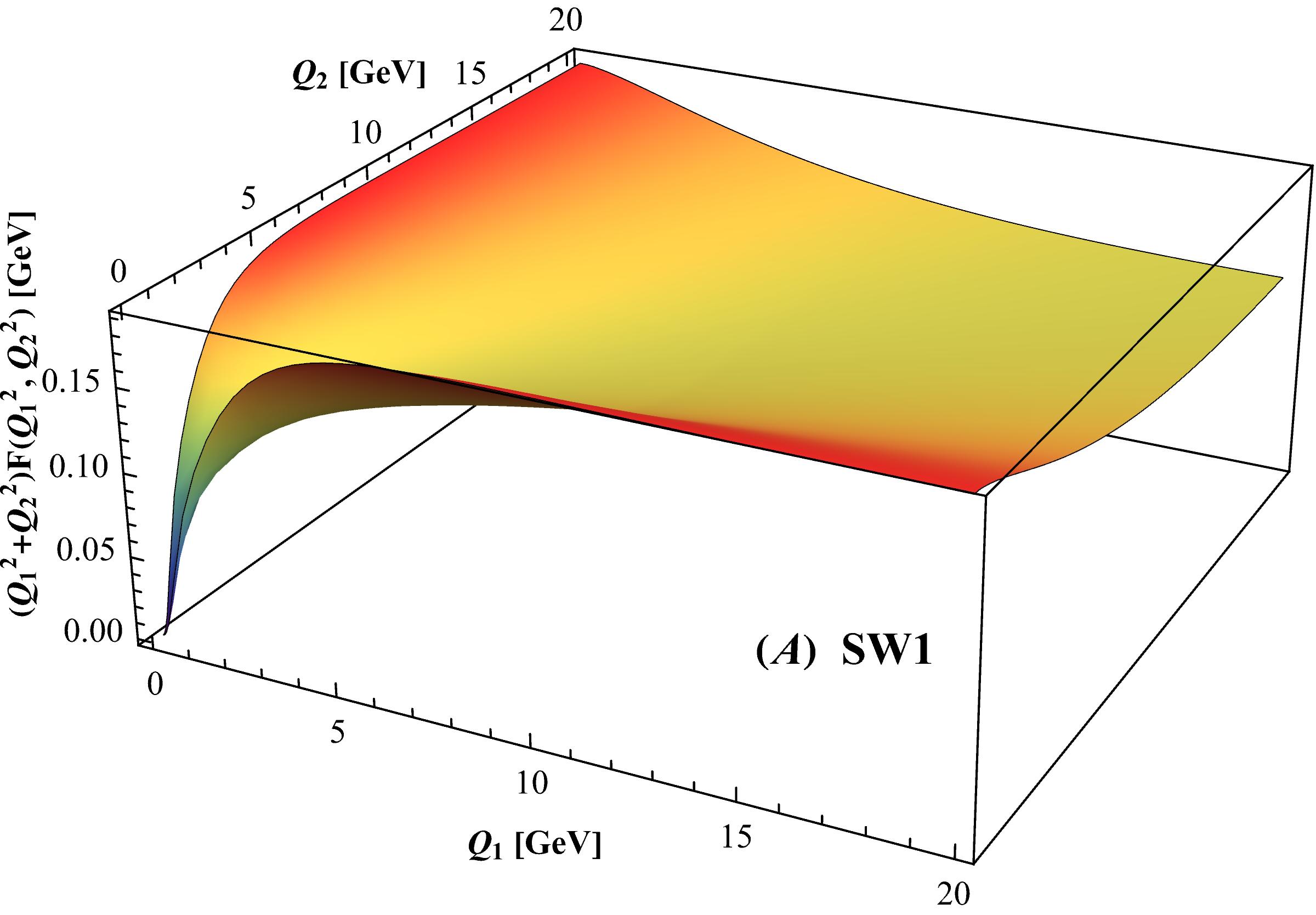}
    \includegraphics[width=0.4\linewidth]{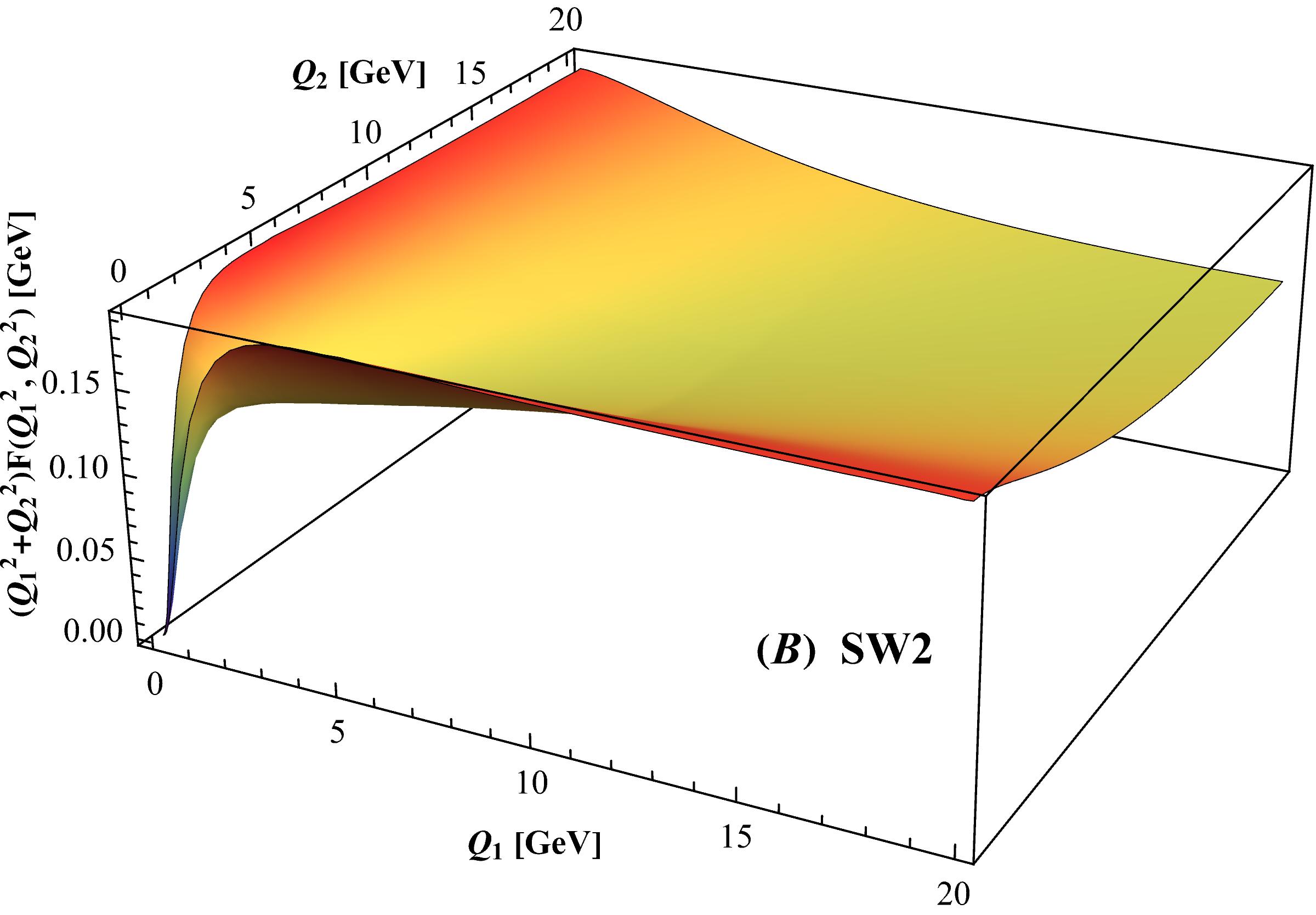} \includegraphics[width=0.4\linewidth]{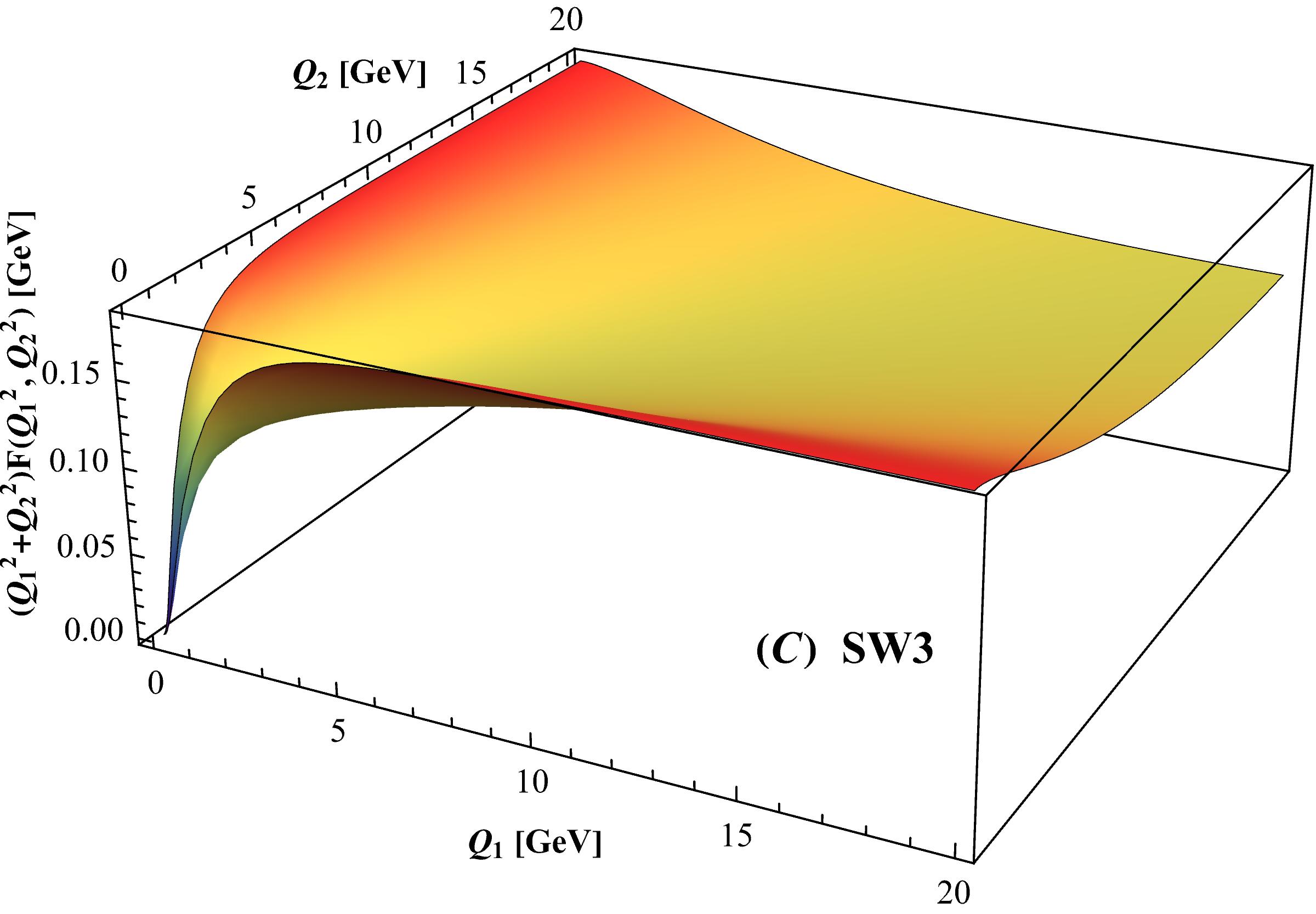}
    \caption{Three-dimensional representation of $(Q_1^2 + Q_2^2)F(Q_1^2, Q_2^2)$ at large momentum $Q$ in different models.}
    \label{fig:6}
\end{figure}

Furthermore, Fig.~\ref{fig:6} presents the three-dimensional representation of $\pi_0$ TFFs obtained through different models in the case of large momentum Q. A prominent feature of the holographic QCD framework is its capacity to extend the calculation into the deep Euclidean region ($Q^2 \gg \Lambda_{\text{QCD}}^2$) with high numerical stability. Unlike some phenomenological models that rely on truncated expansions or limited momentum parameterizations, the holographic approach naturally incorporates the correct high-energy scaling. This capability allows for a more reliable determination of the TFFs over an exceptionally broad range of $Q$. Consequently, this reduces the reliance on phenomenological extrapolations at high momenta.

\subsection{Numerical results for the ground-state pseudoscalar-pole HLbL contribution}

By computing the $\pi^{0}\!\to\!\gamma^\ast\gamma^\ast$ transition form factor within the holographic framework and inserting the result into the standard expression for the pseudoscalar-exchange contribution, we are able to evaluate the pole contribution to $a_{\mu}^{\mathrm{HLbL}}$ in our holographic QCD models. According to Ref.~\cite{Nyffeler:2016gnb}, the expression takes the form
\begin{equation}\label{amup}
\begin{split}
a_{\mu}^{\mathrm{HLbL}}
= &\left( \frac{\alpha}{\pi} \right)^{3}
\int_{0}^{\infty} dQ_{1}
\int_{0}^{\infty} dQ_{2}
\int_{-1}^{1} d\tau \,
\Big(
w_{1}(Q_{1},Q_{2},\tau)\,
F(Q_{1}^{2}, Q_1^2+Q_2^2+2\tau Q_1Q_2)\,
F(Q_{2}^{2}, 0)
\\
&+\, w_{2}(Q_{1},Q_{2},\tau)\,
F(Q_{1}^{2},Q_{2}^{2})\,
F(Q_1^2+Q_2^2+2\tau Q_1Q_2,0)
\Big)\,,
\end{split}
\end{equation}
where \(Q_1\) and \(Q_2\) are scalar Euclidean integration variables and \(\tau\) is the angle cosine. Hence the mixed momentum argument is written explicitly as \(Q_1^2+Q_2^2+2\tau Q_1Q_2\). This is the expression used in the numerical evaluation. The weight functions $w_{1}$ and $w_{2}$ are given in Appendix~A of Ref.~\cite{Nyffeler:2016gnb}. The integral is evaluated numerically in \textit{Mathematica} using the \texttt{LocalAdaptive} integration method, with integration ranges chosen as $Q\in[0,20\,\mathrm{GeV}]$. 
Although the integral is already well saturated for $Q > 5\,\mathrm{GeV}$, we extend the upper limit to $20\,\mathrm{GeV}$ 
 to ensure numerical stability. The resulting $\pi^{0}$-pole contributions to $a_{\mu}^{\mathrm{HLbL}}$ obtained from different holographic models are summarized in Table~\ref{table3}. As shown there, the SW1 and SW3 models yield very similar predictions for $a_\mu^{\pi^0}$, both of which are smaller than the result obtained in the hard-wall model, whereas the SW2 model gives a noticeably larger
 value. For comparison, Table~\ref{table3} also includes the dispersive approach result adopted in the White Paper and the latest determination by the BMW lattice QCD collaboration. Additional lattice calculations by the Mainz group give $a_{\mu}^{\pi^{0}\text{-pole}} = 59.7(3.6)\times 10^{-11}$~\cite{Gerardin:2019vio} and the Canterbury approximant method gives $a_{\mu}^{\pi^{0}\text{-pole}} = 63.6(2.7)\times 10^{-11}$~\cite{Masjuan:2017tvw}. More recently,
 resonance chiral theory has reported $a_{\mu}^{\pi^{0}\text{-pole}} \approx 61.6(1.8)\times 10^{-11}$~\cite{Zhang:2025ijd}. These independent determinations are mutually consistent within uncertainties, indicating that the pion-pole contribution to the HLbL correction is now relatively well constrained.

\begin{table}[htbp]
  \centering
  \begin{tabular}{c   c   c  c  c  c  c   c}
    \hline\hline
     $a_{\mu}^{\mathrm{HLbL}}\times10^{11}$ & SW1 & SW2& SW3& HW& SW& WP(2025) \cite{Aliberti:2025beg}& LQCD \cite{Gerardin:2023naa}\\
    \hline
     $a_{\mu}^{\pi^0}$  & 57.1  & 70.3  & 60.0 & 65.2  & 75.2 & $63.6^{+3.0}_{-2.5}$ & 57.8$\pm$1.8$\pm$ 0.9\\
    \hline\hline
  \end{tabular}
  \caption{The pole contributions of $\pi^0$ to $a_{\mu}^{\mathrm{HLbL}}$. ``WP'' denotes the white paper result~\cite{Aliberti:2025beg}, and ``LQCD'' refers to lattice QCD data from the BMW Collaboration~\cite{Gerardin:2023naa}.}
  \label{table3}
\end{table}

Following the procedure in Ref.~\cite{Leutgeb:2019zpq}, we estimate the contributions from the $\eta$ and $\eta^{\prime}$ mesons via a simple extrapolation based on the universal kernel $K(Q_1^2,Q_2^2)$. Assuming that the TFFs of the $\eta$ and $\eta'$ share the same momentum dependence as that of the $\pi^0$, we perform a global rescaling of the holographic result for $F_{\pi^0}(Q^2,Q^2)$ by matching the experimental central values~\cite{Danilkin:2019mhd}: $F_{\eta}(0,0)/F_{\pi^{0}}(0,0)=0.2736/0.2725$ and $F_{\eta^{\prime}}(0,0)/F_{\pi^{0}}(0,0)=0.3412/0.2725$. Subsequently, the mass of the $\pi^0$ in the numerical integration for the HLbL contributions is replaced with the physical masses of the $\eta$ and $\eta'$ mesons. Our numerical results for $(a_\mu^\eta,\,a_\mu^{\eta'})$ are
$(15.1,\,10.7)$, $(20.0,\,14.6)$, and $(16.1,\,11.5)$
for SW1, SW2, and SW3, respectively, in units of $10^{-11}$. These extrapolated values are consistent in both magnitude and hierarchy with full holographic calculations in hard-wall and soft-wall models~\cite{Leutgeb:2022lqw,Colangelo:2023een}, supporting the validity of this approximation as a reliable first estimate. Nevertheless, this extrapolation is not meant to replace a dedicated $(2+1)$-flavor calculation. Incorporating explicit strange-quark-mass corrections and the $U(1)_A$ anomaly remains a crucial step for a rigorous treatment, which is left for future work.

Compared with these reference results, the predictions obtained from the SW1 and SW3 models are lower than the values quoted in the White Paper and are in closer agreement with the lattice QCD results from the BMW Collaboration and the Mainz group, whereas the SW2 prediction is slightly higher than both the White Paper value and other data-driven estimates. By contrast, the result from the standard soft-wall model is significantly larger than those from the other holographic models as well as from alternative theoretical approaches. These findings clearly indicate that the holographic predictions for the pseudoscalar pole contribution are highly sensitive to the specific realization of the infrared dynamics. Although all three improved models successfully reproduce the ground-state mass and decay constant of the pion, the SW2 model retains the simple dilaton background of the original soft-wall setup, and its more limited parameter freedom leads to a slightly less accurate description of the excited-state mass spectrum compared to the other two models. This limitation is directly reflected in the calculation of the $\pi^0$ transition form factor: in the low- and intermediate-momentum regions that dominate the HLbL loop integral, the transition form factor predicted by SW2 exhibits noticeable deviations from those obtained in the SW1 and SW3 models.

Nevertheless, in comparison with the standard soft-wall model, which yields less satisfactory results, all three improved models demonstrate clear advantages in both the theoretical description of the TFFs and the evaluation of the HLbL contribution. These observations highlight a crucial point: within the framework of holographic soft-wall models, the introduction of a physically well-motivated infrared background and an accurate reproduction of the excited-state spectrum are essential for a reliable determination of the pseudoscalar TFFs and its contribution to the anomalous magnetic moment of the muon.

\subsection{Hadronic Contributions from the Scalar-extended Chern-Simons Term}

In the preceding two subsections, we employed the standard Chern-Simons action to calculate the pion transition form factor and the ground-state pseudoscalar contribution to the anomalous magnetic moment of the muon. Extending this framework to the infinite tower of excited states, however, leads to a convergence problem in soft-wall-type models. Recent studies have shown that the standard soft-wall construction produces unavoidable divergences in calculations of hadronic light-by-light scattering amplitudes~\cite{Leutgeb:2025jmv}. More specifically, the TFFs of highly excited pseudoscalar and axial-vector mesons grow exponentially with their masses. This behavior originates from the dilaton factor $e^{-\Phi}=e^{-c^2z^2}$ in the soft-wall background. After canonical normalization of the wave functions, the dilaton weighting effectively generates an exponentially growing factor $e^{\Phi/2}$ in the large-$z$ region. Consequently, neither the infinite sum over intermediate resonances nor the equivalent diagonal integration of bulk-to-bulk propagators in the five-dimensional formulation converges in the evaluation of the HLbL tensor.

To regulate this infrared behavior, we employ the string-inspired scalar-extended Chern--Simons action motivated by the brane--antibrane construction~\cite{Leutgeb:2024rfs}. In the superconnection formalism, the left- and right-handed flavor gauge fields and the bifundamental open-string tachyon are combined into a single superconnection. The resulting anomalous action contains tachyon-dependent terms together with the infrared suppression factor $e^{-2\pi\alpha'T_0^2}$. The open-string tachyon is the string-inspired counterpart of the bulk scalar field dual to $\bar q_Rq_L$, motivating the identification of $T_0$ with the chiral background: $T_0 \propto X_0$. The infrared growth of $T_0$ then provides the additional damping required to control the transition-form-factor overlap integrals. The corresponding quantitative convergence condition is derived below and verified for all three parameter sets in Appendix~\ref{app:convergence}.

Other infrared completions are also possible. For example, one may restrict the holographic coordinate to a finite interval, modify the scalar, warp-factor, or dilaton backgrounds to produce hard-wall-like endpoint behavior, or introduce sufficiently strong asymptotic suppression through other background fields. Fully dynamical constructions such as V-QCD combine a tachyon extension with backreacted geometries~\cite{Jarvinen:2011qe}. We adopt the scalar extension because it provides the required infrared damping without modifying the two-derivative background equations used to calibrate the SW1--SW3 meson sectors. It should therefore be regarded as one motivated infrared completion among several viable alternatives.

We now evaluate the pion transition form factor using the scalar-extended Chern-Simons term, which gives~\cite{Leutgeb:2024rfs}
\begin{align}
F(Q_1^2,Q_2^2)
&= \frac{N_c}{12\pi^2}\int dz\,\Bigl\{
e^{-2\pi\alpha'T_0^2}\,\partial_z\varphi\,
V(Q_1,z)V(Q_2,z)
\nonumber\\
&\qquad +\partial_z\!\left(e^{-2\pi\alpha'T_0^2}\right) (\varphi-\pi)\,V(Q_1,z)V(Q_2,z)
\Bigr\},
\end{align}
where $\alpha'$ is fixed by the string-inspired normalization $\alpha'=1/6$, and $T_0(z)$ denotes the background open-string tachyon appearing in the scalar-extended Chern--Simons action. In the brane-antibrane superconnection construction, this field is related to the light-flavor component of the scalar vacuum expectation value.

To apply the scalar-extended Chern-Simons construction consistently to all three backgrounds, we introduce the common notation
\begin{equation}\label{uniformX0}
X_0(z)
=
\frac{1}{2}
\operatorname{diag}\!\left(
v_\ell(z),\,v_\ell(z),\,v_s(z)
\right).
\end{equation}
The strange-flavor entry is omitted in the two-flavor models. The relation between this common notation and the conventions of the original model constructions is $v_\ell=\chi$ for SW1 and SW2, whereas $v_\ell=\sqrt{2}\,\chi_u$ and $v_s=\sqrt{2}\,\chi_s$ for SW3. Consequently, the light-flavor component is $X_{0,\ell}=\chi/2$ in SW1 and SW2 and $X_{0,\ell}=\chi_u/\sqrt{2}$ in SW3.

The background tachyon entering the scalar-extended Chern-Simons action is related to the light-flavor scalar component through
\begin{equation}\label{tachyonX0relation}
T_0^2(z) = \frac{g_5^2\pi}{2}\,X_{0,\ell}^2(z) = \frac{g_5^2\pi}{8}\,v_\ell^2(z).
\end{equation}
In all three models, the light-flavor scalar background grows linearly in the deep infrared, $v_\ell(z)\propto z$, and hence $T_0^2(z)\propto z^2$. The tachyon-dependent exponential can thus compensate for the dilaton-induced enhancement. A sufficient strict condition for infrared suppression is~\cite{Leutgeb:2025jmv}
\begin{equation}\label{IRsuppressionCondition}
2\pi\alpha'T_0^2(z) > \frac{\Phi(z)}{2},
\qquad z\to\infty .
\end{equation}
This condition ensures that the net exponential factor $\exp[-2\pi\alpha'T_0^2+\Phi/2]$ decays at large $z$. For the parameter sets used here, the difference $2\pi\alpha'T_0^2-\Phi/2$ has a strictly positive quadratic infrared coefficient in SW1, SW2, and SW3. The model-by-model asymptotic analysis and the proof of Hilbert-space convergence at fixed Euclidean kinematics are presented in Appendix~\ref{app:convergence}. Parameter sets that fail to satisfy this condition are excluded from the excited-state HLbL calculation.

\begin{figure}[htbp]
    \centering   \includegraphics[width=0.5\linewidth]{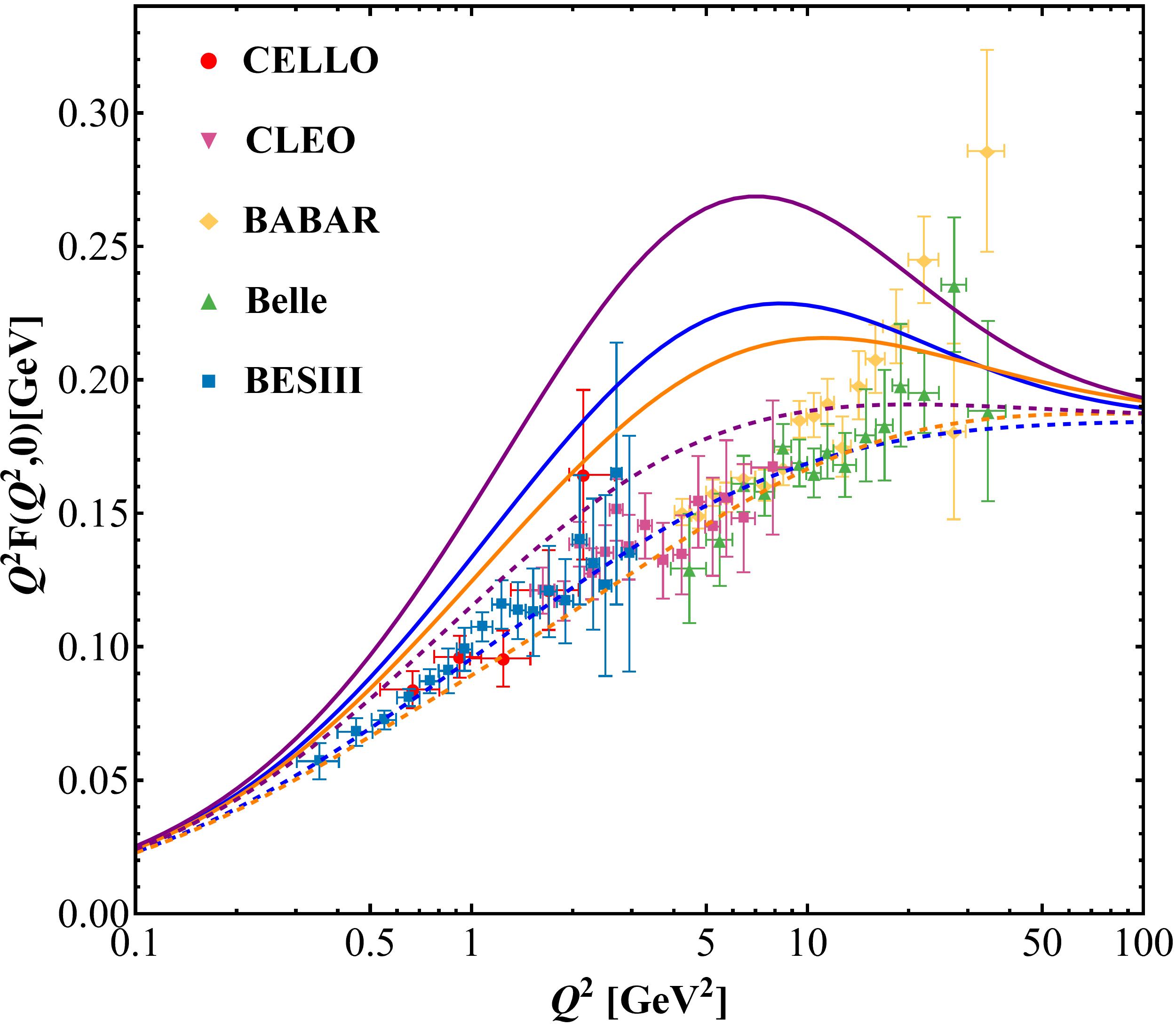}
    \caption{The single-virtual $\pi^0$ TFFs $Q^2F(Q^2,0)$ calculated in different infrared-improved soft-wall models. The solid curves represent the results obtained from the scalar-extended Chern-Simons term, while the dashed curves correspond to those derived from the standard Chern-Simons term. The orange curves correspond to the SW1 model, the purple curves to the SW2 model, and the blue curves to the SW3 model.}
    \label{PTFF1}
\end{figure}

Figure~\ref{PTFF1} shows the comparison of the $\pi^0$ single-virtual TFFs as a function of the momentum transfer squared $Q^2$ obtained from different improved soft-wall models. It can be clearly seen that in the low-momentum region, the results obtained from the two different Chern-Simons terms are very similar, as they are strongly constrained by the chiral anomaly. As $Q^2$ increases, however, the results obtained from the scalar-extended Chern-Simons term begin to deviate from the experimental data and are generally larger than those obtained from the standard Chern-Simons term. Nevertheless, in the large-$Q^2$ region, all results satisfy the Brodsky--Lepage asymptotic behavior, with the TFFs approaching the limiting value $2f_\pi$.

In addition, the transition form factors of axial-vector mesons can also be computed using the scalar-extended Chern-Simons term\footnote{With the standard Chern-Simons term, the infrared pathology affects both axial-vector modes and radially excited pseudoscalar modes. The pseudoscalar ground state is an exception because of its special wave-function structure~\cite{Leutgeb:2025jmv}.}:
\begin{align}
F_{A_n^a \gamma^\star \gamma^\star}(Q_1^2,Q_2^2)= \frac{N_c}{4\pi^2}\text{Tr}(Q^2_{em}t^a)\, A(Q_1^2,Q_2^2).
\end{align}
Here, the function $A(Q_1^2,Q_2^2)$ is given by
\begin{align}
A(Q_1^2,Q_2^2) = \frac{2}{Q_1^2} \int dz \, e^{-2\pi\alpha' T_0^2}\, A_n^{\perp}(z)\, \partial_z V(Q_1,z)\, V(Q_2,z),
\end{align}
where $A_n^{\perp}(z)$ denotes the normalized axial-vector wave function defined according to Eq.~(\ref{Anor}). The index $a=3,8,0$ leads to different values of the factor $\mathrm{Tr}(Q_{em}^2 t^a)$, thereby distinguishing between different flavor states.

From our numerical analysis, the values of the TFFs at zero momentum transfer for the ground-state axial-vector mesons, $A(0,0)$, are found to be
\begin{align}
A_{\mathrm{SW1}}(0,0)=-9.6~\mathrm{GeV}^{-2},\quad
A_{\mathrm{SW2}}(0,0)=-7.4~\mathrm{GeV}^{-2},\quad
A_{\mathrm{SW3}}(0,0)=-8.1~\mathrm{GeV}^{-2}.
\end{align}
Compared with the experimental value $|A(0,0)|^{\mathrm{exp.}}=15(2)~\mathrm{GeV}^{-2}$ quoted in Ref.~\cite{Leutgeb:2019gbz}, all three models exhibit a certain degree of underestimation. The axial-vector results should therefore be interpreted as exploratory estimates within the present scalar-extended framework.

\begin{figure}[htbp]
    \centering   \includegraphics[width=0.5\linewidth]{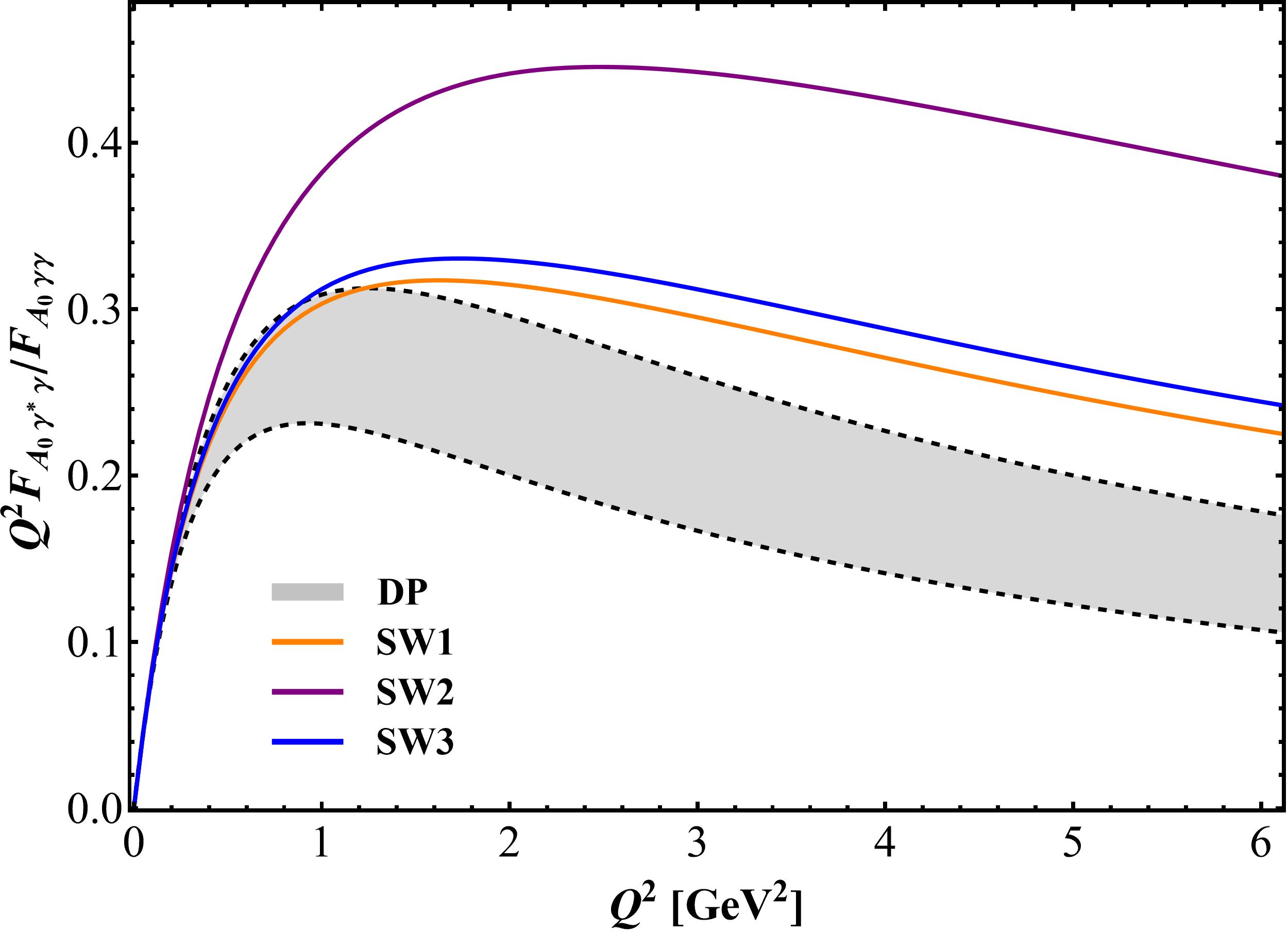}
    \caption{The single-virtual transition form factor of the ground-state axial-vector meson, $Q^2 F_{A_0 \gamma^\ast \gamma} / F_{A_0 \gamma \gamma}$. The gray band denotes the result obtained from the dipole parametrization (DP), with the uncertainty reflecting that of the parameter $\Lambda$.}
    \label{ASVF1}
\end{figure}

\begin{figure}[htbp]
    \centering   \includegraphics[width=0.5\linewidth]{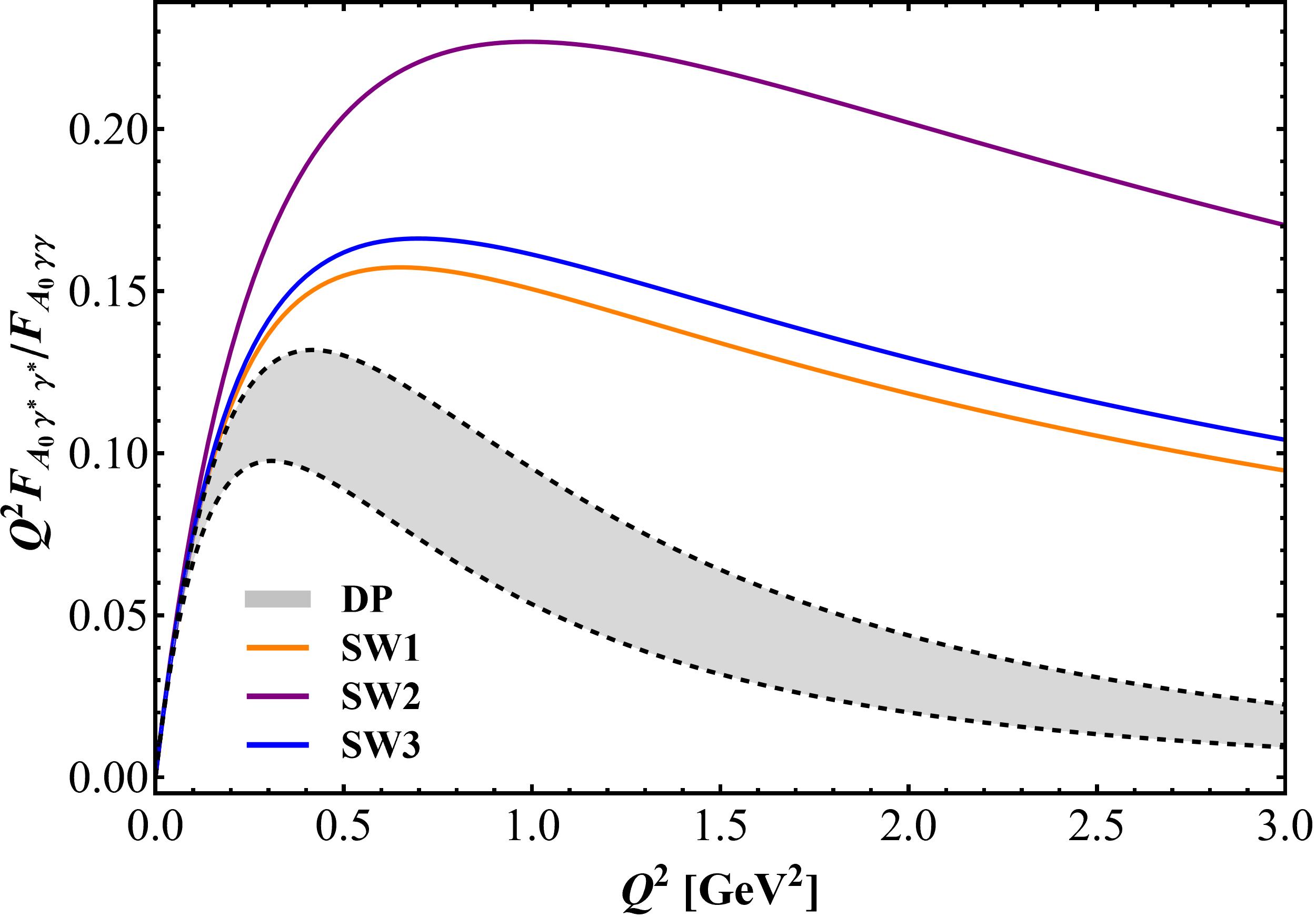}
    \caption{The double-virtual transition form factor of the ground-state axial-vector meson, $Q^2 F_{A_0 \gamma^\ast \gamma^\ast} / F_{A_0 \gamma \gamma}$.}
    \label{ATVF1}
\end{figure}

Figure~\ref{ASVF1} shows the single-virtual TFFs of the ground-state axial-vector meson. The shaded band corresponds to the dipole parametrization~\cite{Pauk:2014rta},
\begin{align}
\frac{F_{A\gamma\gamma}(Q_1^2, Q_2^2)}{F_{A\gamma\gamma}(0,0)}
= \frac{1}{\left(1 + Q_1^2 / \Lambda^2\right)^2}
\frac{1}{\left(1 + Q_2^2 / \Lambda^2\right)^2},
\end{align}
where the parameter $\Lambda=1040\pm78~\mathrm{MeV}$ is determined phenomenologically from the $f_1(1285)$ meson~\cite{L3:2001cyf}. In addition, Fig.~\ref{ATVF1} displays the double-virtual TFFs of the ground-state axial-vector meson.

After computing the transition form factors using the scalar-extended Chern-Simons term, one can proceed to evaluate the contributions to the hadronic light-by-light scattering. For pseudoscalar mesons, this is achieved by directly inserting the TFFs into Eq.~(\ref{amup}) and performing the corresponding integrations. For axial-vector mesons, we employ the following representation:
\begin{align}
a_{\mu}^{\mathrm{HLbL}}
= \int_{0}^{\infty} dQ_{1}
  \int_{0}^{\infty} dQ_{2}
  \int_{-1}^{1} d\tau \,
  \rho_{a}(Q_{1}, Q_{2}, \tau)\,,
\end{align}
where the definition of $\rho_{a}$ can be found in Appendix~B of Ref.~\cite{Leutgeb:2019gbz}. The numerical results for the contributions of the $\pi$ meson excited states to $a_\mu^{\mathrm{HLbL}}$ in different models are summarized in Table~\ref{exstate1}. It is evident that, after introducing the scalar-extended Chern-Simons term, the ground-state $\pi$ pole contribution obtained in the improved soft-wall models is significantly enhanced compared to the results computed with the standard Chern-Simons term (see Table~\ref{table3}). This behavior is consistent with the findings of Refs.~\cite{Leutgeb:2024rfs,Leutgeb:2025jmv}, where the corresponding results increase from $66.6\times10^{-11}$ to $73.3\times10^{-11}$ in the hard-wall model and from $75.2\times10^{-11}$ to $89\times10^{-11}$ in the standard soft-wall model. However, when compared with data-driven approaches and lattice QCD results, these holographic predictions tend to be systematically larger.

\begin{table}[htbp]
  \centering
  \begin{tabular}{c c  c   c  c  c  c}
    \toprule
     $a_{\mu}^{\mathrm{HLbL}}\times10^{11}$ & Model & 0 & 1 & 2 & 3 & 4 \\
\hline\hline
\multirow{3}{*}{$a_{\mu}^{\pi}$}
& SW1  & 73.4  & 0.0345  & 0.0061 & 0.0056  & 0.0049 \\
& SW2  & 89.6  & 0.2685  & 0.0446 & 0.0139  & 0.0049 \\
& SW3  & 79   & 0.1008  & 0.0102 & 0.003   & 0.0041 \\
\hline
\multirow{3}{*}{$\underset{(L+T)}{a_{\mu}^{\text{AV}}}$}
& SW1  & $\underset{(7.66+5)}{12.66}$  & $\underset{(2.48+1.85)}{4.33}$  & $\underset{(1.5+1.16)}{2.66}$ & $\underset{(1.06+0.82)}{1.88}$  & $\underset{(0.78+0.61)}{1.39}$ \\
& SW2  & $\underset{(5.14+3.58)}{8.72}$  & $\underset{(1.26+0.93)}{2.19}$  & $\underset{(0.94+0.71)}{1.65}$ & $\underset{(0.73+0.56)}{1.29}$  & $\underset{(0.58+0.44)}{1.02}$ \\
& SW3  & $\underset{(5.35+3.52)}{8.87}$  & $\underset{(3.33+2.44)}{5.77}$  & $\underset{(1.73+1.3)}{3.03}$ & $\underset{(0.98+0.74)}{1.72}$  & $\underset{(0.66+0.5)}{1.16}$ \\
\hline\hline
  \end{tabular}
  \caption{The contributions of the pseudoscalar $\pi$ meson and axial-vector meson (AV) to $a_{\mu}^{\text{HLbL}}$ in different excited states.}
  \label{exstate1}
\end{table}

The finite-mode entries in Table~\ref{exstate1} illustrate the strong numerical suppression of the displayed pseudoscalar excitations but should not be interpreted as a proof of convergence. Absolute convergence of the resonance representation at each fixed Euclidean momentum configuration is instead established by the infrared condition and the Hilbert-space argument presented in Appendix~\ref{app:convergence}.

The entries labeled $a_\mu^{A_n}$ in Table~\ref{exstate1} are sums of the contributions from the three neutral flavor directions $a=3,8,0$. In the flavor-symmetric approximation adopted here, each radial level represents the neutral $(a_1+f_1+f_1^\prime)_n$ multiplet. The individual channel contributions scale with the squared flavor factors according to $a_\mu^{A_n^3}:a_\mu^{A_n^8}:a_\mu^{A_n^0}=3:1:8$, and therefore $a_\mu^{A_n}=a_\mu^{A_n^3}+a_\mu^{A_n^8}+a_\mu^{A_n^0}=4a_\mu^{A_n^3}$. The updated White Paper quotes $12.2(2.3)\times10^{-11}$ for the lowest physical multiplet $a_1(1260)+f_1(1285)+f_1^\prime(1420)$ in the common low-energy region $Q_1,Q_2,Q_3\leq1.5~\mathrm{GeV}$~\cite{Aliberti:2025beg}. Applying the same kinematic restriction to the $n=0$ contributions gives $7.19$, $4.16$, and $4.88$ in units of $10^{-11}$ for SW1, SW2, and SW3, respectively. These values correspond to approximately $59\%$, $34\%$, and $40\%$ of the White-Paper central value. The smaller contributions are consistent with the reduced-TFF normalizations $|A_0(0,0)|=9.6$, $7.4$, and $8.1~\mathrm{GeV}^{-2}$, respectively, compared with the phenomenological estimate $15(2)~\mathrm{GeV}^{-2}$~\cite{Leutgeb:2019gbz}. This comparison is restricted to the lowest flavor multiplet and excludes radial excitations. Moreover, our calculation assumes degenerate masses and reduced TFFs in the three neutral channels; physical $f_1$-$f_1^\prime$ mixing and channel-dependent mass splittings lie beyond the scope of the present analysis.

Table~\ref{exstate1} also reveals a marked contrast between the radial behavior of the axial-vector and pseudoscalar contributions. Unlike the rapid suppression observed in the pion channel, the axial-vector contributions decrease only gradually with increasing radial quantum number. In particular, the first several excited states contribute at $\mathcal{O}(10^{-11})$ and therefore remain relevant to the total result. The longitudinal ($L$) and transverse ($T$) components are also comparable in magnitude at each excitation level and contribute jointly to the final value. This relatively weak radial suppression reflects the structure of the axial-vector TFFs and is consistent with the important role played by the full resonance tower in implementing the short-distance constraints of holographic QCD. A ground-state truncation is therefore inadequate in the axial-vector channel, and a controlled treatment of the excited states is essential for reliably assessing its contribution.

The strict infrared condition and the Hilbert-space estimates presented in Appendix~\ref{app:convergence} establish absolute convergence of both the excited-pseudoscalar and axial-vector resonance representations at fixed Euclidean kinematics. Numerically, the displayed pseudoscalar excitations are strongly suppressed, whereas the axial-vector contributions decrease more gradually.
The resulting predictions for $a_\mu^{\mathrm{HLbL}}$, however, remain systematically larger than those obtained from data-driven analyses and lattice QCD. This discrepancy may arise from the overall enhancement of the transition form factors, the choice of infrared background, and the determination of the five-dimensional coupling. Some model dependence also remains in the implementation of the short-distance constraints. Thus, although the scalar extension improves the mathematical consistency of the holographic framework, further refinements are required to achieve closer quantitative agreement with other determinations. We emphasize that the scalar-extended Chern-Simons term represents one effective infrared completion motivated by the brane-antibrane construction and should not be regarded as a unique modification of the holographic action.

\section{Summary and discussion}\label{sec5}

In this work, we have carried out a comparative analysis of hadronic contributions to the anomalous magnetic moment of the muon within three representative infrared-improved soft-wall AdS/QCD models (SW1, SW2, and SW3). The scope is deliberately restricted: it is not a systematic scan of holographic model space, but a test of how these improved soft-wall backgrounds perform for the LO-HVP contribution, the pion transition form factor, and selected HLbL channels.

For the LO-HVP contribution, we find that all improved soft-wall models still yield results below those obtained from data-driven dispersive approaches. This discrepancy can be traced back to the imperfect reproduction of the $\rho$-meson decay constant, to which the HVP contribution is particularly sensitive. When the vector decay constant is adjusted to its experimental value, the model predictions become consistent with dispersive results, indicating that the discrepancy is closely related to the vector sector.

For the HLbL contribution, we computed the transition form factors of pseudoscalar and axial-vector mesons and evaluated their contributions to $a_\mu^{\mathrm{HLbL}}$. Within the standard Chern-Simons framework we quote only the ground-state pseudoscalar-pole results, because the extension of this framework to the full tower of excited pseudoscalars and axial-vector mesons is divergent in soft-wall models. In this restricted ground-state calculation, the pseudoscalar-pole contributions from the SW1 and SW3 models are in reasonable agreement with dispersive analyses, lattice QCD, and white-paper reference values, while the SW2 model yields somewhat larger results; nevertheless, all of them improve upon the standard soft-wall model. This suggests that, within the soft-wall framework, models that better reproduce the experimental meson mass spectra tend to provide more realistic TFFs, leading to more reliable predictions for the ground-state pseudoscalar-pole HLbL contribution.

To resolve the divergence generated by the standard Chern--Simons action in the soft-wall background, we introduced a scalar-extended formulation. For all three parameter sets, the tachyon-induced damping satisfies the strict infrared condition with a positive margin. As shown in Appendix~\ref{app:convergence}, this margin renders the relevant transition vertices bounded functionals in the normal-mode Hilbert spaces and establishes absolute convergence of the excited-pseudoscalar and axial-vector resonance representations at fixed Euclidean kinematics. Numerically, the displayed pseudoscalar excitations are strongly suppressed, whereas the axial-vector contributions decrease more gradually, with longitudinal and transverse components of comparable magnitude. Despite the improved mathematical consistency provided by the scalar-extended Chern--Simons action, the predicted pseudoscalar-pole contribution remains systematically larger than current data-driven and lattice-QCD estimates. This discrepancy may be associated with the modeling of the scalar background, the normalization of the TFFs, and the implementation of short-distance constraints, indicating that further refinements of the model are required.

In summary, infrared-improved soft-wall models are useful phenomenological laboratories for studying hadronic contributions to the muon \(g-2\), but the present analysis also makes clear that the standard soft-wall Chern-Simons construction is insufficient for a finite full HLbL amplitude. The scalar-extended Chern-Simons term provides one effective repair and allows us to include pseudoscalar excitations and axial-vector states, but the resulting TFFs and \(a_\mu^{\mathrm{HLbL}}\) remain quantitatively sensitive to the infrared completion. Future developments should therefore focus on a more constrained implementation of the scalar extension, a complete treatment of \(\eta\), \(\eta^\prime\), and additional hadronic channels, and a systematic enforcement of QCD short-distance constraints.

\appendix
\section{Infrared criterion and convergence of the resonance representation}
\label{app:convergence}

This appendix verifies the infrared condition used in the scalar-extended Chern--Simons sector and proves absolute convergence of the corresponding resonance representations at fixed Euclidean kinematics. Define
\begin{equation}\label{appDeltaDefinition}
D(z)\equiv2\pi\alpha'T_0^2(z)
=\alpha'g_5^2\pi^2X_{0,\ell}^2(z),
\qquad
\Delta_{\mathrm{IR}}(z)\equiv
\frac{D(z)-\Phi(z)/2}{z^2},
\end{equation}
where $g_5^2=12\pi^2/N_c=4\pi^2$, $\alpha'=1/6$, and the uniform scalar convention of Eq.~(\ref{uniformX0}) has been used. Write
\begin{equation}\label{appIRSlopes}
X_{0,\ell}(z)=\kappa_Xz+o(z),
\qquad
\Phi(z)=\kappa_\Phi z^2+o(z^2).
\end{equation}
Then
\begin{equation}\label{appDeltaInfinity}
\Delta_{\mathrm{IR}}(z)\longrightarrow
\Delta_\infty\equiv
\alpha'g_5^2\pi^2\kappa_X^2-\frac{\kappa_\Phi}{2}.
\end{equation}
The infrared condition in the main text holds with a strict quadratic margin when $\Delta_\infty>0$.

For SW1, the analytic scalar profile gives
\begin{equation}\label{appSW1Slope}
\chi(z)=\frac{Az+Bz^3}{1+Cz^2}
=v_1z+O(z^{-1}),
\qquad
v_1=\frac{B}{C}=\frac{\sigma}{\mu_c^2}+m_q\zeta,
\qquad
\kappa_X=\frac{v_1}{2}.
\end{equation}
For SW2, substituting $\chi(z)=v_2z+o(z)$, $m_5^2(z)=-3-\mu_c^2z^2$, and $\Phi'(z)=2\mu_g^2z$ into the scalar-background EOM and retaining the coefficient of $z$ yields
\begin{equation}\label{appSW2Slope}
v_2\left(\mu_c^2-2\mu_g^2-\frac{\lambda}{2}v_2^2\right)=0,
\qquad
v_2^2=\frac{2(\mu_c^2-2\mu_g^2)}{\lambda},
\qquad
\kappa_X=\frac{v_2}{2}.
\end{equation}
In SW3, the light and strange scalar backgrounds have the same leading linear slope, whereas the determinant term is subleading by one power of $z$. The $O(z)$ part of the scalar EOM therefore gives
\begin{equation}\label{appSW3Slope}
v_3\left(\mu_c^2-2\mu_g^2-\lambda v_3^2\right)=0,
\qquad
v_3^2=\frac{\mu_c^2-2\mu_g^2}{\lambda},
\qquad
\kappa_X=\frac{v_3}{\sqrt2}.
\end{equation}
The nonzero real linear branches of SW2 and SW3 require
\begin{equation}\label{appLinearBranch}
\lambda>0,
\qquad
\mu_c^2>2\mu_g^2,
\end{equation}
which is satisfied by both selected parameter sets. In all three models $\kappa_\Phi=\mu_g^2$: the second term in the SW1 dilaton is $O(z^{-2})$, and the exponential correction in SW3 vanishes in the infrared. The resulting margins are listed in Table~\ref{tab:appIRMargins}.

\begin{table}[htbp]
\centering
\small
\setlength{\tabcolsep}{4.5pt}
\begin{tabular}{c|ccccc}
Model & $v_i$ & $\kappa_X$ & $\kappa_\Phi/2$
& $\alpha'g_5^2\pi^2\kappa_X^2$ & $\Delta_\infty$ \\
& \multicolumn{2}{c}{[GeV]} & \multicolumn{3}{c}{[GeV$^2$]} \\ \hline
SW1 & 0.172812 & 0.086406 & 0.111865 & 0.484840 & 0.372975 \\
SW2 & 0.241521 & 0.120761 & 0.075078 & 0.947021 & 0.871942 \\
SW3 & 0.093307 & 0.065978 & 0.115200 & 0.282687 & 0.167487
\end{tabular}
\caption{Infrared scalar slopes and strict convergence margins. Here $v_i$ is the slope of $\chi$ in SW1--SW2 and of $\chi_u$ in SW3; the different factors in $\kappa_X$ follow from $X_{0,\ell}=\chi/2$ and $X_{0,\ell}=\chi_u/\sqrt2$, respectively.}
\label{tab:appIRMargins}
\end{table}

All three values of $\Delta_\infty$ are strictly positive. Hence, for each model there exist $z_0$ and $\delta>0$ such that
\begin{equation}\label{appStrictIRBound}
D(z)-\frac{\Phi(z)}{2}\geq\delta z^2,
\qquad z\geq z_0.
\end{equation}
Only this eventual bound is required; nonmonotonic behavior of $\Delta_{\mathrm{IR}}(z)$ on a finite interval, including that of SW3 at intermediate $z$, changes finite overlap integrals but cannot produce an infrared divergence.

We next prove convergence of the axial-vector resonance representation. Let $p(z)=e^{-\Phi(z)}/z$ and equip the normalizable axial modes with
\begin{equation}\label{appAxialProduct}
\langle f,g\rangle_A=\int_0^\infty dz\,\frac{p(z)}{g_5^2}f(z)g(z),
\qquad
\langle A_n,A_m\rangle_A=\delta_{nm}.
\end{equation}
At fixed Euclidean photon virtualities define
\begin{equation}\label{appAxialSource}
J_{12}(z)=\frac{2}{Q_1^2}e^{-D(z)}
\partial_zV(Q_1,z)V(Q_2,z),
\qquad
f_{12}(z)=\frac{g_5^2}{p(z)}J_{12}(z).
\end{equation}
The apparent $Q_1^{-2}$ singularity is removable. The regular bulk-to-boundary solution is analytic in $Q_1^2$ near the origin and satisfies
\begin{equation}\label{appSmallQ}
V(Q_1,z)=1+Q_1^2\mathcal V_1(z)+O(Q_1^4),
\qquad
\partial_zV(Q_1,z)=Q_1^2\mathcal V_1'(z)+O(Q_1^4).
\end{equation}
Thus $\partial_zV(Q_1,z)/Q_1^2$ has a finite $Q_1\to0$ limit. The reduced transition form factor is the Fourier coefficient
\begin{equation}\label{appTFFCoefficient}
\mathcal A_n(Q_1^2,Q_2^2)
=\int_0^\infty dz\,J_{12}(z)A_n(z)
=\langle f_{12},A_n\rangle_A.
\end{equation}
Its source norm is
\begin{equation}\label{appAxialSourceNorm}
\|f_{12}\|_A^2
=\frac{4g_5^2}{Q_1^4}\int_0^\infty dz\,
z e^{\Phi(z)-2D(z)}
\left|\partial_zV(Q_1,z)V(Q_2,z)\right|^2.
\end{equation}
The Euclidean vector equation can be written as
\begin{equation}\label{appVectorLargeZ}
V''-\left(\Phi'+\frac1z\right)V'-Q^2V=0.
\end{equation}
For $\Phi(z)=\kappa_\Phi z^2+o(z^2)$, its infrared-regular branch behaves as $V(Q,z) = O\!\left(z^{-Q^2/2\kappa_\Phi}\right)$; the independent branch grows as $e^{\kappa_\Phi z^2}$ times a power and is excluded by infrared regularity. Therefore $V$ and $\partial_zV$ grow at most polynomially at fixed Euclidean $Q$. Equation~(\ref{appStrictIRBound}) then implies $\Phi-2D\leq-2\delta z^2$, and the norm in Eq.~(\ref{appAxialSourceNorm}) is finite. Bessel's inequality gives
\begin{equation}\label{appBesselAxial}
\sum_{n=0}^\infty
|\mathcal A_n(Q_1^2,Q_2^2)|^2
\leq\|f_{12}\|_A^2<\infty.
\end{equation}
For either spectral denominator appearing in the axial exchange, the Cauchy--Schwarz inequality therefore yields
\begin{align}
\sum_{n=0}^\infty
\left|\frac{\mathcal A_n^L\mathcal A_n^R}{Q^2+m_{A_n}^2}\right|
&\leq\frac{\|f_L\|_A\|f_R\|_A}{Q^2+m_{A_0}^2}<\infty,
\label{appAbsoluteAxialT}\\
\sum_{n=0}^\infty
\left|\frac{\mathcal A_n^L\mathcal A_n^R}{m_{A_n}^2}\right|
&\leq\frac{\|f_L\|_A\|f_R\|_A}{m_{A_0}^2}<\infty.
\label{appAbsoluteAxialL}
\end{align}
Moreover, because the discrete confining spectrum satisfies $m_{A_n}^2\to\infty$, the remainder after $N$ modes obeys
\begin{equation}\label{appAxialTail}
\left|R_Q-R_Q^{(N)}\right|
\leq\frac{\|f_L\|_A\|f_R\|_A}{Q^2+m_{A_{N+1}}^2}
\xrightarrow[N\to\infty]{}0.
\end{equation}

The coupled pseudoscalar problem is treated analogously. After removal of the residual gauge redundancy, its physical positive-norm space has norm
\begin{equation}\label{appPseudoscalarNorm}
\|\Psi\|_P^2=\int_0^\infty dz\left[
\frac{e^{-\Phi}}{g_5^2z}|\varphi'|^2
+\frac{e^{-\Phi}v_\ell^2}{z^3}|\pi-\varphi|^2\right].
\end{equation}
The scalar-extended transition vertex is the linear functional
\begin{equation}\label{appPseudoscalarFunctional}
\ell_{12}(\Psi)=\int_0^\infty dz\,V(Q_1,z)V(Q_2,z)
\left[e^{-D}\varphi'+(e^{-D})'(\varphi-\pi)\right].
\end{equation}
Cauchy-Schwarz bounds its two terms by $\|\Psi\|_P$ times the square roots of
\begin{equation}\label{appPseudoscalarDualNorms}
g_5^2\int_0^\infty dz\,z e^{\Phi-2D}|V_1V_2|^2,
\qquad
\int_0^\infty dz\,\frac{z^3e^\Phi}{v_\ell^2}
\left|(e^{-D})'V_1V_2\right|^2.
\end{equation}
Both integrals are finite because $v_\ell\sim z$, $D'\sim z$, and $\Phi-2D\leq-2\delta z^2$ in the infrared. Thus $\ell_{12}$ is bounded. The Riesz representation theorem and Bessel's inequality imply
\begin{equation}
\sum_{n=0}^\infty
|F_{P_n\gamma^*\gamma^*}(Q_1^2,Q_2^2)|^2<\infty,
\end{equation}
and the same Cauchy-Schwarz estimates as in Eqs.~(\ref{appAbsoluteAxialT})-(\ref{appAxialTail}) establish absolute convergence of the excited-pseudoscalar pole representation at fixed Euclidean kinematics. The pseudoscalar ground state is treated separately in the main text. This Hilbert-space argument applies directly to the self-adjoint spectral problem whose resolvent defines the corresponding Green function and is consistent with the infrared analysis of Ref.~\cite{Leutgeb:2025jmv}.

\section*{Acknowledgments}
The authors are grateful to  Xin-Yi Liu for the technical assistance regarding numerical algorithms, and to Yi-Hao Zhang  for valuable discussions. The authors would also like to thank Jonas Mager, Josef Leutgeb, and Anton Rebhan for valuable comments on this work. This work is supported by Hunan Provincial Natural Science Foundation of China (Grants No. 2023JJ30115 and No. 2024JJ3004). L.-Y. Dai thanks the support from National Natural Science Foundation of China (NSFC) with Grants No.12322502, 12335002.

\bibliography{ref}

\begin{thebibliography}{104}%
\makeatletter
\providecommand \@ifxundefined [1]{%
 \@ifx{#1\undefined}
}%
\providecommand \@ifnum [1]{%
 \ifnum #1\expandafter \@firstoftwo
 \else \expandafter \@secondoftwo
 \fi
}%
\providecommand \@ifx [1]{%
 \ifx #1\expandafter \@firstoftwo
 \else \expandafter \@secondoftwo
 \fi
}%
\providecommand \natexlab [1]{#1}%
\providecommand \enquote  [1]{``#1''}%
\providecommand \bibnamefont  [1]{#1}%
\providecommand \bibfnamefont [1]{#1}%
\providecommand \citenamefont [1]{#1}%
\providecommand \href@noop [0]{\@secondoftwo}%
\providecommand \href [0]{\begingroup \@sanitize@url \@href}%
\providecommand \@href[1]{\@@startlink{#1}\@@href}%
\providecommand \@@href[1]{\endgroup#1\@@endlink}%
\providecommand \@sanitize@url [0]{\catcode `\\12\catcode `\$12\catcode
  `\&12\catcode `\#12\catcode `\^12\catcode `\_12\catcode `\%12\relax}%
\providecommand \@@startlink[1]{}%
\providecommand \@@endlink[0]{}%
\providecommand \url  [0]{\begingroup\@sanitize@url \@url }%
\providecommand \@url [1]{\endgroup\@href {#1}{\urlprefix }}%
\providecommand \urlprefix  [0]{URL }%
\providecommand \Eprint [0]{\href }%
\providecommand \doibase [0]{https://doi.org/}%
\providecommand \selectlanguage [0]{\@gobble}%
\providecommand \bibinfo  [0]{\@secondoftwo}%
\providecommand \bibfield  [0]{\@secondoftwo}%
\providecommand \translation [1]{[#1]}%
\providecommand \BibitemOpen [0]{}%
\providecommand \bibitemStop [0]{}%
\providecommand \bibitemNoStop [0]{.\EOS\space}%
\providecommand \EOS [0]{\spacefactor3000\relax}%
\providecommand \BibitemShut  [1]{\csname bibitem#1\endcsname}%
\let\auto@bib@innerbib\@empty
\bibitem [{\citenamefont {Schwinger}(1948)}]{Schwinger:1948iu}%
  \BibitemOpen
  \bibfield  {author} {\bibinfo {author} {\bibfnamefont {J.~S.}\ \bibnamefont
  {Schwinger}},\ }\bibfield  {title} {\bibinfo {title} {{On Quantum
  electrodynamics and the magnetic moment of the electron}},\ }\href
  {https://doi.org/10.1103/PhysRev.73.416} {\bibfield  {journal} {\bibinfo
  {journal} {Phys. Rev.}\ }\textbf {\bibinfo {volume} {73}},\ \bibinfo {pages}
  {416} (\bibinfo {year} {1948})}\BibitemShut {NoStop}%
\bibitem [{\citenamefont {Bailey}\ \emph {et~al.}(1979)\citenamefont {Bailey}
  \emph {et~al.}}]{CERN-Mainz-Daresbury:1978ccd}%
  \BibitemOpen
  \bibfield  {author} {\bibinfo {author} {\bibfnamefont {J.}~\bibnamefont
  {Bailey}} \emph {et~al.} (\bibinfo {collaboration} {CERN-Mainz-Daresbury}),\
  }\bibfield  {title} {\bibinfo {title} {{Final Report on the CERN Muon Storage
  Ring Including the Anomalous Magnetic Moment and the Electric Dipole Moment
  of the Muon, and a Direct Test of Relativistic Time Dilation}},\ }\href
  {https://doi.org/10.1016/0550-3213(79)90292-X} {\bibfield  {journal}
  {\bibinfo  {journal} {Nucl. Phys. B}\ }\textbf {\bibinfo {volume} {150}},\
  \bibinfo {pages} {1} (\bibinfo {year} {1979})}\BibitemShut {NoStop}%
\bibitem [{\citenamefont {Bennett}\ \emph {et~al.}(2006)\citenamefont {Bennett}
  \emph {et~al.}}]{Muong-2:2006rrc}%
  \BibitemOpen
  \bibfield  {author} {\bibinfo {author} {\bibfnamefont {G.~W.}\ \bibnamefont
  {Bennett}} \emph {et~al.} (\bibinfo {collaboration} {Muon g-2}),\ }\bibfield
  {title} {\bibinfo {title} {{Final Report of the Muon E821 Anomalous Magnetic
  Moment Measurement at BNL}},\ }\href
  {https://doi.org/10.1103/PhysRevD.73.072003} {\bibfield  {journal} {\bibinfo
  {journal} {Phys. Rev. D}\ }\textbf {\bibinfo {volume} {73}},\ \bibinfo
  {pages} {072003} (\bibinfo {year} {2006})},\ \Eprint
  {https://arxiv.org/abs/hep-ex/0602035} {arXiv:hep-ex/0602035} \BibitemShut
  {NoStop}%
\bibitem [{\citenamefont {Abi}\ \emph {et~al.}(2021)\citenamefont {Abi} \emph
  {et~al.}}]{Muong-2:2021ojo}%
  \BibitemOpen
  \bibfield  {author} {\bibinfo {author} {\bibfnamefont {B.}~\bibnamefont
  {Abi}} \emph {et~al.} (\bibinfo {collaboration} {Muon g-2}),\ }\bibfield
  {title} {\bibinfo {title} {{Measurement of the Positive Muon Anomalous
  Magnetic Moment to 0.46 ppm}},\ }\href
  {https://doi.org/10.1103/PhysRevLett.126.141801} {\bibfield  {journal}
  {\bibinfo  {journal} {Phys. Rev. Lett.}\ }\textbf {\bibinfo {volume} {126}},\
  \bibinfo {pages} {141801} (\bibinfo {year} {2021})},\ \Eprint
  {https://arxiv.org/abs/2104.03281} {arXiv:2104.03281 [hep-ex]} \BibitemShut
  {NoStop}%
\bibitem [{\citenamefont {Aguillard}\ \emph {et~al.}(2023)\citenamefont
  {Aguillard} \emph {et~al.}}]{Muong-2:2023cdq}%
  \BibitemOpen
  \bibfield  {author} {\bibinfo {author} {\bibfnamefont {D.~P.}\ \bibnamefont
  {Aguillard}} \emph {et~al.} (\bibinfo {collaboration} {Muon g-2}),\
  }\bibfield  {title} {\bibinfo {title} {{Measurement of the Positive Muon
  Anomalous Magnetic Moment to 0.20~ppm}},\ }\href
  {https://doi.org/10.1103/PhysRevLett.131.161802} {\bibfield  {journal}
  {\bibinfo  {journal} {Phys. Rev. Lett.}\ }\textbf {\bibinfo {volume} {131}},\
  \bibinfo {pages} {161802} (\bibinfo {year} {2023})},\ \Eprint
  {https://arxiv.org/abs/2308.06230} {arXiv:2308.06230 [hep-ex]} \BibitemShut
  {NoStop}%
\bibitem [{\citenamefont {Aguillard}\ \emph {et~al.}(2025)\citenamefont
  {Aguillard} \emph {et~al.}}]{Muong-2:2025xyk}%
  \BibitemOpen
  \bibfield  {author} {\bibinfo {author} {\bibfnamefont {D.~P.}\ \bibnamefont
  {Aguillard}} \emph {et~al.} (\bibinfo {collaboration} {Muon g-2}),\
  }\bibfield  {title} {\bibinfo {title} {{Measurement of the Positive Muon
  Anomalous Magnetic Moment to 127~ppb}},\ }\href
  {https://doi.org/10.1103/7clf-sm2v} {\bibfield  {journal} {\bibinfo
  {journal} {Phys. Rev. Lett.}\ }\textbf {\bibinfo {volume} {135}},\ \bibinfo
  {pages} {101802} (\bibinfo {year} {2025})},\ \Eprint
  {https://arxiv.org/abs/2506.03069} {arXiv:2506.03069 [hep-ex]} \BibitemShut
  {NoStop}%
\bibitem [{\citenamefont {Aoyama}\ \emph {et~al.}(2020)\citenamefont {Aoyama}
  \emph {et~al.}}]{Aoyama:2020ynm}%
  \BibitemOpen
  \bibfield  {author} {\bibinfo {author} {\bibfnamefont {T.}~\bibnamefont
  {Aoyama}} \emph {et~al.},\ }\bibfield  {title} {\bibinfo {title} {{The
  anomalous magnetic moment of the muon in the Standard Model}},\ }\href
  {https://doi.org/10.1016/j.physrep.2020.07.006} {\bibfield  {journal}
  {\bibinfo  {journal} {Phys. Rept.}\ }\textbf {\bibinfo {volume} {887}},\
  \bibinfo {pages} {1} (\bibinfo {year} {2020})},\ \Eprint
  {https://arxiv.org/abs/2006.04822} {arXiv:2006.04822 [hep-ph]} \BibitemShut
  {NoStop}%
\bibitem [{\citenamefont {Aliberti}\ \emph {et~al.}(2025)\citenamefont
  {Aliberti} \emph {et~al.}}]{Aliberti:2025beg}%
  \BibitemOpen
  \bibfield  {author} {\bibinfo {author} {\bibfnamefont {R.}~\bibnamefont
  {Aliberti}} \emph {et~al.},\ }\bibfield  {title} {\bibinfo {title} {{The
  anomalous magnetic moment of the muon in the Standard Model: an update}},\
  }\href {https://doi.org/10.1016/j.physrep.2025.08.002} {\bibfield  {journal}
  {\bibinfo  {journal} {Phys. Rept.}\ }\textbf {\bibinfo {volume} {1143}},\
  \bibinfo {pages} {1} (\bibinfo {year} {2025})},\ \Eprint
  {https://arxiv.org/abs/2505.21476} {arXiv:2505.21476 [hep-ph]} \BibitemShut
  {NoStop}%
\bibitem [{\citenamefont {Davier}\ \emph {et~al.}(2017)\citenamefont {Davier},
  \citenamefont {Hoecker}, \citenamefont {Malaescu},\ and\ \citenamefont
  {Zhang}}]{Davier:2017zfy}%
  \BibitemOpen
  \bibfield  {author} {\bibinfo {author} {\bibfnamefont {M.}~\bibnamefont
  {Davier}}, \bibinfo {author} {\bibfnamefont {A.}~\bibnamefont {Hoecker}},
  \bibinfo {author} {\bibfnamefont {B.}~\bibnamefont {Malaescu}},\ and\
  \bibinfo {author} {\bibfnamefont {Z.}~\bibnamefont {Zhang}},\ }\bibfield
  {title} {\bibinfo {title} {{Reevaluation of the hadronic vacuum polarisation
  contributions to the Standard Model predictions of the muon $g-2$ and
  ${\alpha (m_Z^2)}$ using newest hadronic cross-section data}},\ }\href
  {https://doi.org/10.1140/epjc/s10052-017-5161-6} {\bibfield  {journal}
  {\bibinfo  {journal} {Eur. Phys. J. C}\ }\textbf {\bibinfo {volume} {77}},\
  \bibinfo {pages} {827} (\bibinfo {year} {2017})},\ \Eprint
  {https://arxiv.org/abs/1706.09436} {arXiv:1706.09436 [hep-ph]} \BibitemShut
  {NoStop}%
\bibitem [{\citenamefont {Hoid}\ \emph {et~al.}(2020)\citenamefont {Hoid},
  \citenamefont {Hoferichter},\ and\ \citenamefont {Kubis}}]{Hoid:2020xjs}%
  \BibitemOpen
  \bibfield  {author} {\bibinfo {author} {\bibfnamefont {B.-L.}\ \bibnamefont
  {Hoid}}, \bibinfo {author} {\bibfnamefont {M.}~\bibnamefont {Hoferichter}},\
  and\ \bibinfo {author} {\bibfnamefont {B.}~\bibnamefont {Kubis}},\ }\bibfield
   {title} {\bibinfo {title} {{Hadronic vacuum polarization and vector-meson
  resonance parameters from $e^+e^-\rightarrow \pi ^0\gamma$}},\ }\href
  {https://doi.org/10.1140/epjc/s10052-020-08550-2} {\bibfield  {journal}
  {\bibinfo  {journal} {Eur. Phys. J. C}\ }\textbf {\bibinfo {volume} {80}},\
  \bibinfo {pages} {988} (\bibinfo {year} {2020})},\ \Eprint
  {https://arxiv.org/abs/2007.12696} {arXiv:2007.12696 [hep-ph]} \BibitemShut
  {NoStop}%
\bibitem [{\citenamefont {Hoferichter}\ and\ \citenamefont
  {Teubner}(2022)}]{Hoferichter:2021wyj}%
  \BibitemOpen
  \bibfield  {author} {\bibinfo {author} {\bibfnamefont {M.}~\bibnamefont
  {Hoferichter}}\ and\ \bibinfo {author} {\bibfnamefont {T.}~\bibnamefont
  {Teubner}},\ }\bibfield  {title} {\bibinfo {title} {{Mixed Leptonic and
  Hadronic Corrections to the Anomalous Magnetic Moment of the Muon}},\ }\href
  {https://doi.org/10.1103/PhysRevLett.128.112002} {\bibfield  {journal}
  {\bibinfo  {journal} {Phys. Rev. Lett.}\ }\textbf {\bibinfo {volume} {128}},\
  \bibinfo {pages} {112002} (\bibinfo {year} {2022})},\ \Eprint
  {https://arxiv.org/abs/2112.06929} {arXiv:2112.06929 [hep-ph]} \BibitemShut
  {NoStop}%
\bibitem [{\citenamefont {Benayoun}\ \emph {et~al.}(2022)\citenamefont
  {Benayoun}, \citenamefont {DelBuono},\ and\ \citenamefont
  {Jegerlehner}}]{Benayoun:2021ody}%
  \BibitemOpen
  \bibfield  {author} {\bibinfo {author} {\bibfnamefont {M.}~\bibnamefont
  {Benayoun}}, \bibinfo {author} {\bibfnamefont {L.}~\bibnamefont {DelBuono}},\
  and\ \bibinfo {author} {\bibfnamefont {F.}~\bibnamefont {Jegerlehner}},\
  }\bibfield  {title} {\bibinfo {title} {{BHLS$_2$ upgrade: $\tau $ spectra,
  muon HVP and the [$\pi ^0,~\eta ,~{\eta ^\prime }$] system}},\ }\href
  {https://doi.org/10.1140/epjc/s10052-022-10096-4} {\bibfield  {journal}
  {\bibinfo  {journal} {Eur. Phys. J. C}\ }\textbf {\bibinfo {volume} {82}},\
  \bibinfo {pages} {184} (\bibinfo {year} {2022})},\ \Eprint
  {https://arxiv.org/abs/2105.13018} {arXiv:2105.13018 [hep-ph]} \BibitemShut
  {NoStop}%
\bibitem [{\citenamefont {Yi}\ \emph {et~al.}(2021)\citenamefont {Yi},
  \citenamefont {Wang},\ and\ \citenamefont {Xiao}}]{Yi:2021ccc}%
  \BibitemOpen
  \bibfield  {author} {\bibinfo {author} {\bibfnamefont {J.-Y.}\ \bibnamefont
  {Yi}}, \bibinfo {author} {\bibfnamefont {Z.-Y.}\ \bibnamefont {Wang}},\ and\
  \bibinfo {author} {\bibfnamefont {C.~W.}\ \bibnamefont {Xiao}},\ }\bibfield
  {title} {\bibinfo {title} {{Study of the pion vector form factor and its
  contribution to the muon $g-2$}},\ }\href
  {https://doi.org/10.1103/PhysRevD.104.116017} {\bibfield  {journal} {\bibinfo
   {journal} {Phys. Rev. D}\ }\textbf {\bibinfo {volume} {104}},\ \bibinfo
  {pages} {116017} (\bibinfo {year} {2021})},\ \Eprint
  {https://arxiv.org/abs/2107.09535} {arXiv:2107.09535 [hep-ph]} \BibitemShut
  {NoStop}%
\bibitem [{\citenamefont {Qin}\ \emph {et~al.}(2021)\citenamefont {Qin},
  \citenamefont {Dai},\ and\ \citenamefont {Portoles}}]{Qin:2020udp}%
  \BibitemOpen
  \bibfield  {author} {\bibinfo {author} {\bibfnamefont {W.}~\bibnamefont
  {Qin}}, \bibinfo {author} {\bibfnamefont {L.-Y.}\ \bibnamefont {Dai}},\ and\
  \bibinfo {author} {\bibfnamefont {J.}~\bibnamefont {Portoles}},\ }\bibfield
  {title} {\bibinfo {title} {{Two and three pseudoscalar production in
  e$^{+}$e$^{−}$ annihilation and their contributions to ($g-2$)$_{\mu}$}},\
  }\href {https://doi.org/10.1007/JHEP03(2021)092} {\bibfield  {journal}
  {\bibinfo  {journal} {JHEP}\ }\textbf {\bibinfo {volume} {03}},\ \bibinfo
  {pages} {092}},\ \Eprint {https://arxiv.org/abs/2011.09618} {arXiv:2011.09618
  [hep-ph]} \BibitemShut {NoStop}%
\bibitem [{\citenamefont {Wang}\ \emph {et~al.}(2023)\citenamefont {Wang},
  \citenamefont {Fang},\ and\ \citenamefont {Dai}}]{Wang:2023njt}%
  \BibitemOpen
  \bibfield  {author} {\bibinfo {author} {\bibfnamefont {S.-J.}\ \bibnamefont
  {Wang}}, \bibinfo {author} {\bibfnamefont {Z.}~\bibnamefont {Fang}},\ and\
  \bibinfo {author} {\bibfnamefont {L.-Y.}\ \bibnamefont {Dai}},\ }\bibfield
  {title} {\bibinfo {title} {{Two body final states production in
  electron-positron annihilation and their contributions to ($g-2$)$_{\mu}$}},\
  }\href {https://doi.org/10.1007/JHEP07(2023)037} {\bibfield  {journal}
  {\bibinfo  {journal} {JHEP}\ }\textbf {\bibinfo {volume} {07}},\ \bibinfo
  {pages} {037}},\ \Eprint {https://arxiv.org/abs/2302.08859} {arXiv:2302.08859
  [hep-ph]} \BibitemShut {NoStop}%
\bibitem [{\citenamefont {Qin}\ \emph {et~al.}(2025)\citenamefont {Qin},
  \citenamefont {Qin},\ and\ \citenamefont {Dai}}]{Qin:2024ulb}%
  \BibitemOpen
  \bibfield  {author} {\bibinfo {author} {\bibfnamefont {B.-H.}\ \bibnamefont
  {Qin}}, \bibinfo {author} {\bibfnamefont {W.}~\bibnamefont {Qin}},\ and\
  \bibinfo {author} {\bibfnamefont {L.-Y.}\ \bibnamefont {Dai}},\ }\bibfield
  {title} {\bibinfo {title} {{Study of electron-positron annihilation into
  $K\bar{K}\pi$ within resonance chiral theory}},\ }\href
  {https://doi.org/10.1103/PhysRevD.111.034025} {\bibfield  {journal} {\bibinfo
   {journal} {Phys. Rev. D}\ }\textbf {\bibinfo {volume} {111}},\ \bibinfo
  {pages} {034025} (\bibinfo {year} {2025})},\ \Eprint
  {https://arxiv.org/abs/2403.14294} {arXiv:2403.14294 [hep-ph]} \BibitemShut
  {NoStop}%
\bibitem [{\citenamefont {Hayakawa}\ and\ \citenamefont
  {Kinoshita}(1998)}]{Hayakawa:1997rq}%
  \BibitemOpen
  \bibfield  {author} {\bibinfo {author} {\bibfnamefont {M.}~\bibnamefont
  {Hayakawa}}\ and\ \bibinfo {author} {\bibfnamefont {T.}~\bibnamefont
  {Kinoshita}},\ }\bibfield  {title} {\bibinfo {title} {{Pseudoscalar pole
  terms in the hadronic light by light scattering contribution to muon
  $g-2$}},\ }\href {https://doi.org/10.1103/PhysRevD.57.465} {\bibfield
  {journal} {\bibinfo  {journal} {Phys. Rev. D}\ }\textbf {\bibinfo {volume}
  {57}},\ \bibinfo {pages} {465} (\bibinfo {year} {1998})},\ \bibinfo {note}
  {[Erratum: Phys.Rev.D 66, 019902 (2002)]},\ \Eprint
  {https://arxiv.org/abs/hep-ph/9708227} {arXiv:hep-ph/9708227} \BibitemShut
  {NoStop}%
\bibitem [{\citenamefont {Guevara}\ \emph {et~al.}(2018)\citenamefont
  {Guevara}, \citenamefont {Roig},\ and\ \citenamefont
  {Sanz-Cillero}}]{Guevara:2018rhj}%
  \BibitemOpen
  \bibfield  {author} {\bibinfo {author} {\bibfnamefont {A.}~\bibnamefont
  {Guevara}}, \bibinfo {author} {\bibfnamefont {P.}~\bibnamefont {Roig}},\ and\
  \bibinfo {author} {\bibfnamefont {J.~J.}\ \bibnamefont {Sanz-Cillero}},\
  }\bibfield  {title} {\bibinfo {title} {{Pseudoscalar pole light-by-light
  contributions to the muon $(g-2)$ in Resonance Chiral Theory}},\ }\href
  {https://doi.org/10.1007/JHEP06(2018)160} {\bibfield  {journal} {\bibinfo
  {journal} {JHEP}\ }\textbf {\bibinfo {volume} {06}},\ \bibinfo {pages}
  {160}},\ \Eprint {https://arxiv.org/abs/1803.08099} {arXiv:1803.08099
  [hep-ph]} \BibitemShut {NoStop}%
\bibitem [{\citenamefont {Raya}\ \emph {et~al.}(2020)\citenamefont {Raya},
  \citenamefont {Bashir},\ and\ \citenamefont {Roig}}]{Raya:2019dnh}%
  \BibitemOpen
  \bibfield  {author} {\bibinfo {author} {\bibfnamefont {K.}~\bibnamefont
  {Raya}}, \bibinfo {author} {\bibfnamefont {A.}~\bibnamefont {Bashir}},\ and\
  \bibinfo {author} {\bibfnamefont {P.}~\bibnamefont {Roig}},\ }\bibfield
  {title} {\bibinfo {title} {{Contribution of neutral pseudoscalar mesons to
  $a_\mu^{\text{HLbL}}$ within a Schwinger-Dyson equations approach to QCD}},\
  }\href {https://doi.org/10.1103/PhysRevD.101.074021} {\bibfield  {journal}
  {\bibinfo  {journal} {Phys. Rev. D}\ }\textbf {\bibinfo {volume} {101}},\
  \bibinfo {pages} {074021} (\bibinfo {year} {2020})},\ \Eprint
  {https://arxiv.org/abs/1910.05960} {arXiv:1910.05960 [hep-ph]} \BibitemShut
  {NoStop}%
\bibitem [{\citenamefont {Estrada}\ \emph {et~al.}(2024)\citenamefont
  {Estrada}, \citenamefont {Gonz{\`a}lez-Sol{\'\i}s}, \citenamefont {Guevara},\
  and\ \citenamefont {Roig}}]{Estrada:2024cfy}%
  \BibitemOpen
  \bibfield  {author} {\bibinfo {author} {\bibfnamefont {E.~J.}\ \bibnamefont
  {Estrada}}, \bibinfo {author} {\bibfnamefont {S.}~\bibnamefont
  {Gonz{\`a}lez-Sol{\'\i}s}}, \bibinfo {author} {\bibfnamefont
  {A.}~\bibnamefont {Guevara}},\ and\ \bibinfo {author} {\bibfnamefont
  {P.}~\bibnamefont {Roig}},\ }\bibfield  {title} {\bibinfo {title} {{Improved
  $\pi^0$, $\eta$, $\eta'$ transition form factors in resonance chiral theory
  and their $ {a}_{\mu}^{\textrm{HLbL}} $ contribution}},\ }\href
  {https://doi.org/10.1007/JHEP12(2024)203} {\bibfield  {journal} {\bibinfo
  {journal} {JHEP}\ }\textbf {\bibinfo {volume} {12}},\ \bibinfo {pages}
  {203}},\ \Eprint {https://arxiv.org/abs/2409.10503} {arXiv:2409.10503
  [hep-ph]} \BibitemShut {NoStop}%
\bibitem [{\citenamefont {Estrada}\ and\ \citenamefont
  {Roig}(2026)}]{Estrada:2025bty}%
  \BibitemOpen
  \bibfield  {author} {\bibinfo {author} {\bibfnamefont {E.~J.}\ \bibnamefont
  {Estrada}}\ and\ \bibinfo {author} {\bibfnamefont {P.}~\bibnamefont {Roig}},\
  }\bibfield  {title} {\bibinfo {title} {{Tensor meson pole contributions to
  the HLbL piece of ${a}_{\mu }^{\text{HLbL}}$ within R{\ensuremath{\chi}}T}},\
  }\href {https://doi.org/10.1007/JHEP01(2026)070} {\bibfield  {journal}
  {\bibinfo  {journal} {JHEP}\ }\textbf {\bibinfo {volume} {01}},\ \bibinfo
  {pages} {070}},\ \Eprint {https://arxiv.org/abs/2504.00448} {arXiv:2504.00448
  [hep-ph]} \BibitemShut {NoStop}%
\bibitem [{\citenamefont {Colangelo}\ \emph
  {et~al.}(2014{\natexlab{a}})\citenamefont {Colangelo}, \citenamefont
  {Hoferichter}, \citenamefont {Procura},\ and\ \citenamefont
  {Stoffer}}]{Colangelo:2014dfa}%
  \BibitemOpen
  \bibfield  {author} {\bibinfo {author} {\bibfnamefont {G.}~\bibnamefont
  {Colangelo}}, \bibinfo {author} {\bibfnamefont {M.}~\bibnamefont
  {Hoferichter}}, \bibinfo {author} {\bibfnamefont {M.}~\bibnamefont
  {Procura}},\ and\ \bibinfo {author} {\bibfnamefont {P.}~\bibnamefont
  {Stoffer}},\ }\bibfield  {title} {\bibinfo {title} {{Dispersive approach to
  hadronic light-by-light scattering}},\ }\href
  {https://doi.org/10.1007/JHEP09(2014)091} {\bibfield  {journal} {\bibinfo
  {journal} {JHEP}\ }\textbf {\bibinfo {volume} {09}},\ \bibinfo {pages}
  {091}},\ \Eprint {https://arxiv.org/abs/1402.7081} {arXiv:1402.7081 [hep-ph]}
  \BibitemShut {NoStop}%
\bibitem [{\citenamefont {Hoferichter}\ \emph {et~al.}(2018)\citenamefont
  {Hoferichter}, \citenamefont {Hoid}, \citenamefont {Kubis}, \citenamefont
  {Leupold},\ and\ \citenamefont {Schneider}}]{Hoferichter:2018kwz}%
  \BibitemOpen
  \bibfield  {author} {\bibinfo {author} {\bibfnamefont {M.}~\bibnamefont
  {Hoferichter}}, \bibinfo {author} {\bibfnamefont {B.-L.}\ \bibnamefont
  {Hoid}}, \bibinfo {author} {\bibfnamefont {B.}~\bibnamefont {Kubis}},
  \bibinfo {author} {\bibfnamefont {S.}~\bibnamefont {Leupold}},\ and\ \bibinfo
  {author} {\bibfnamefont {S.~P.}\ \bibnamefont {Schneider}},\ }\bibfield
  {title} {\bibinfo {title} {{Dispersion relation for hadronic light-by-light
  scattering: pion pole}},\ }\href {https://doi.org/10.1007/JHEP10(2018)141}
  {\bibfield  {journal} {\bibinfo  {journal} {JHEP}\ }\textbf {\bibinfo
  {volume} {10}},\ \bibinfo {pages} {141}},\ \Eprint
  {https://arxiv.org/abs/1808.04823} {arXiv:1808.04823 [hep-ph]} \BibitemShut
  {NoStop}%
\bibitem [{\citenamefont {Danilkin}\ \emph {et~al.}(2019)\citenamefont
  {Danilkin}, \citenamefont {Redmer},\ and\ \citenamefont
  {Vanderhaeghen}}]{Danilkin:2019mhd}%
  \BibitemOpen
  \bibfield  {author} {\bibinfo {author} {\bibfnamefont {I.}~\bibnamefont
  {Danilkin}}, \bibinfo {author} {\bibfnamefont {C.~F.}\ \bibnamefont
  {Redmer}},\ and\ \bibinfo {author} {\bibfnamefont {M.}~\bibnamefont
  {Vanderhaeghen}},\ }\bibfield  {title} {\bibinfo {title} {{The hadronic
  light-by-light contribution to the muon{\textquoteright}s anomalous magnetic
  moment}},\ }\href {https://doi.org/10.1016/j.ppnp.2019.05.002} {\bibfield
  {journal} {\bibinfo  {journal} {Prog. Part. Nucl. Phys.}\ }\textbf {\bibinfo
  {volume} {107}},\ \bibinfo {pages} {20} (\bibinfo {year} {2019})},\ \Eprint
  {https://arxiv.org/abs/1901.10346} {arXiv:1901.10346 [hep-ph]} \BibitemShut
  {NoStop}%
\bibitem [{\citenamefont {Alexandrou}\ \emph
  {et~al.}(2023{\natexlab{a}})\citenamefont {Alexandrou} \emph
  {et~al.}}]{ExtendedTwistedMass:2022ofm}%
  \BibitemOpen
  \bibfield  {author} {\bibinfo {author} {\bibfnamefont {C.}~\bibnamefont
  {Alexandrou}} \emph {et~al.} (\bibinfo {collaboration} {Extended Twisted
  Mass}),\ }\bibfield  {title} {\bibinfo {title} {{$\eta\to\gamma^*\gamma^*$
  transition form factor and the hadronic light-by-light
  {\ensuremath{\eta}}-pole contribution to the muon $g-2$ from lattice QCD}},\
  }\href {https://doi.org/10.1103/PhysRevD.108.054509} {\bibfield  {journal}
  {\bibinfo  {journal} {Phys. Rev. D}\ }\textbf {\bibinfo {volume} {108}},\
  \bibinfo {pages} {054509} (\bibinfo {year} {2023}{\natexlab{a}})},\ \Eprint
  {https://arxiv.org/abs/2212.06704} {arXiv:2212.06704 [hep-lat]} \BibitemShut
  {NoStop}%
\bibitem [{\citenamefont {Alexandrou}\ \emph
  {et~al.}(2023{\natexlab{b}})\citenamefont {Alexandrou} \emph
  {et~al.}}]{ExtendedTwistedMass:2023hin}%
  \BibitemOpen
  \bibfield  {author} {\bibinfo {author} {\bibfnamefont {C.}~\bibnamefont
  {Alexandrou}} \emph {et~al.} (\bibinfo {collaboration} {Extended Twisted
  Mass}),\ }\bibfield  {title} {\bibinfo {title} {{Pion transition form factor
  from twisted-mass lattice QCD and the hadronic light-by-light
  {\ensuremath{\pi}}0-pole contribution to the muon $g-2$}},\ }\href
  {https://doi.org/10.1103/PhysRevD.108.094514} {\bibfield  {journal} {\bibinfo
   {journal} {Phys. Rev. D}\ }\textbf {\bibinfo {volume} {108}},\ \bibinfo
  {pages} {094514} (\bibinfo {year} {2023}{\natexlab{b}})},\ \Eprint
  {https://arxiv.org/abs/2308.12458} {arXiv:2308.12458 [hep-lat]} \BibitemShut
  {NoStop}%
\bibitem [{\citenamefont {Holz}\ \emph {et~al.}(2025)\citenamefont {Holz},
  \citenamefont {Hoferichter}, \citenamefont {Hoid},\ and\ \citenamefont
  {Kubis}}]{Holz:2024lom}%
  \BibitemOpen
  \bibfield  {author} {\bibinfo {author} {\bibfnamefont {S.}~\bibnamefont
  {Holz}}, \bibinfo {author} {\bibfnamefont {M.}~\bibnamefont {Hoferichter}},
  \bibinfo {author} {\bibfnamefont {B.-L.}\ \bibnamefont {Hoid}},\ and\
  \bibinfo {author} {\bibfnamefont {B.}~\bibnamefont {Kubis}},\ }\bibfield
  {title} {\bibinfo {title} {{Precision Evaluation of the {\ensuremath{\eta}}-
  and $\eta'$-Pole Contributions to Hadronic Light-by-Light Scattering in the
  Anomalous Magnetic Moment of the Muon}},\ }\href
  {https://doi.org/10.1103/PhysRevLett.134.171902} {\bibfield  {journal}
  {\bibinfo  {journal} {Phys. Rev. Lett.}\ }\textbf {\bibinfo {volume} {134}},\
  \bibinfo {pages} {171902} (\bibinfo {year} {2025})},\ \Eprint
  {https://arxiv.org/abs/2411.08098} {arXiv:2411.08098 [hep-ph]} \BibitemShut
  {NoStop}%
\bibitem [{\citenamefont {Zhang}\ \emph {et~al.}(2025)\citenamefont {Zhang},
  \citenamefont {Jiang},\ and\ \citenamefont {Dai}}]{Zhang:2025ijd}%
  \BibitemOpen
  \bibfield  {author} {\bibinfo {author} {\bibfnamefont {Y.-H.}\ \bibnamefont
  {Zhang}}, \bibinfo {author} {\bibfnamefont {S.-Z.}\ \bibnamefont {Jiang}},\
  and\ \bibinfo {author} {\bibfnamefont {L.-Y.}\ \bibnamefont {Dai}},\
  }\bibfield  {title} {\bibinfo {title} {{Study of transition form factors of
  the lightest pseudoscalars}},\ }\href@noop {} {\  (\bibinfo {year} {2025})},\
  \Eprint {https://arxiv.org/abs/2509.06471} {arXiv:2509.06471 [hep-ph]}
  \BibitemShut {NoStop}%
\bibitem [{\citenamefont {Jegerlehner}(2017)}]{Jegerlehner:2017gek}%
  \BibitemOpen
  \bibfield  {author} {\bibinfo {author} {\bibfnamefont {F.}~\bibnamefont
  {Jegerlehner}},\ }\href {https://doi.org/10.1007/978-3-319-63577-4} {\emph
  {\bibinfo {title} {{The Anomalous Magnetic Moment of the Muon}}}},\ Vol.\
  \bibinfo {volume} {274}\ (\bibinfo  {publisher} {Springer},\ \bibinfo
  {address} {Cham},\ \bibinfo {year} {2017})\BibitemShut {NoStop}%
\bibitem [{\citenamefont {Ignatov}\ \emph {et~al.}(2024)\citenamefont {Ignatov}
  \emph {et~al.}}]{CMD-3:2023alj}%
  \BibitemOpen
  \bibfield  {author} {\bibinfo {author} {\bibfnamefont {F.~V.}\ \bibnamefont
  {Ignatov}} \emph {et~al.} (\bibinfo {collaboration} {CMD-3}),\ }\bibfield
  {title} {\bibinfo {title} {{Measurement of the $e^+e^-\to\pi^+\pi^-$ cross
  section from threshold to 1.2~GeV with the CMD-3 detector}},\ }\href
  {https://doi.org/10.1103/PhysRevD.109.112002} {\bibfield  {journal} {\bibinfo
   {journal} {Phys. Rev. D}\ }\textbf {\bibinfo {volume} {109}},\ \bibinfo
  {pages} {112002} (\bibinfo {year} {2024})},\ \Eprint
  {https://arxiv.org/abs/2302.08834} {arXiv:2302.08834 [hep-ex]} \BibitemShut
  {NoStop}%
\bibitem [{\citenamefont {Blum}\ \emph {et~al.}(2020)\citenamefont {Blum},
  \citenamefont {Christ}, \citenamefont {Hayakawa}, \citenamefont {Izubuchi},
  \citenamefont {Jin}, \citenamefont {Jung},\ and\ \citenamefont
  {Lehner}}]{Blum:2019ugy}%
  \BibitemOpen
  \bibfield  {author} {\bibinfo {author} {\bibfnamefont {T.}~\bibnamefont
  {Blum}}, \bibinfo {author} {\bibfnamefont {N.}~\bibnamefont {Christ}},
  \bibinfo {author} {\bibfnamefont {M.}~\bibnamefont {Hayakawa}}, \bibinfo
  {author} {\bibfnamefont {T.}~\bibnamefont {Izubuchi}}, \bibinfo {author}
  {\bibfnamefont {L.}~\bibnamefont {Jin}}, \bibinfo {author} {\bibfnamefont
  {C.}~\bibnamefont {Jung}},\ and\ \bibinfo {author} {\bibfnamefont
  {C.}~\bibnamefont {Lehner}},\ }\bibfield  {title} {\bibinfo {title}
  {{Hadronic Light-by-Light Scattering Contribution to the Muon Anomalous
  Magnetic Moment from Lattice QCD}},\ }\href
  {https://doi.org/10.1103/PhysRevLett.124.132002} {\bibfield  {journal}
  {\bibinfo  {journal} {Phys. Rev. Lett.}\ }\textbf {\bibinfo {volume} {124}},\
  \bibinfo {pages} {132002} (\bibinfo {year} {2020})},\ \Eprint
  {https://arxiv.org/abs/1911.08123} {arXiv:1911.08123 [hep-lat]} \BibitemShut
  {NoStop}%
\bibitem [{\citenamefont {Colangelo}\ \emph
  {et~al.}(2014{\natexlab{b}})\citenamefont {Colangelo}, \citenamefont
  {Hoferichter}, \citenamefont {Kubis}, \citenamefont {Procura},\ and\
  \citenamefont {Stoffer}}]{Colangelo:2014pva}%
  \BibitemOpen
  \bibfield  {author} {\bibinfo {author} {\bibfnamefont {G.}~\bibnamefont
  {Colangelo}}, \bibinfo {author} {\bibfnamefont {M.}~\bibnamefont
  {Hoferichter}}, \bibinfo {author} {\bibfnamefont {B.}~\bibnamefont {Kubis}},
  \bibinfo {author} {\bibfnamefont {M.}~\bibnamefont {Procura}},\ and\ \bibinfo
  {author} {\bibfnamefont {P.}~\bibnamefont {Stoffer}},\ }\bibfield  {title}
  {\bibinfo {title} {{Towards a data-driven analysis of hadronic light-by-light
  scattering}},\ }\href {https://doi.org/10.1016/j.physletb.2014.09.021}
  {\bibfield  {journal} {\bibinfo  {journal} {Phys. Lett. B}\ }\textbf
  {\bibinfo {volume} {738}},\ \bibinfo {pages} {6} (\bibinfo {year}
  {2014}{\natexlab{b}})},\ \Eprint {https://arxiv.org/abs/1408.2517}
  {arXiv:1408.2517 [hep-ph]} \BibitemShut {NoStop}%
\bibitem [{\citenamefont {Borsanyi}\ \emph {et~al.}(2021)\citenamefont
  {Borsanyi} \emph {et~al.}}]{Borsanyi:2020mff}%
  \BibitemOpen
  \bibfield  {author} {\bibinfo {author} {\bibfnamefont {S.}~\bibnamefont
  {Borsanyi}} \emph {et~al.},\ }\bibfield  {title} {\bibinfo {title} {{Leading
  hadronic contribution to the muon magnetic moment from lattice QCD}},\ }\href
  {https://doi.org/10.1038/s41586-021-03418-1} {\bibfield  {journal} {\bibinfo
  {journal} {Nature}\ }\textbf {\bibinfo {volume} {593}},\ \bibinfo {pages}
  {51} (\bibinfo {year} {2021})},\ \Eprint {https://arxiv.org/abs/2002.12347}
  {arXiv:2002.12347 [hep-lat]} \BibitemShut {NoStop}%
\bibitem [{\citenamefont {Blum}\ \emph {et~al.}(2023)\citenamefont {Blum} \emph
  {et~al.}}]{RBC:2023pvn}%
  \BibitemOpen
  \bibfield  {author} {\bibinfo {author} {\bibfnamefont {T.}~\bibnamefont
  {Blum}} \emph {et~al.} (\bibinfo {collaboration} {RBC, UKQCD}),\ }\bibfield
  {title} {\bibinfo {title} {{Update of Euclidean windows of the hadronic
  vacuum polarization}},\ }\href {https://doi.org/10.1103/PhysRevD.108.054507}
  {\bibfield  {journal} {\bibinfo  {journal} {Phys. Rev. D}\ }\textbf {\bibinfo
  {volume} {108}},\ \bibinfo {pages} {054507} (\bibinfo {year} {2023})},\
  \Eprint {https://arxiv.org/abs/2301.08696} {arXiv:2301.08696 [hep-lat]}
  \BibitemShut {NoStop}%
\bibitem [{\citenamefont {Boccaletti}\ \emph {et~al.}(2026)\citenamefont
  {Boccaletti} \emph {et~al.}}]{Boccaletti:2024guq}%
  \BibitemOpen
  \bibfield  {author} {\bibinfo {author} {\bibfnamefont {A.}~\bibnamefont
  {Boccaletti}} \emph {et~al.},\ }\bibfield  {title} {\bibinfo {title} {{Hybrid
  calculation of hadronic vacuum polarization in muon $g-2$ to 0.48{\%}}},\
  }\href {https://doi.org/10.1038/s41586-026-10449-z} {\bibfield  {journal}
  {\bibinfo  {journal} {Nature}\ }\textbf {\bibinfo {volume} {653}},\ \bibinfo
  {pages} {373} (\bibinfo {year} {2026})},\ \Eprint
  {https://arxiv.org/abs/2407.10913} {arXiv:2407.10913 [hep-lat]} \BibitemShut
  {NoStop}%
\bibitem [{\citenamefont {Witten}(1998)}]{Witten:1998qj}%
  \BibitemOpen
  \bibfield  {author} {\bibinfo {author} {\bibfnamefont {E.}~\bibnamefont
  {Witten}},\ }\bibfield  {title} {\bibinfo {title} {{Anti de Sitter space and
  holography}},\ }\href {https://doi.org/10.4310/ATMP.1998.v2.n2.a2} {\bibfield
   {journal} {\bibinfo  {journal} {Adv. Theor. Math. Phys.}\ }\textbf {\bibinfo
  {volume} {2}},\ \bibinfo {pages} {253} (\bibinfo {year} {1998})},\ \Eprint
  {https://arxiv.org/abs/hep-th/9802150} {arXiv:hep-th/9802150} \BibitemShut
  {NoStop}%
\bibitem [{\citenamefont {Gubser}\ \emph {et~al.}(1998)\citenamefont {Gubser},
  \citenamefont {Klebanov},\ and\ \citenamefont {Polyakov}}]{Gubser:1998bc}%
  \BibitemOpen
  \bibfield  {author} {\bibinfo {author} {\bibfnamefont {S.~S.}\ \bibnamefont
  {Gubser}}, \bibinfo {author} {\bibfnamefont {I.~R.}\ \bibnamefont
  {Klebanov}},\ and\ \bibinfo {author} {\bibfnamefont {A.~M.}\ \bibnamefont
  {Polyakov}},\ }\bibfield  {title} {\bibinfo {title} {{Gauge theory
  correlators from noncritical string theory}},\ }\href
  {https://doi.org/10.1016/S0370-2693(98)00377-3} {\bibfield  {journal}
  {\bibinfo  {journal} {Phys. Lett. B}\ }\textbf {\bibinfo {volume} {428}},\
  \bibinfo {pages} {105} (\bibinfo {year} {1998})},\ \Eprint
  {https://arxiv.org/abs/hep-th/9802109} {arXiv:hep-th/9802109} \BibitemShut
  {NoStop}%
\bibitem [{\citenamefont {Maldacena}(1998)}]{Maldacena:1997re}%
  \BibitemOpen
  \bibfield  {author} {\bibinfo {author} {\bibfnamefont {J.~M.}\ \bibnamefont
  {Maldacena}},\ }\bibfield  {title} {\bibinfo {title} {{The Large $N$ limit of
  superconformal field theories and supergravity}},\ }\href
  {https://doi.org/10.4310/ATMP.1998.v2.n2.a1} {\bibfield  {journal} {\bibinfo
  {journal} {Adv. Theor. Math. Phys.}\ }\textbf {\bibinfo {volume} {2}},\
  \bibinfo {pages} {231} (\bibinfo {year} {1998})},\ \Eprint
  {https://arxiv.org/abs/hep-th/9711200} {arXiv:hep-th/9711200} \BibitemShut
  {NoStop}%
\bibitem [{\citenamefont {Erlich}\ \emph {et~al.}(2005)\citenamefont {Erlich},
  \citenamefont {Katz}, \citenamefont {Son},\ and\ \citenamefont
  {Stephanov}}]{Erlich:2005qh}%
  \BibitemOpen
  \bibfield  {author} {\bibinfo {author} {\bibfnamefont {J.}~\bibnamefont
  {Erlich}}, \bibinfo {author} {\bibfnamefont {E.}~\bibnamefont {Katz}},
  \bibinfo {author} {\bibfnamefont {D.~T.}\ \bibnamefont {Son}},\ and\ \bibinfo
  {author} {\bibfnamefont {M.~A.}\ \bibnamefont {Stephanov}},\ }\bibfield
  {title} {\bibinfo {title} {{QCD and a holographic model of hadrons}},\ }\href
  {https://doi.org/10.1103/PhysRevLett.95.261602} {\bibfield  {journal}
  {\bibinfo  {journal} {Phys. Rev. Lett.}\ }\textbf {\bibinfo {volume} {95}},\
  \bibinfo {pages} {261602} (\bibinfo {year} {2005})},\ \Eprint
  {https://arxiv.org/abs/hep-ph/0501128} {arXiv:hep-ph/0501128} \BibitemShut
  {NoStop}%
\bibitem [{\citenamefont {Karch}\ \emph {et~al.}(2006)\citenamefont {Karch},
  \citenamefont {Katz}, \citenamefont {Son},\ and\ \citenamefont
  {Stephanov}}]{Karch:2006pv}%
  \BibitemOpen
  \bibfield  {author} {\bibinfo {author} {\bibfnamefont {A.}~\bibnamefont
  {Karch}}, \bibinfo {author} {\bibfnamefont {E.}~\bibnamefont {Katz}},
  \bibinfo {author} {\bibfnamefont {D.~T.}\ \bibnamefont {Son}},\ and\ \bibinfo
  {author} {\bibfnamefont {M.~A.}\ \bibnamefont {Stephanov}},\ }\bibfield
  {title} {\bibinfo {title} {{Linear confinement and AdS/QCD}},\ }\href
  {https://doi.org/10.1103/PhysRevD.74.015005} {\bibfield  {journal} {\bibinfo
  {journal} {Phys. Rev. D}\ }\textbf {\bibinfo {volume} {74}},\ \bibinfo
  {pages} {015005} (\bibinfo {year} {2006})},\ \Eprint
  {https://arxiv.org/abs/hep-ph/0602229} {arXiv:hep-ph/0602229} \BibitemShut
  {NoStop}%
\bibitem [{\citenamefont {Da~Rold}\ and\ \citenamefont
  {Pomarol}(2005)}]{DaRold:2005mxj}%
  \BibitemOpen
  \bibfield  {author} {\bibinfo {author} {\bibfnamefont {L.}~\bibnamefont
  {Da~Rold}}\ and\ \bibinfo {author} {\bibfnamefont {A.}~\bibnamefont
  {Pomarol}},\ }\bibfield  {title} {\bibinfo {title} {{Chiral symmetry breaking
  from five dimensional spaces}},\ }\href
  {https://doi.org/10.1016/j.nuclphysb.2005.05.009} {\bibfield  {journal}
  {\bibinfo  {journal} {Nucl. Phys. B}\ }\textbf {\bibinfo {volume} {721}},\
  \bibinfo {pages} {79} (\bibinfo {year} {2005})},\ \Eprint
  {https://arxiv.org/abs/hep-ph/0501218} {arXiv:hep-ph/0501218} \BibitemShut
  {NoStop}%
\bibitem [{\citenamefont {Jarvinen}\ and\ \citenamefont
  {Kiritsis}(2012)}]{Jarvinen:2011qe}%
  \BibitemOpen
  \bibfield  {author} {\bibinfo {author} {\bibfnamefont {M.}~\bibnamefont
  {Jarvinen}}\ and\ \bibinfo {author} {\bibfnamefont {E.}~\bibnamefont
  {Kiritsis}},\ }\bibfield  {title} {\bibinfo {title} {{Holographic Models for
  QCD in the Veneziano Limit}},\ }\href
  {https://doi.org/10.1007/JHEP03(2012)002} {\bibfield  {journal} {\bibinfo
  {journal} {JHEP}\ }\textbf {\bibinfo {volume} {03}},\ \bibinfo {pages}
  {002}},\ \Eprint {https://arxiv.org/abs/1112.1261} {arXiv:1112.1261 [hep-ph]}
  \BibitemShut {NoStop}%
\bibitem [{\citenamefont {Arean}\ \emph {et~al.}(2013)\citenamefont {Arean},
  \citenamefont {Iatrakis}, \citenamefont {J{\"a}rvinen},\ and\ \citenamefont
  {Kiritsis}}]{Arean:2012mq}%
  \BibitemOpen
  \bibfield  {author} {\bibinfo {author} {\bibfnamefont {D.}~\bibnamefont
  {Arean}}, \bibinfo {author} {\bibfnamefont {I.}~\bibnamefont {Iatrakis}},
  \bibinfo {author} {\bibfnamefont {M.}~\bibnamefont {J{\"a}rvinen}},\ and\
  \bibinfo {author} {\bibfnamefont {E.}~\bibnamefont {Kiritsis}},\ }\bibfield
  {title} {\bibinfo {title} {{V-QCD: Spectra, the dilaton and the
  S-parameter}},\ }\href {https://doi.org/10.1016/j.physletb.2013.01.070}
  {\bibfield  {journal} {\bibinfo  {journal} {Phys. Lett. B}\ }\textbf
  {\bibinfo {volume} {720}},\ \bibinfo {pages} {219} (\bibinfo {year}
  {2013})},\ \Eprint {https://arxiv.org/abs/1211.6125} {arXiv:1211.6125
  [hep-ph]} \BibitemShut {NoStop}%
\bibitem [{\citenamefont {Li}\ and\ \citenamefont {Huang}(2013)}]{Li:2013oda}%
  \BibitemOpen
  \bibfield  {author} {\bibinfo {author} {\bibfnamefont {D.}~\bibnamefont
  {Li}}\ and\ \bibinfo {author} {\bibfnamefont {M.}~\bibnamefont {Huang}},\
  }\bibfield  {title} {\bibinfo {title} {{Dynamical holographic QCD model for
  glueball and light meson spectra}},\ }\href
  {https://doi.org/10.1007/JHEP11(2013)088} {\bibfield  {journal} {\bibinfo
  {journal} {JHEP}\ }\textbf {\bibinfo {volume} {11}},\ \bibinfo {pages}
  {088}},\ \Eprint {https://arxiv.org/abs/1303.6929} {arXiv:1303.6929 [hep-ph]}
  \BibitemShut {NoStop}%
\bibitem [{\citenamefont {Folco~Capossoli}\ \emph {et~al.}(2020)\citenamefont
  {Folco~Capossoli}, \citenamefont {Mart{\'\i}n~Contreras}, \citenamefont {Li},
  \citenamefont {Vega},\ and\ \citenamefont
  {Boschi-Filho}}]{FolcoCapossoli:2019imm}%
  \BibitemOpen
  \bibfield  {author} {\bibinfo {author} {\bibfnamefont {E.}~\bibnamefont
  {Folco~Capossoli}}, \bibinfo {author} {\bibfnamefont {M.~A.}\ \bibnamefont
  {Mart{\'\i}n~Contreras}}, \bibinfo {author} {\bibfnamefont {D.}~\bibnamefont
  {Li}}, \bibinfo {author} {\bibfnamefont {A.}~\bibnamefont {Vega}},\ and\
  \bibinfo {author} {\bibfnamefont {H.}~\bibnamefont {Boschi-Filho}},\
  }\bibfield  {title} {\bibinfo {title} {{Hadronic spectra from deformed AdS
  backgrounds}},\ }\href {https://doi.org/10.1088/1674-1137/44/6/064104}
  {\bibfield  {journal} {\bibinfo  {journal} {Chin. Phys. C}\ }\textbf
  {\bibinfo {volume} {44}},\ \bibinfo {pages} {064104} (\bibinfo {year}
  {2020})},\ \Eprint {https://arxiv.org/abs/1903.06269} {arXiv:1903.06269
  [hep-ph]} \BibitemShut {NoStop}%
\bibitem [{\citenamefont {Li}\ and\ \citenamefont {Huang}(2017)}]{Li:2016smq}%
  \BibitemOpen
  \bibfield  {author} {\bibinfo {author} {\bibfnamefont {D.}~\bibnamefont
  {Li}}\ and\ \bibinfo {author} {\bibfnamefont {M.}~\bibnamefont {Huang}},\
  }\bibfield  {title} {\bibinfo {title} {{Chiral phase transition of QCD with
  $N_f=2+1$ flavors from holography}},\ }\href
  {https://doi.org/10.1007/JHEP02(2017)042} {\bibfield  {journal} {\bibinfo
  {journal} {JHEP}\ }\textbf {\bibinfo {volume} {02}},\ \bibinfo {pages}
  {042}},\ \Eprint {https://arxiv.org/abs/1610.09814} {arXiv:1610.09814
  [hep-ph]} \BibitemShut {NoStop}%
\bibitem [{\citenamefont {Chen}\ \emph {et~al.}(2021)\citenamefont {Chen},
  \citenamefont {Zhang}, \citenamefont {Li}, \citenamefont {Hou},\ and\
  \citenamefont {Huang}}]{Chen:2020ath}%
  \BibitemOpen
  \bibfield  {author} {\bibinfo {author} {\bibfnamefont {X.}~\bibnamefont
  {Chen}}, \bibinfo {author} {\bibfnamefont {L.}~\bibnamefont {Zhang}},
  \bibinfo {author} {\bibfnamefont {D.}~\bibnamefont {Li}}, \bibinfo {author}
  {\bibfnamefont {D.}~\bibnamefont {Hou}},\ and\ \bibinfo {author}
  {\bibfnamefont {M.}~\bibnamefont {Huang}},\ }\bibfield  {title} {\bibinfo
  {title} {{Gluodynamics and deconfinement phase transition under rotation from
  holography}},\ }\href {https://doi.org/10.1007/JHEP07(2021)132} {\bibfield
  {journal} {\bibinfo  {journal} {JHEP}\ }\textbf {\bibinfo {volume} {07}},\
  \bibinfo {pages} {132}},\ \Eprint {https://arxiv.org/abs/2010.14478}
  {arXiv:2010.14478 [hep-ph]} \BibitemShut {NoStop}%
\bibitem [{\citenamefont {Li}\ \emph {et~al.}(2013)\citenamefont {Li},
  \citenamefont {Huang},\ and\ \citenamefont {Yan}}]{Li:2012ay}%
  \BibitemOpen
  \bibfield  {author} {\bibinfo {author} {\bibfnamefont {D.}~\bibnamefont
  {Li}}, \bibinfo {author} {\bibfnamefont {M.}~\bibnamefont {Huang}},\ and\
  \bibinfo {author} {\bibfnamefont {Q.-S.}\ \bibnamefont {Yan}},\ }\bibfield
  {title} {\bibinfo {title} {{A dynamical soft-wall holographic QCD model for
  chiral symmetry breaking and linear confinement}},\ }\href
  {https://doi.org/10.1140/epjc/s10052-013-2615-3} {\bibfield  {journal}
  {\bibinfo  {journal} {Eur. Phys. J. C}\ }\textbf {\bibinfo {volume} {73}},\
  \bibinfo {pages} {2615} (\bibinfo {year} {2013})},\ \Eprint
  {https://arxiv.org/abs/1206.2824} {arXiv:1206.2824 [hep-th]} \BibitemShut
  {NoStop}%
\bibitem [{\citenamefont {Chen}\ and\ \citenamefont
  {Huang}(2024)}]{Chen:2024ckb}%
  \BibitemOpen
  \bibfield  {author} {\bibinfo {author} {\bibfnamefont {X.}~\bibnamefont
  {Chen}}\ and\ \bibinfo {author} {\bibfnamefont {M.}~\bibnamefont {Huang}},\
  }\bibfield  {title} {\bibinfo {title} {{Machine learning holographic black
  hole from lattice QCD equation of state}},\ }\href
  {https://doi.org/10.1103/PhysRevD.109.L051902} {\bibfield  {journal}
  {\bibinfo  {journal} {Phys. Rev. D}\ }\textbf {\bibinfo {volume} {109}},\
  \bibinfo {pages} {L051902} (\bibinfo {year} {2024})},\ \Eprint
  {https://arxiv.org/abs/2401.06417} {arXiv:2401.06417 [hep-ph]} \BibitemShut
  {NoStop}%
\bibitem [{\citenamefont {He}\ \emph {et~al.}(2013)\citenamefont {He},
  \citenamefont {Wu}, \citenamefont {Yang},\ and\ \citenamefont
  {Yuan}}]{He:2013qq}%
  \BibitemOpen
  \bibfield  {author} {\bibinfo {author} {\bibfnamefont {S.}~\bibnamefont
  {He}}, \bibinfo {author} {\bibfnamefont {S.-Y.}\ \bibnamefont {Wu}}, \bibinfo
  {author} {\bibfnamefont {Y.}~\bibnamefont {Yang}},\ and\ \bibinfo {author}
  {\bibfnamefont {P.-H.}\ \bibnamefont {Yuan}},\ }\bibfield  {title} {\bibinfo
  {title} {{Phase Structure in a Dynamical Soft-Wall Holographic QCD Model}},\
  }\href {https://doi.org/10.1007/JHEP04(2013)093} {\bibfield  {journal}
  {\bibinfo  {journal} {JHEP}\ }\textbf {\bibinfo {volume} {04}},\ \bibinfo
  {pages} {093}},\ \Eprint {https://arxiv.org/abs/1301.0385} {arXiv:1301.0385
  [hep-th]} \BibitemShut {NoStop}%
\bibitem [{\citenamefont {Fang}\ \emph
  {et~al.}(2016{\natexlab{a}})\citenamefont {Fang}, \citenamefont {He},\ and\
  \citenamefont {Li}}]{Fang:2015ytf}%
  \BibitemOpen
  \bibfield  {author} {\bibinfo {author} {\bibfnamefont {Z.}~\bibnamefont
  {Fang}}, \bibinfo {author} {\bibfnamefont {S.}~\bibnamefont {He}},\ and\
  \bibinfo {author} {\bibfnamefont {D.}~\bibnamefont {Li}},\ }\bibfield
  {title} {\bibinfo {title} {{Chiral and Deconfining Phase Transitions from
  Holographic QCD Study}},\ }\href
  {https://doi.org/10.1016/j.nuclphysb.2016.04.003} {\bibfield  {journal}
  {\bibinfo  {journal} {Nucl. Phys. B}\ }\textbf {\bibinfo {volume} {907}},\
  \bibinfo {pages} {187} (\bibinfo {year} {2016}{\natexlab{a}})},\ \Eprint
  {https://arxiv.org/abs/1512.04062} {arXiv:1512.04062 [hep-ph]} \BibitemShut
  {NoStop}%
\bibitem [{\citenamefont {de~Teramond}\ and\ \citenamefont
  {Brodsky}(2005)}]{deTeramond:2005su}%
  \BibitemOpen
  \bibfield  {author} {\bibinfo {author} {\bibfnamefont {G.~F.}\ \bibnamefont
  {de~Teramond}}\ and\ \bibinfo {author} {\bibfnamefont {S.~J.}\ \bibnamefont
  {Brodsky}},\ }\bibfield  {title} {\bibinfo {title} {{Hadronic spectrum of a
  holographic dual of QCD}},\ }\href
  {https://doi.org/10.1103/PhysRevLett.94.201601} {\bibfield  {journal}
  {\bibinfo  {journal} {Phys. Rev. Lett.}\ }\textbf {\bibinfo {volume} {94}},\
  \bibinfo {pages} {201601} (\bibinfo {year} {2005})},\ \Eprint
  {https://arxiv.org/abs/hep-th/0501022} {arXiv:hep-th/0501022} \BibitemShut
  {NoStop}%
\bibitem [{\citenamefont {Sui}\ \emph {et~al.}(2010)\citenamefont {Sui},
  \citenamefont {Wu}, \citenamefont {Xie},\ and\ \citenamefont
  {Yang}}]{Sui:2009xe}%
  \BibitemOpen
  \bibfield  {author} {\bibinfo {author} {\bibfnamefont {Y.-Q.}\ \bibnamefont
  {Sui}}, \bibinfo {author} {\bibfnamefont {Y.-L.}\ \bibnamefont {Wu}},
  \bibinfo {author} {\bibfnamefont {Z.-F.}\ \bibnamefont {Xie}},\ and\ \bibinfo
  {author} {\bibfnamefont {Y.-B.}\ \bibnamefont {Yang}},\ }\bibfield  {title}
  {\bibinfo {title} {{Prediction for the Mass Spectra of Resonance Mesons in
  the Soft-Wall AdS/QCD with a Modified 5D Metric}},\ }\href
  {https://doi.org/10.1103/PhysRevD.81.014024} {\bibfield  {journal} {\bibinfo
  {journal} {Phys. Rev. D}\ }\textbf {\bibinfo {volume} {81}},\ \bibinfo
  {pages} {014024} (\bibinfo {year} {2010})},\ \Eprint
  {https://arxiv.org/abs/0909.3887} {arXiv:0909.3887 [hep-ph]} \BibitemShut
  {NoStop}%
\bibitem [{\citenamefont {Gherghetta}\ \emph {et~al.}(2009)\citenamefont
  {Gherghetta}, \citenamefont {Kapusta},\ and\ \citenamefont
  {Kelley}}]{Gherghetta:2009ac}%
  \BibitemOpen
  \bibfield  {author} {\bibinfo {author} {\bibfnamefont {T.}~\bibnamefont
  {Gherghetta}}, \bibinfo {author} {\bibfnamefont {J.~I.}\ \bibnamefont
  {Kapusta}},\ and\ \bibinfo {author} {\bibfnamefont {T.~M.}\ \bibnamefont
  {Kelley}},\ }\bibfield  {title} {\bibinfo {title} {{Chiral symmetry breaking
  in the soft-wall AdS/QCD model}},\ }\href
  {https://doi.org/10.1103/PhysRevD.79.076003} {\bibfield  {journal} {\bibinfo
  {journal} {Phys. Rev. D}\ }\textbf {\bibinfo {volume} {79}},\ \bibinfo
  {pages} {076003} (\bibinfo {year} {2009})},\ \Eprint
  {https://arxiv.org/abs/0902.1998} {arXiv:0902.1998 [hep-ph]} \BibitemShut
  {NoStop}%
\bibitem [{\citenamefont {Chelabi}\ \emph {et~al.}(2016)\citenamefont
  {Chelabi}, \citenamefont {Fang}, \citenamefont {Huang}, \citenamefont {Li},\
  and\ \citenamefont {Wu}}]{Chelabi:2015gpc}%
  \BibitemOpen
  \bibfield  {author} {\bibinfo {author} {\bibfnamefont {K.}~\bibnamefont
  {Chelabi}}, \bibinfo {author} {\bibfnamefont {Z.}~\bibnamefont {Fang}},
  \bibinfo {author} {\bibfnamefont {M.}~\bibnamefont {Huang}}, \bibinfo
  {author} {\bibfnamefont {D.}~\bibnamefont {Li}},\ and\ \bibinfo {author}
  {\bibfnamefont {Y.-L.}\ \bibnamefont {Wu}},\ }\bibfield  {title} {\bibinfo
  {title} {{Chiral Phase Transition in the Soft-Wall Model of AdS/QCD}},\
  }\href {https://doi.org/10.1007/JHEP04(2016)036} {\bibfield  {journal}
  {\bibinfo  {journal} {JHEP}\ }\textbf {\bibinfo {volume} {04}},\ \bibinfo
  {pages} {036}},\ \Eprint {https://arxiv.org/abs/1512.06493} {arXiv:1512.06493
  [hep-ph]} \BibitemShut {NoStop}%
\bibitem [{\citenamefont {Boschi-Filho}\ \emph {et~al.}(2006)\citenamefont
  {Boschi-Filho}, \citenamefont {Braga},\ and\ \citenamefont
  {Carrion}}]{Boschi-Filho:2005xct}%
  \BibitemOpen
  \bibfield  {author} {\bibinfo {author} {\bibfnamefont {H.}~\bibnamefont
  {Boschi-Filho}}, \bibinfo {author} {\bibfnamefont {N.~R.~F.}\ \bibnamefont
  {Braga}},\ and\ \bibinfo {author} {\bibfnamefont {H.~L.}\ \bibnamefont
  {Carrion}},\ }\bibfield  {title} {\bibinfo {title} {{Glueball Regge
  trajectories from gauge/string duality and the Pomeron}},\ }\href
  {https://doi.org/10.1103/PhysRevD.73.047901} {\bibfield  {journal} {\bibinfo
  {journal} {Phys. Rev. D}\ }\textbf {\bibinfo {volume} {73}},\ \bibinfo
  {pages} {047901} (\bibinfo {year} {2006})},\ \Eprint
  {https://arxiv.org/abs/hep-th/0507063} {arXiv:hep-th/0507063} \BibitemShut
  {NoStop}%
\bibitem [{\citenamefont {Shen}\ \emph {et~al.}(2025)\citenamefont {Shen},
  \citenamefont {Liu}, \citenamefont {Wu}, \citenamefont {Wu},\ and\
  \citenamefont {Fang}}]{Shen:2025zkj}%
  \BibitemOpen
  \bibfield  {author} {\bibinfo {author} {\bibfnamefont {J.-Y.}\ \bibnamefont
  {Shen}}, \bibinfo {author} {\bibfnamefont {X.-Y.}\ \bibnamefont {Liu}},
  \bibinfo {author} {\bibfnamefont {J.-R.}\ \bibnamefont {Wu}}, \bibinfo
  {author} {\bibfnamefont {Y.-L.}\ \bibnamefont {Wu}},\ and\ \bibinfo {author}
  {\bibfnamefont {Z.}~\bibnamefont {Fang}},\ }\bibfield  {title} {\bibinfo
  {title} {{Toward a holographic realization of the 2+1-flavor QCD phase
  structure}},\ }\href {https://doi.org/10.1103/3plt-wlt7} {\bibfield
  {journal} {\bibinfo  {journal} {Phys. Rev. D}\ }\textbf {\bibinfo {volume}
  {112}},\ \bibinfo {pages} {L111504} (\bibinfo {year} {2025})},\ \Eprint
  {https://arxiv.org/abs/2511.11273} {arXiv:2511.11273 [hep-ph]} \BibitemShut
  {NoStop}%
\bibitem [{\citenamefont {Liu}\ \emph {et~al.}(2026)\citenamefont {Liu},
  \citenamefont {Wu},\ and\ \citenamefont {Fang}}]{Liu:2026cpr}%
  \BibitemOpen
  \bibfield  {author} {\bibinfo {author} {\bibfnamefont {X.-Y.}\ \bibnamefont
  {Liu}}, \bibinfo {author} {\bibfnamefont {Y.-L.}\ \bibnamefont {Wu}},\ and\
  \bibinfo {author} {\bibfnamefont {Z.}~\bibnamefont {Fang}},\ }\bibfield
  {title} {\bibinfo {title} {{Axial-anomaly effects and chiral phase structure
  in holographic QCD}},\ }\href@noop {} {\  (\bibinfo {year} {2026})},\ \Eprint
  {https://arxiv.org/abs/2603.12616} {arXiv:2603.12616 [hep-ph]} \BibitemShut
  {NoStop}%
\bibitem [{\citenamefont {Cappiello}\ \emph {et~al.}(2011)\citenamefont
  {Cappiello}, \citenamefont {Cata},\ and\ \citenamefont
  {D'Ambrosio}}]{Cappiello:2010uy}%
  \BibitemOpen
  \bibfield  {author} {\bibinfo {author} {\bibfnamefont {L.}~\bibnamefont
  {Cappiello}}, \bibinfo {author} {\bibfnamefont {O.}~\bibnamefont {Cata}},\
  and\ \bibinfo {author} {\bibfnamefont {G.}~\bibnamefont {D'Ambrosio}},\
  }\bibfield  {title} {\bibinfo {title} {{The hadronic light by light
  contribution to the $(g-2)_{\mu}$ with holographic models of QCD}},\ }\href
  {https://doi.org/10.1103/PhysRevD.83.093006} {\bibfield  {journal} {\bibinfo
  {journal} {Phys. Rev. D}\ }\textbf {\bibinfo {volume} {83}},\ \bibinfo
  {pages} {093006} (\bibinfo {year} {2011})},\ \Eprint
  {https://arxiv.org/abs/1009.1161} {arXiv:1009.1161 [hep-ph]} \BibitemShut
  {NoStop}%
\bibitem [{\citenamefont {Leutgeb}\ \emph {et~al.}(2019)\citenamefont
  {Leutgeb}, \citenamefont {Mager},\ and\ \citenamefont
  {Rebhan}}]{Leutgeb:2019zpq}%
  \BibitemOpen
  \bibfield  {author} {\bibinfo {author} {\bibfnamefont {J.}~\bibnamefont
  {Leutgeb}}, \bibinfo {author} {\bibfnamefont {J.}~\bibnamefont {Mager}},\
  and\ \bibinfo {author} {\bibfnamefont {A.}~\bibnamefont {Rebhan}},\
  }\bibfield  {title} {\bibinfo {title} {{Pseudoscalar transition form factors
  and the hadronic light-by-light contribution to the anomalous magnetic moment
  of the muon from holographic QCD}},\ }\href
  {https://doi.org/10.1103/PhysRevD.104.059903} {\bibfield  {journal} {\bibinfo
   {journal} {Phys. Rev. D}\ }\textbf {\bibinfo {volume} {100}},\ \bibinfo
  {pages} {094038} (\bibinfo {year} {2019})},\ \bibinfo {note} {[Erratum:
  Phys.Rev.D 104, 059903 (2021)]},\ \Eprint {https://arxiv.org/abs/1906.11795}
  {arXiv:1906.11795 [hep-ph]} \BibitemShut {NoStop}%
\bibitem [{\citenamefont {Leutgeb}\ and\ \citenamefont
  {Rebhan}(2020)}]{Leutgeb:2019gbz}%
  \BibitemOpen
  \bibfield  {author} {\bibinfo {author} {\bibfnamefont {J.}~\bibnamefont
  {Leutgeb}}\ and\ \bibinfo {author} {\bibfnamefont {A.}~\bibnamefont
  {Rebhan}},\ }\bibfield  {title} {\bibinfo {title} {{Axial vector transition
  form factors in holographic QCD and their contribution to the anomalous
  magnetic moment of the muon}},\ }\href
  {https://doi.org/10.1103/PhysRevD.101.114015} {\bibfield  {journal} {\bibinfo
   {journal} {Phys. Rev. D}\ }\textbf {\bibinfo {volume} {101}},\ \bibinfo
  {pages} {114015} (\bibinfo {year} {2020})},\ \Eprint
  {https://arxiv.org/abs/1912.01596} {arXiv:1912.01596 [hep-ph]} \BibitemShut
  {NoStop}%
\bibitem [{\citenamefont {Leutgeb}\ and\ \citenamefont
  {Rebhan}(2021)}]{Leutgeb:2021mpu}%
  \BibitemOpen
  \bibfield  {author} {\bibinfo {author} {\bibfnamefont {J.}~\bibnamefont
  {Leutgeb}}\ and\ \bibinfo {author} {\bibfnamefont {A.}~\bibnamefont
  {Rebhan}},\ }\bibfield  {title} {\bibinfo {title} {{Hadronic light-by-light
  contribution to the muon $g-2$ from holographic QCD with massive pions}},\
  }\href {https://doi.org/10.1103/PhysRevD.104.094017} {\bibfield  {journal}
  {\bibinfo  {journal} {Phys. Rev. D}\ }\textbf {\bibinfo {volume} {104}},\
  \bibinfo {pages} {094017} (\bibinfo {year} {2021})},\ \Eprint
  {https://arxiv.org/abs/2108.12345} {arXiv:2108.12345 [hep-ph]} \BibitemShut
  {NoStop}%
\bibitem [{\citenamefont {Leutgeb}\ \emph {et~al.}(2023)\citenamefont
  {Leutgeb}, \citenamefont {Mager},\ and\ \citenamefont
  {Rebhan}}]{Leutgeb:2022lqw}%
  \BibitemOpen
  \bibfield  {author} {\bibinfo {author} {\bibfnamefont {J.}~\bibnamefont
  {Leutgeb}}, \bibinfo {author} {\bibfnamefont {J.}~\bibnamefont {Mager}},\
  and\ \bibinfo {author} {\bibfnamefont {A.}~\bibnamefont {Rebhan}},\
  }\bibfield  {title} {\bibinfo {title} {{Hadronic light-by-light contribution
  to the muon $g-2$ from holographic QCD with solved $U(1)_A$ problem}},\
  }\href {https://doi.org/10.1103/PhysRevD.107.054021} {\bibfield  {journal}
  {\bibinfo  {journal} {Phys. Rev. D}\ }\textbf {\bibinfo {volume} {107}},\
  \bibinfo {pages} {054021} (\bibinfo {year} {2023})},\ \Eprint
  {https://arxiv.org/abs/2211.16562} {arXiv:2211.16562 [hep-ph]} \BibitemShut
  {NoStop}%
\bibitem [{\citenamefont {Mager}\ \emph {et~al.}(2025)\citenamefont {Mager},
  \citenamefont {Cappiello}, \citenamefont {Leutgeb},\ and\ \citenamefont
  {Rebhan}}]{Mager:2025pvz}%
  \BibitemOpen
  \bibfield  {author} {\bibinfo {author} {\bibfnamefont {J.}~\bibnamefont
  {Mager}}, \bibinfo {author} {\bibfnamefont {L.}~\bibnamefont {Cappiello}},
  \bibinfo {author} {\bibfnamefont {J.}~\bibnamefont {Leutgeb}},\ and\ \bibinfo
  {author} {\bibfnamefont {A.}~\bibnamefont {Rebhan}},\ }\bibfield  {title}
  {\bibinfo {title} {{Longitudinal Short-Distance Constraints on Hadronic
  Light-by-Light Scattering and Tensor-Meson Contributions to the Muon
  $g-2$}},\ }\href {https://doi.org/10.1103/dxwr-gpsl} {\bibfield  {journal}
  {\bibinfo  {journal} {Phys. Rev. Lett.}\ }\textbf {\bibinfo {volume} {135}},\
  \bibinfo {pages} {091901} (\bibinfo {year} {2025})},\ \Eprint
  {https://arxiv.org/abs/2501.19293} {arXiv:2501.19293 [hep-ph]} \BibitemShut
  {NoStop}%
\bibitem [{\citenamefont {Colangelo}\ \emph {et~al.}(2024)\citenamefont
  {Colangelo}, \citenamefont {Giannuzzi},\ and\ \citenamefont
  {Nicotri}}]{Colangelo:2024xfh}%
  \BibitemOpen
  \bibfield  {author} {\bibinfo {author} {\bibfnamefont {P.}~\bibnamefont
  {Colangelo}}, \bibinfo {author} {\bibfnamefont {F.}~\bibnamefont
  {Giannuzzi}},\ and\ \bibinfo {author} {\bibfnamefont {S.}~\bibnamefont
  {Nicotri}},\ }\bibfield  {title} {\bibinfo {title} {{Hadronic light-by-light
  scattering contributions to $(g-2)_\mu$ from axial-vector and tensor mesons
  in the holographic soft-wall model}},\ }\href
  {https://doi.org/10.1103/PhysRevD.109.094036} {\bibfield  {journal} {\bibinfo
   {journal} {Phys. Rev. D}\ }\textbf {\bibinfo {volume} {109}},\ \bibinfo
  {pages} {094036} (\bibinfo {year} {2024})},\ \Eprint
  {https://arxiv.org/abs/2402.07579} {arXiv:2402.07579 [hep-ph]} \BibitemShut
  {NoStop}%
\bibitem [{\citenamefont {Cappiello}\ \emph {et~al.}(2022)\citenamefont
  {Cappiello}, \citenamefont {Cat{\`a}},\ and\ \citenamefont
  {D'Ambrosio}}]{Cappiello:2021vzi}%
  \BibitemOpen
  \bibfield  {author} {\bibinfo {author} {\bibfnamefont {L.}~\bibnamefont
  {Cappiello}}, \bibinfo {author} {\bibfnamefont {O.}~\bibnamefont
  {Cat{\`a}}},\ and\ \bibinfo {author} {\bibfnamefont {G.}~\bibnamefont
  {D'Ambrosio}},\ }\bibfield  {title} {\bibinfo {title} {{Scalar resonances in
  the hadronic light-by-light contribution to the muon (g-2)}},\ }\href
  {https://doi.org/10.1103/PhysRevD.105.056020} {\bibfield  {journal} {\bibinfo
   {journal} {Phys. Rev. D}\ }\textbf {\bibinfo {volume} {105}},\ \bibinfo
  {pages} {056020} (\bibinfo {year} {2022})},\ \Eprint
  {https://arxiv.org/abs/2110.05962} {arXiv:2110.05962 [hep-ph]} \BibitemShut
  {NoStop}%
\bibitem [{\citenamefont {Cappiello}\ \emph {et~al.}(2020)\citenamefont
  {Cappiello}, \citenamefont {Cat{\`a}}, \citenamefont {D'Ambrosio},
  \citenamefont {Greynat},\ and\ \citenamefont {Iyer}}]{Cappiello:2019hwh}%
  \BibitemOpen
  \bibfield  {author} {\bibinfo {author} {\bibfnamefont {L.}~\bibnamefont
  {Cappiello}}, \bibinfo {author} {\bibfnamefont {O.}~\bibnamefont {Cat{\`a}}},
  \bibinfo {author} {\bibfnamefont {G.}~\bibnamefont {D'Ambrosio}}, \bibinfo
  {author} {\bibfnamefont {D.}~\bibnamefont {Greynat}},\ and\ \bibinfo {author}
  {\bibfnamefont {A.}~\bibnamefont {Iyer}},\ }\bibfield  {title} {\bibinfo
  {title} {{Axial-vector and pseudoscalar mesons in the hadronic light-by-light
  contribution to the muon $(g-2)$}},\ }\href
  {https://doi.org/10.1103/PhysRevD.102.016009} {\bibfield  {journal} {\bibinfo
   {journal} {Phys. Rev. D}\ }\textbf {\bibinfo {volume} {102}},\ \bibinfo
  {pages} {016009} (\bibinfo {year} {2020})},\ \Eprint
  {https://arxiv.org/abs/1912.02779} {arXiv:1912.02779 [hep-ph]} \BibitemShut
  {NoStop}%
\bibitem [{\citenamefont {Cappiello}\ \emph {et~al.}(2025)\citenamefont
  {Cappiello}, \citenamefont {Leutgeb}, \citenamefont {Mager},\ and\
  \citenamefont {Rebhan}}]{Cappiello:2025fyf}%
  \BibitemOpen
  \bibfield  {author} {\bibinfo {author} {\bibfnamefont {L.}~\bibnamefont
  {Cappiello}}, \bibinfo {author} {\bibfnamefont {J.}~\bibnamefont {Leutgeb}},
  \bibinfo {author} {\bibfnamefont {J.}~\bibnamefont {Mager}},\ and\ \bibinfo
  {author} {\bibfnamefont {A.}~\bibnamefont {Rebhan}},\ }\bibfield  {title}
  {\bibinfo {title} {{Tensor meson transition form factors in holographic QCD
  and the muon $g-2$}},\ }\href {https://doi.org/10.1007/JHEP07(2025)033}
  {\bibfield  {journal} {\bibinfo  {journal} {JHEP}\ }\textbf {\bibinfo
  {volume} {07}},\ \bibinfo {pages} {033}},\ \Eprint
  {https://arxiv.org/abs/2501.09699} {arXiv:2501.09699 [hep-ph]} \BibitemShut
  {NoStop}%
\bibitem [{\citenamefont {Cui}\ \emph {et~al.}(2016)\citenamefont {Cui},
  \citenamefont {Fang},\ and\ \citenamefont {Wu}}]{Cui:2013xva}%
  \BibitemOpen
  \bibfield  {author} {\bibinfo {author} {\bibfnamefont {L.-X.}\ \bibnamefont
  {Cui}}, \bibinfo {author} {\bibfnamefont {Z.}~\bibnamefont {Fang}},\ and\
  \bibinfo {author} {\bibfnamefont {Y.-L.}\ \bibnamefont {Wu}},\ }\bibfield
  {title} {\bibinfo {title} {{Infrared-Improved Soft-wall AdS/QCD Model for
  Mesons}},\ }\href {https://doi.org/10.1140/epjc/s10052-015-3866-y} {\bibfield
   {journal} {\bibinfo  {journal} {Eur. Phys. J. C}\ }\textbf {\bibinfo
  {volume} {76}},\ \bibinfo {pages} {22} (\bibinfo {year} {2016})},\ \Eprint
  {https://arxiv.org/abs/1310.6487} {arXiv:1310.6487 [hep-ph]} \BibitemShut
  {NoStop}%
\bibitem [{\citenamefont {Fang}\ \emph
  {et~al.}(2016{\natexlab{b}})\citenamefont {Fang}, \citenamefont {Wu},\ and\
  \citenamefont {Zhang}}]{Fang:2016nfj}%
  \BibitemOpen
  \bibfield  {author} {\bibinfo {author} {\bibfnamefont {Z.}~\bibnamefont
  {Fang}}, \bibinfo {author} {\bibfnamefont {Y.-L.}\ \bibnamefont {Wu}},\ and\
  \bibinfo {author} {\bibfnamefont {L.}~\bibnamefont {Zhang}},\ }\bibfield
  {title} {\bibinfo {title} {{Chiral phase transition and meson spectrum in
  improved soft-wall AdS/QCD}},\ }\href
  {https://doi.org/10.1016/j.physletb.2016.09.009} {\bibfield  {journal}
  {\bibinfo  {journal} {Phys. Lett. B}\ }\textbf {\bibinfo {volume} {762}},\
  \bibinfo {pages} {86} (\bibinfo {year} {2016}{\natexlab{b}})},\ \Eprint
  {https://arxiv.org/abs/1604.02571} {arXiv:1604.02571 [hep-ph]} \BibitemShut
  {NoStop}%
\bibitem [{\citenamefont {Fang}\ \emph {et~al.}(2019)\citenamefont {Fang},
  \citenamefont {Wu},\ and\ \citenamefont {Zhang}}]{Fang:2019lmd}%
  \BibitemOpen
  \bibfield  {author} {\bibinfo {author} {\bibfnamefont {Z.}~\bibnamefont
  {Fang}}, \bibinfo {author} {\bibfnamefont {Y.-L.}\ \bibnamefont {Wu}},\ and\
  \bibinfo {author} {\bibfnamefont {L.}~\bibnamefont {Zhang}},\ }\bibfield
  {title} {\bibinfo {title} {{Octet meson spectra and chiral phase diagram in
  the improved soft-wall AdS/QCD model}},\ }\href
  {https://doi.org/10.1103/PhysRevD.100.054008} {\bibfield  {journal} {\bibinfo
   {journal} {Phys. Rev. D}\ }\textbf {\bibinfo {volume} {100}},\ \bibinfo
  {pages} {054008} (\bibinfo {year} {2019})},\ \Eprint
  {https://arxiv.org/abs/1904.04695} {arXiv:1904.04695 [hep-ph]} \BibitemShut
  {NoStop}%
\bibitem [{\citenamefont {Leutgeb}\ \emph
  {et~al.}(2025{\natexlab{a}})\citenamefont {Leutgeb}, \citenamefont {Mager},\
  and\ \citenamefont {Rebhan}}]{Leutgeb:2025jmv}%
  \BibitemOpen
  \bibfield  {author} {\bibinfo {author} {\bibfnamefont {J.}~\bibnamefont
  {Leutgeb}}, \bibinfo {author} {\bibfnamefont {J.}~\bibnamefont {Mager}},\
  and\ \bibinfo {author} {\bibfnamefont {A.}~\bibnamefont {Rebhan}},\
  }\bibfield  {title} {\bibinfo {title} {{Divergences in the hadronic
  light-by-light amplitude of the holographic soft-wall model}},\ }\href@noop
  {} {\  (\bibinfo {year} {2025}{\natexlab{a}})},\ \Eprint
  {https://arxiv.org/abs/2511.11797} {arXiv:2511.11797 [hep-ph]} \BibitemShut
  {NoStop}%
\bibitem [{\citenamefont {Leutgeb}\ \emph
  {et~al.}(2025{\natexlab{b}})\citenamefont {Leutgeb}, \citenamefont {Mager},\
  and\ \citenamefont {Rebhan}}]{Leutgeb:2024rfs}%
  \BibitemOpen
  \bibfield  {author} {\bibinfo {author} {\bibfnamefont {J.}~\bibnamefont
  {Leutgeb}}, \bibinfo {author} {\bibfnamefont {J.}~\bibnamefont {Mager}},\
  and\ \bibinfo {author} {\bibfnamefont {A.}~\bibnamefont {Rebhan}},\
  }\bibfield  {title} {\bibinfo {title} {{Superconnections in AdS/QCD and the
  hadronic light-by-light contribution to the muon $g-2$}},\ }\href
  {https://doi.org/10.1103/PhysRevD.111.114001} {\bibfield  {journal} {\bibinfo
   {journal} {Phys. Rev. D}\ }\textbf {\bibinfo {volume} {111}},\ \bibinfo
  {pages} {114001} (\bibinfo {year} {2025}{\natexlab{b}})},\ \Eprint
  {https://arxiv.org/abs/2411.10432} {arXiv:2411.10432 [hep-ph]} \BibitemShut
  {NoStop}%
\bibitem [{\citenamefont {Colangelo}\ \emph {et~al.}(2008)\citenamefont
  {Colangelo}, \citenamefont {De~Fazio}, \citenamefont {Giannuzzi},
  \citenamefont {Jugeau},\ and\ \citenamefont {Nicotri}}]{Colangelo:2008us}%
  \BibitemOpen
  \bibfield  {author} {\bibinfo {author} {\bibfnamefont {P.}~\bibnamefont
  {Colangelo}}, \bibinfo {author} {\bibfnamefont {F.}~\bibnamefont {De~Fazio}},
  \bibinfo {author} {\bibfnamefont {F.}~\bibnamefont {Giannuzzi}}, \bibinfo
  {author} {\bibfnamefont {F.}~\bibnamefont {Jugeau}},\ and\ \bibinfo {author}
  {\bibfnamefont {S.}~\bibnamefont {Nicotri}},\ }\bibfield  {title} {\bibinfo
  {title} {{Light scalar mesons in the soft-wall model of AdS/QCD}},\ }\href
  {https://doi.org/10.1103/PhysRevD.78.055009} {\bibfield  {journal} {\bibinfo
  {journal} {Phys. Rev. D}\ }\textbf {\bibinfo {volume} {78}},\ \bibinfo
  {pages} {055009} (\bibinfo {year} {2008})},\ \Eprint
  {https://arxiv.org/abs/0807.1054} {arXiv:0807.1054 [hep-ph]} \BibitemShut
  {NoStop}%
\bibitem [{\citenamefont {Shifman}\ and\ \citenamefont
  {Vainshtein}(2008)}]{Shifman:2007xn}%
  \BibitemOpen
  \bibfield  {author} {\bibinfo {author} {\bibfnamefont {M.}~\bibnamefont
  {Shifman}}\ and\ \bibinfo {author} {\bibfnamefont {A.}~\bibnamefont
  {Vainshtein}},\ }\bibfield  {title} {\bibinfo {title} {{Highly Excited
  Mesons, Linear Regge Trajectories and the Pattern of the Chiral Symmetry
  Realization}},\ }\href {https://doi.org/10.1103/PhysRevD.77.034002}
  {\bibfield  {journal} {\bibinfo  {journal} {Phys. Rev. D}\ }\textbf {\bibinfo
  {volume} {77}},\ \bibinfo {pages} {034002} (\bibinfo {year} {2008})},\
  \Eprint {https://arxiv.org/abs/0710.0863} {arXiv:0710.0863 [hep-ph]}
  \BibitemShut {NoStop}%
\bibitem [{\citenamefont {Leutgeb}\ \emph {et~al.}(2022)\citenamefont
  {Leutgeb}, \citenamefont {Rebhan},\ and\ \citenamefont
  {Stadlbauer}}]{Leutgeb:2022cvg}%
  \BibitemOpen
  \bibfield  {author} {\bibinfo {author} {\bibfnamefont {J.}~\bibnamefont
  {Leutgeb}}, \bibinfo {author} {\bibfnamefont {A.}~\bibnamefont {Rebhan}},\
  and\ \bibinfo {author} {\bibfnamefont {M.}~\bibnamefont {Stadlbauer}},\
  }\bibfield  {title} {\bibinfo {title} {{Hadronic vacuum polarization
  contribution to the muon $g-2$ in holographic QCD}},\ }\href
  {https://doi.org/10.1103/PhysRevD.105.094032} {\bibfield  {journal} {\bibinfo
   {journal} {Phys. Rev. D}\ }\textbf {\bibinfo {volume} {105}},\ \bibinfo
  {pages} {094032} (\bibinfo {year} {2022})},\ \Eprint
  {https://arxiv.org/abs/2203.16508} {arXiv:2203.16508 [hep-ph]} \BibitemShut
  {NoStop}%
\bibitem [{\citenamefont {Navas}\ \emph {et~al.}(2024)\citenamefont {Navas}
  \emph {et~al.}}]{ParticleDataGroup:2024cfk}%
  \BibitemOpen
  \bibfield  {author} {\bibinfo {author} {\bibfnamefont {S.}~\bibnamefont
  {Navas}} \emph {et~al.} (\bibinfo {collaboration} {Particle Data Group}),\
  }\bibfield  {title} {\bibinfo {title} {{Review of particle physics}},\ }\href
  {https://doi.org/10.1103/PhysRevD.110.030001} {\bibfield  {journal} {\bibinfo
   {journal} {Phys. Rev. D}\ }\textbf {\bibinfo {volume} {110}},\ \bibinfo
  {pages} {030001} (\bibinfo {year} {2024})}\BibitemShut {NoStop}%
\bibitem [{\citenamefont {Blum}(2003)}]{Blum:2002ii}%
  \BibitemOpen
  \bibfield  {author} {\bibinfo {author} {\bibfnamefont {T.}~\bibnamefont
  {Blum}},\ }\bibfield  {title} {\bibinfo {title} {{Lattice calculation of the
  lowest order hadronic contribution to the muon anomalous magnetic moment}},\
  }\href {https://doi.org/10.1103/PhysRevLett.91.052001} {\bibfield  {journal}
  {\bibinfo  {journal} {Phys. Rev. Lett.}\ }\textbf {\bibinfo {volume} {91}},\
  \bibinfo {pages} {052001} (\bibinfo {year} {2003})},\ \Eprint
  {https://arxiv.org/abs/hep-lat/0212018} {arXiv:hep-lat/0212018} \BibitemShut
  {NoStop}%
\bibitem [{\citenamefont {Aubin}\ and\ \citenamefont
  {Blum}(2007)}]{Aubin:2006xv}%
  \BibitemOpen
  \bibfield  {author} {\bibinfo {author} {\bibfnamefont {C.}~\bibnamefont
  {Aubin}}\ and\ \bibinfo {author} {\bibfnamefont {T.}~\bibnamefont {Blum}},\
  }\bibfield  {title} {\bibinfo {title} {{Calculating the hadronic vacuum
  polarization and leading hadronic contribution to the muon anomalous magnetic
  moment with improved staggered quarks}},\ }\href
  {https://doi.org/10.1103/PhysRevD.75.114502} {\bibfield  {journal} {\bibinfo
  {journal} {Phys. Rev. D}\ }\textbf {\bibinfo {volume} {75}},\ \bibinfo
  {pages} {114502} (\bibinfo {year} {2007})},\ \Eprint
  {https://arxiv.org/abs/hep-lat/0608011} {arXiv:hep-lat/0608011} \BibitemShut
  {NoStop}%
\bibitem [{\citenamefont {Hong}\ \emph {et~al.}(2010)\citenamefont {Hong},
  \citenamefont {Kim},\ and\ \citenamefont {Matsuzaki}}]{Hong:2009jv}%
  \BibitemOpen
  \bibfield  {author} {\bibinfo {author} {\bibfnamefont {D.~K.}\ \bibnamefont
  {Hong}}, \bibinfo {author} {\bibfnamefont {D.}~\bibnamefont {Kim}},\ and\
  \bibinfo {author} {\bibfnamefont {S.}~\bibnamefont {Matsuzaki}},\ }\bibfield
  {title} {\bibinfo {title} {{Holographic calculation of hadronic contributions
  to muon $g-2$}},\ }\href {https://doi.org/10.1103/PhysRevD.81.073005}
  {\bibfield  {journal} {\bibinfo  {journal} {Phys. Rev. D}\ }\textbf {\bibinfo
  {volume} {81}},\ \bibinfo {pages} {073005} (\bibinfo {year} {2010})},\
  \Eprint {https://arxiv.org/abs/0911.0560} {arXiv:0911.0560 [hep-ph]}
  \BibitemShut {NoStop}%
\bibitem [{\citenamefont {Grigoryan}\ and\ \citenamefont
  {Radyushkin}(2007{\natexlab{a}})}]{Grigoryan:2007my}%
  \BibitemOpen
  \bibfield  {author} {\bibinfo {author} {\bibfnamefont {H.~R.}\ \bibnamefont
  {Grigoryan}}\ and\ \bibinfo {author} {\bibfnamefont {A.~V.}\ \bibnamefont
  {Radyushkin}},\ }\bibfield  {title} {\bibinfo {title} {{Structure of vector
  mesons in holographic model with linear confinement}},\ }\href
  {https://doi.org/10.1103/PhysRevD.76.095007} {\bibfield  {journal} {\bibinfo
  {journal} {Phys. Rev. D}\ }\textbf {\bibinfo {volume} {76}},\ \bibinfo
  {pages} {095007} (\bibinfo {year} {2007}{\natexlab{a}})},\ \Eprint
  {https://arxiv.org/abs/0706.1543} {arXiv:0706.1543 [hep-ph]} \BibitemShut
  {NoStop}%
\bibitem [{\citenamefont {Casero}\ \emph {et~al.}(2007)\citenamefont {Casero},
  \citenamefont {Kiritsis},\ and\ \citenamefont {Paredes}}]{Casero:2007ae}%
  \BibitemOpen
  \bibfield  {author} {\bibinfo {author} {\bibfnamefont {R.}~\bibnamefont
  {Casero}}, \bibinfo {author} {\bibfnamefont {E.}~\bibnamefont {Kiritsis}},\
  and\ \bibinfo {author} {\bibfnamefont {A.}~\bibnamefont {Paredes}},\
  }\bibfield  {title} {\bibinfo {title} {{Chiral symmetry breaking as open
  string tachyon condensation}},\ }\href
  {https://doi.org/10.1016/j.nuclphysb.2007.07.009} {\bibfield  {journal}
  {\bibinfo  {journal} {Nucl. Phys. B}\ }\textbf {\bibinfo {volume} {787}},\
  \bibinfo {pages} {98} (\bibinfo {year} {2007})},\ \Eprint
  {https://arxiv.org/abs/hep-th/0702155} {arXiv:hep-th/0702155} \BibitemShut
  {NoStop}%
\bibitem [{\citenamefont {Iatrakis}\ \emph
  {et~al.}(2010{\natexlab{a}})\citenamefont {Iatrakis}, \citenamefont
  {Kiritsis},\ and\ \citenamefont {Paredes}}]{Iatrakis:2010zf}%
  \BibitemOpen
  \bibfield  {author} {\bibinfo {author} {\bibfnamefont {I.}~\bibnamefont
  {Iatrakis}}, \bibinfo {author} {\bibfnamefont {E.}~\bibnamefont {Kiritsis}},\
  and\ \bibinfo {author} {\bibfnamefont {A.}~\bibnamefont {Paredes}},\
  }\bibfield  {title} {\bibinfo {title} {{An AdS/QCD model from Sen's tachyon
  action}},\ }\href {https://doi.org/10.1103/PhysRevD.81.115004} {\bibfield
  {journal} {\bibinfo  {journal} {Phys. Rev. D}\ }\textbf {\bibinfo {volume}
  {81}},\ \bibinfo {pages} {115004} (\bibinfo {year} {2010}{\natexlab{a}})},\
  \Eprint {https://arxiv.org/abs/1003.2377} {arXiv:1003.2377 [hep-ph]}
  \BibitemShut {NoStop}%
\bibitem [{\citenamefont {Iatrakis}\ \emph
  {et~al.}(2010{\natexlab{b}})\citenamefont {Iatrakis}, \citenamefont
  {Kiritsis},\ and\ \citenamefont {Paredes}}]{Iatrakis:2010jb}%
  \BibitemOpen
  \bibfield  {author} {\bibinfo {author} {\bibfnamefont {I.}~\bibnamefont
  {Iatrakis}}, \bibinfo {author} {\bibfnamefont {E.}~\bibnamefont {Kiritsis}},\
  and\ \bibinfo {author} {\bibfnamefont {A.}~\bibnamefont {Paredes}},\
  }\bibfield  {title} {\bibinfo {title} {{An AdS/QCD model from tachyon
  condensation: II}},\ }\href {https://doi.org/10.1007/JHEP11(2010)123}
  {\bibfield  {journal} {\bibinfo  {journal} {JHEP}\ }\textbf {\bibinfo
  {volume} {11}},\ \bibinfo {pages} {123}},\ \Eprint
  {https://arxiv.org/abs/1010.1364} {arXiv:1010.1364 [hep-ph]} \BibitemShut
  {NoStop}%
\bibitem [{\citenamefont {Donoghue}\ \emph {et~al.}(2014)\citenamefont
  {Donoghue}, \citenamefont {Golowich},\ and\ \citenamefont
  {Holstein}}]{Donoghue_Golowich_Holstein_2014}%
  \BibitemOpen
  \bibfield  {author} {\bibinfo {author} {\bibfnamefont {J.~F.}\ \bibnamefont
  {Donoghue}}, \bibinfo {author} {\bibfnamefont {E.}~\bibnamefont {Golowich}},\
  and\ \bibinfo {author} {\bibfnamefont {B.~R.}\ \bibnamefont {Holstein}},\
  }\href@noop {} {\emph {\bibinfo {title} {Dynamics of the Standard Model}}},\
  \bibinfo {edition} {2nd}\ ed.,\ Cambridge Monographs on Particle Physics,
  Nuclear Physics and Cosmology\ (\bibinfo  {publisher} {Cambridge University
  Press},\ \bibinfo {year} {2014})\BibitemShut {NoStop}%
\bibitem [{\citenamefont {Davier}\ \emph {et~al.}(2020)\citenamefont {Davier},
  \citenamefont {Hoecker}, \citenamefont {Malaescu},\ and\ \citenamefont
  {Zhang}}]{Davier:2019can}%
  \BibitemOpen
  \bibfield  {author} {\bibinfo {author} {\bibfnamefont {M.}~\bibnamefont
  {Davier}}, \bibinfo {author} {\bibfnamefont {A.}~\bibnamefont {Hoecker}},
  \bibinfo {author} {\bibfnamefont {B.}~\bibnamefont {Malaescu}},\ and\
  \bibinfo {author} {\bibfnamefont {Z.}~\bibnamefont {Zhang}},\ }\bibfield
  {title} {\bibinfo {title} {{A new evaluation of the hadronic vacuum
  polarisation contributions to the muon anomalous magnetic moment and to
  $\mathbf{\boldsymbol\alpha(m_Z^2)}$}},\ }\href
  {https://doi.org/10.1140/epjc/s10052-020-7792-2} {\bibfield  {journal}
  {\bibinfo  {journal} {Eur. Phys. J. C}\ }\textbf {\bibinfo {volume} {80}},\
  \bibinfo {pages} {241} (\bibinfo {year} {2020})},\ \bibinfo {note} {[Erratum:
  Eur.Phys.J.C 80, 410 (2020)]},\ \Eprint {https://arxiv.org/abs/1908.00921}
  {arXiv:1908.00921 [hep-ph]} \BibitemShut {NoStop}%
\bibitem [{\citenamefont {Grigoryan}\ and\ \citenamefont
  {Radyushkin}(2007{\natexlab{b}})}]{Grigoryan:2007wn}%
  \BibitemOpen
  \bibfield  {author} {\bibinfo {author} {\bibfnamefont {H.~R.}\ \bibnamefont
  {Grigoryan}}\ and\ \bibinfo {author} {\bibfnamefont {A.~V.}\ \bibnamefont
  {Radyushkin}},\ }\bibfield  {title} {\bibinfo {title} {{Pion form-factor in
  chiral limit of hard-wall AdS/QCD model}},\ }\href
  {https://doi.org/10.1103/PhysRevD.76.115007} {\bibfield  {journal} {\bibinfo
  {journal} {Phys. Rev. D}\ }\textbf {\bibinfo {volume} {76}},\ \bibinfo
  {pages} {115007} (\bibinfo {year} {2007}{\natexlab{b}})},\ \Eprint
  {https://arxiv.org/abs/0709.0500} {arXiv:0709.0500 [hep-ph]} \BibitemShut
  {NoStop}%
\bibitem [{\citenamefont {Kwee}\ and\ \citenamefont
  {Lebed}(2008)}]{Kwee:2007dd}%
  \BibitemOpen
  \bibfield  {author} {\bibinfo {author} {\bibfnamefont {H.~J.}\ \bibnamefont
  {Kwee}}\ and\ \bibinfo {author} {\bibfnamefont {R.~F.}\ \bibnamefont
  {Lebed}},\ }\bibfield  {title} {\bibinfo {title} {{Pion form-factors in
  holographic QCD}},\ }\href {https://doi.org/10.1088/1126-6708/2008/01/027}
  {\bibfield  {journal} {\bibinfo  {journal} {JHEP}\ }\textbf {\bibinfo
  {volume} {01}},\ \bibinfo {pages} {027}},\ \Eprint
  {https://arxiv.org/abs/0708.4054} {arXiv:0708.4054 [hep-ph]} \BibitemShut
  {NoStop}%
\bibitem [{\citenamefont {Grigoryan}\ and\ \citenamefont
  {Radyushkin}(2008)}]{Grigoryan:2008up}%
  \BibitemOpen
  \bibfield  {author} {\bibinfo {author} {\bibfnamefont {H.~R.}\ \bibnamefont
  {Grigoryan}}\ and\ \bibinfo {author} {\bibfnamefont {A.~V.}\ \bibnamefont
  {Radyushkin}},\ }\bibfield  {title} {\bibinfo {title} {{Anomalous Form Factor
  of the Neutral Pion in Extended AdS/QCD Model with Chern-Simons Term}},\
  }\href {https://doi.org/10.1103/PhysRevD.77.115024} {\bibfield  {journal}
  {\bibinfo  {journal} {Phys. Rev. D}\ }\textbf {\bibinfo {volume} {77}},\
  \bibinfo {pages} {115024} (\bibinfo {year} {2008})},\ \Eprint
  {https://arxiv.org/abs/0803.1143} {arXiv:0803.1143 [hep-ph]} \BibitemShut
  {NoStop}%
\bibitem [{\citenamefont {Zuo}\ and\ \citenamefont {Huang}(2012)}]{Zuo:2011sk}%
  \BibitemOpen
  \bibfield  {author} {\bibinfo {author} {\bibfnamefont {F.}~\bibnamefont
  {Zuo}}\ and\ \bibinfo {author} {\bibfnamefont {T.}~\bibnamefont {Huang}},\
  }\bibfield  {title} {\bibinfo {title} {{Photon-to-pion transition form factor
  and pion distribution amplitude from holographic QCD}},\ }\href
  {https://doi.org/10.1140/epjc/s10052-011-1813-0} {\bibfield  {journal}
  {\bibinfo  {journal} {Eur. Phys. J. C}\ }\textbf {\bibinfo {volume} {72}},\
  \bibinfo {pages} {1813} (\bibinfo {year} {2012})},\ \Eprint
  {https://arxiv.org/abs/1105.6008} {arXiv:1105.6008 [hep-ph]} \BibitemShut
  {NoStop}%
\bibitem [{\citenamefont {Brodsky}\ \emph {et~al.}(2011)\citenamefont
  {Brodsky}, \citenamefont {Cao},\ and\ \citenamefont
  {de~Teramond}}]{Brodsky:2011xx}%
  \BibitemOpen
  \bibfield  {author} {\bibinfo {author} {\bibfnamefont {S.~J.}\ \bibnamefont
  {Brodsky}}, \bibinfo {author} {\bibfnamefont {F.-G.}\ \bibnamefont {Cao}},\
  and\ \bibinfo {author} {\bibfnamefont {G.~F.}\ \bibnamefont {de~Teramond}},\
  }\bibfield  {title} {\bibinfo {title} {{Meson Transition Form Factors in
  Light-Front Holographic QCD}},\ }\href
  {https://doi.org/10.1103/PhysRevD.84.075012} {\bibfield  {journal} {\bibinfo
  {journal} {Phys. Rev. D}\ }\textbf {\bibinfo {volume} {84}},\ \bibinfo
  {pages} {075012} (\bibinfo {year} {2011})},\ \Eprint
  {https://arxiv.org/abs/1105.3999} {arXiv:1105.3999 [hep-ph]} \BibitemShut
  {NoStop}%
\bibitem [{\citenamefont {Colangelo}\ \emph {et~al.}(2023)\citenamefont
  {Colangelo}, \citenamefont {Giannuzzi},\ and\ \citenamefont
  {Nicotri}}]{Colangelo:2023een}%
  \BibitemOpen
  \bibfield  {author} {\bibinfo {author} {\bibfnamefont {P.}~\bibnamefont
  {Colangelo}}, \bibinfo {author} {\bibfnamefont {F.}~\bibnamefont
  {Giannuzzi}},\ and\ \bibinfo {author} {\bibfnamefont {S.}~\bibnamefont
  {Nicotri}},\ }\bibfield  {title} {\bibinfo {title} {{$\pi^0$, $\eta$, $\eta'$
  two-photon transition form factors in the holographic soft-wall model and
  contributions to $(g-2)_\mu$}},\ }\href
  {https://doi.org/10.1016/j.physletb.2023.137878} {\bibfield  {journal}
  {\bibinfo  {journal} {Phys. Lett. B}\ }\textbf {\bibinfo {volume} {840}},\
  \bibinfo {pages} {137878} (\bibinfo {year} {2023})},\ \Eprint
  {https://arxiv.org/abs/2301.06456} {arXiv:2301.06456 [hep-ph]} \BibitemShut
  {NoStop}%
\bibitem [{\citenamefont {Lepage}\ and\ \citenamefont
  {Brodsky}(1980)}]{Lepage:1980fj}%
  \BibitemOpen
  \bibfield  {author} {\bibinfo {author} {\bibfnamefont {G.~P.}\ \bibnamefont
  {Lepage}}\ and\ \bibinfo {author} {\bibfnamefont {S.~J.}\ \bibnamefont
  {Brodsky}},\ }\bibfield  {title} {\bibinfo {title} {{Exclusive Processes in
  Perturbative Quantum Chromodynamics}},\ }\href
  {https://doi.org/10.1103/PhysRevD.22.2157} {\bibfield  {journal} {\bibinfo
  {journal} {Phys. Rev. D}\ }\textbf {\bibinfo {volume} {22}},\ \bibinfo
  {pages} {2157} (\bibinfo {year} {1980})}\BibitemShut {NoStop}%
\bibitem [{\citenamefont {Behrend}\ \emph {et~al.}(1991)\citenamefont {Behrend}
  \emph {et~al.}}]{CELLO:1990klc}%
  \BibitemOpen
  \bibfield  {author} {\bibinfo {author} {\bibfnamefont {H.~J.}\ \bibnamefont
  {Behrend}} \emph {et~al.} (\bibinfo {collaboration} {CELLO}),\ }\bibfield
  {title} {\bibinfo {title} {{A Measurement of the $\pi_0$, $\eta$ and $\eta'$
  electromagnetic form-factors}},\ }\href {https://doi.org/10.1007/BF01549692}
  {\bibfield  {journal} {\bibinfo  {journal} {Z. Phys. C}\ }\textbf {\bibinfo
  {volume} {49}},\ \bibinfo {pages} {401} (\bibinfo {year} {1991})}\BibitemShut
  {NoStop}%
\bibitem [{\citenamefont {Gronberg}\ \emph {et~al.}(1998)\citenamefont
  {Gronberg} \emph {et~al.}}]{CLEO:1997fho}%
  \BibitemOpen
  \bibfield  {author} {\bibinfo {author} {\bibfnamefont {J.}~\bibnamefont
  {Gronberg}} \emph {et~al.} (\bibinfo {collaboration} {CLEO}),\ }\bibfield
  {title} {\bibinfo {title} {{Measurements of the meson - photon transition
  form-factors of light pseudoscalar mesons at large momentum transfer}},\
  }\href {https://doi.org/10.1103/PhysRevD.57.33} {\bibfield  {journal}
  {\bibinfo  {journal} {Phys. Rev. D}\ }\textbf {\bibinfo {volume} {57}},\
  \bibinfo {pages} {33} (\bibinfo {year} {1998})},\ \Eprint
  {https://arxiv.org/abs/hep-ex/9707031} {arXiv:hep-ex/9707031} \BibitemShut
  {NoStop}%
\bibitem [{\citenamefont {Aubert}\ \emph {et~al.}(2009)\citenamefont {Aubert}
  \emph {et~al.}}]{BaBar:2009rrj}%
  \BibitemOpen
  \bibfield  {author} {\bibinfo {author} {\bibfnamefont {B.}~\bibnamefont
  {Aubert}} \emph {et~al.} (\bibinfo {collaboration} {BaBar}),\ }\bibfield
  {title} {\bibinfo {title} {{Measurement of the $\gamma \gamma^* \to \pi^0$
  transition form factor}},\ }\href
  {https://doi.org/10.1103/PhysRevD.80.052002} {\bibfield  {journal} {\bibinfo
  {journal} {Phys. Rev. D}\ }\textbf {\bibinfo {volume} {80}},\ \bibinfo
  {pages} {052002} (\bibinfo {year} {2009})},\ \Eprint
  {https://arxiv.org/abs/0905.4778} {arXiv:0905.4778 [hep-ex]} \BibitemShut
  {NoStop}%
\bibitem [{\citenamefont {Uehara}\ \emph {et~al.}(2012)\citenamefont {Uehara}
  \emph {et~al.}}]{Belle:2012wwz}%
  \BibitemOpen
  \bibfield  {author} {\bibinfo {author} {\bibfnamefont {S.}~\bibnamefont
  {Uehara}} \emph {et~al.} (\bibinfo {collaboration} {Belle}),\ }\bibfield
  {title} {\bibinfo {title} {{Measurement of $\gamma \gamma^* \to \pi^0$
  transition form factor at Belle}},\ }\href
  {https://doi.org/10.1103/PhysRevD.86.092007} {\bibfield  {journal} {\bibinfo
  {journal} {Phys. Rev. D}\ }\textbf {\bibinfo {volume} {86}},\ \bibinfo
  {pages} {092007} (\bibinfo {year} {2012})},\ \Eprint
  {https://arxiv.org/abs/1205.3249} {arXiv:1205.3249 [hep-ex]} \BibitemShut
  {NoStop}%
\bibitem [{\citenamefont {Redmer}(2018)}]{Redmer:2018uew}%
  \BibitemOpen
  \bibfield  {author} {\bibinfo {author} {\bibfnamefont {C.~F.}\ \bibnamefont
  {Redmer}} (\bibinfo {collaboration} {BESIII}),\ }\bibfield  {title} {\bibinfo
  {title} {{Measurement of meson transition form factors at BESIII}},\ }in\
  \href@noop {} {\emph {\bibinfo {booktitle} {{13th Conference on the
  Intersections of Particle and Nuclear Physics}}}}\ (\bibinfo {year} {2018})\
  \Eprint {https://arxiv.org/abs/1810.00654} {arXiv:1810.00654 [hep-ex]}
  \BibitemShut {NoStop}%
\bibitem [{\citenamefont {Nyffeler}(2016)}]{Nyffeler:2016gnb}%
  \BibitemOpen
  \bibfield  {author} {\bibinfo {author} {\bibfnamefont {A.}~\bibnamefont
  {Nyffeler}},\ }\bibfield  {title} {\bibinfo {title} {{Precision of a
  data-driven estimate of hadronic light-by-light scattering in the muon $g-2$:
  Pseudoscalar-pole contribution}},\ }\href
  {https://doi.org/10.1103/PhysRevD.94.053006} {\bibfield  {journal} {\bibinfo
  {journal} {Phys. Rev. D}\ }\textbf {\bibinfo {volume} {94}},\ \bibinfo
  {pages} {053006} (\bibinfo {year} {2016})},\ \Eprint
  {https://arxiv.org/abs/1602.03398} {arXiv:1602.03398 [hep-ph]} \BibitemShut
  {NoStop}%
\bibitem [{\citenamefont {G{\'e}rardin}\ \emph {et~al.}(2019)\citenamefont
  {G{\'e}rardin}, \citenamefont {Meyer},\ and\ \citenamefont
  {Nyffeler}}]{Gerardin:2019vio}%
  \BibitemOpen
  \bibfield  {author} {\bibinfo {author} {\bibfnamefont {A.}~\bibnamefont
  {G{\'e}rardin}}, \bibinfo {author} {\bibfnamefont {H.~B.}\ \bibnamefont
  {Meyer}},\ and\ \bibinfo {author} {\bibfnamefont {A.}~\bibnamefont
  {Nyffeler}},\ }\bibfield  {title} {\bibinfo {title} {{Lattice calculation of
  the pion transition form factor with $N_f=2+1$ Wilson quarks}},\ }\href
  {https://doi.org/10.1103/PhysRevD.100.034520} {\bibfield  {journal} {\bibinfo
   {journal} {Phys. Rev. D}\ }\textbf {\bibinfo {volume} {100}},\ \bibinfo
  {pages} {034520} (\bibinfo {year} {2019})},\ \Eprint
  {https://arxiv.org/abs/1903.09471} {arXiv:1903.09471 [hep-lat]} \BibitemShut
  {NoStop}%
\bibitem [{\citenamefont {Masjuan}\ and\ \citenamefont
  {Sanchez-Puertas}(2017)}]{Masjuan:2017tvw}%
  \BibitemOpen
  \bibfield  {author} {\bibinfo {author} {\bibfnamefont {P.}~\bibnamefont
  {Masjuan}}\ and\ \bibinfo {author} {\bibfnamefont {P.}~\bibnamefont
  {Sanchez-Puertas}},\ }\bibfield  {title} {\bibinfo {title}
  {{Pseudoscalar-pole contribution to the $(g_{\mu}-2)$: a rational
  approach}},\ }\href {https://doi.org/10.1103/PhysRevD.95.054026} {\bibfield
  {journal} {\bibinfo  {journal} {Phys. Rev. D}\ }\textbf {\bibinfo {volume}
  {95}},\ \bibinfo {pages} {054026} (\bibinfo {year} {2017})},\ \Eprint
  {https://arxiv.org/abs/1701.05829} {arXiv:1701.05829 [hep-ph]} \BibitemShut
  {NoStop}%
\bibitem [{\citenamefont {G{\'e}rardin}\ \emph {et~al.}(2025)\citenamefont
  {G{\'e}rardin}, \citenamefont {Verplanke}, \citenamefont {Wang},
  \citenamefont {Fodor}, \citenamefont {Guenther}, \citenamefont {Lellouch},
  \citenamefont {Szabo},\ and\ \citenamefont {Varnhorst}}]{Gerardin:2023naa}%
  \BibitemOpen
  \bibfield  {author} {\bibinfo {author} {\bibfnamefont {A.}~\bibnamefont
  {G{\'e}rardin}}, \bibinfo {author} {\bibfnamefont {W.~E.~A.}\ \bibnamefont
  {Verplanke}}, \bibinfo {author} {\bibfnamefont {G.}~\bibnamefont {Wang}},
  \bibinfo {author} {\bibfnamefont {Z.}~\bibnamefont {Fodor}}, \bibinfo
  {author} {\bibfnamefont {J.~N.}\ \bibnamefont {Guenther}}, \bibinfo {author}
  {\bibfnamefont {L.}~\bibnamefont {Lellouch}}, \bibinfo {author}
  {\bibfnamefont {K.~K.}\ \bibnamefont {Szabo}},\ and\ \bibinfo {author}
  {\bibfnamefont {L.}~\bibnamefont {Varnhorst}},\ }\bibfield  {title} {\bibinfo
  {title} {{Lattice calculation of the $\pi^0$, $\eta$ and $\eta'$ transition
  form factors and the hadronic light-by-light contribution to the muon
  $g-2$}},\ }\href {https://doi.org/10.1103/PhysRevD.111.054511} {\bibfield
  {journal} {\bibinfo  {journal} {Phys. Rev. D}\ }\textbf {\bibinfo {volume}
  {111}},\ \bibinfo {pages} {054511} (\bibinfo {year} {2025})},\ \Eprint
  {https://arxiv.org/abs/2305.04570} {arXiv:2305.04570 [hep-lat]} \BibitemShut
  {NoStop}%
\bibitem [{\citenamefont {Pauk}\ and\ \citenamefont
  {Vanderhaeghen}(2014)}]{Pauk:2014rta}%
  \BibitemOpen
  \bibfield  {author} {\bibinfo {author} {\bibfnamefont {V.}~\bibnamefont
  {Pauk}}\ and\ \bibinfo {author} {\bibfnamefont {M.}~\bibnamefont
  {Vanderhaeghen}},\ }\bibfield  {title} {\bibinfo {title} {{Single meson
  contributions to the muon`s anomalous magnetic moment}},\ }\href
  {https://doi.org/10.1140/epjc/s10052-014-3008-y} {\bibfield  {journal}
  {\bibinfo  {journal} {Eur. Phys. J. C}\ }\textbf {\bibinfo {volume} {74}},\
  \bibinfo {pages} {3008} (\bibinfo {year} {2014})},\ \Eprint
  {https://arxiv.org/abs/1401.0832} {arXiv:1401.0832 [hep-ph]} \BibitemShut
  {NoStop}%
\bibitem [{\citenamefont {Achard}\ \emph {et~al.}(2002)\citenamefont {Achard}
  \emph {et~al.}}]{L3:2001cyf}%
  \BibitemOpen
  \bibfield  {author} {\bibinfo {author} {\bibfnamefont {P.}~\bibnamefont
  {Achard}} \emph {et~al.} (\bibinfo {collaboration} {L3}),\ }\bibfield
  {title} {\bibinfo {title} {{$f_1(1285)$ formation in two photon collisions at
  LEP}},\ }\href {https://doi.org/10.1016/S0370-2693(01)01477-0} {\bibfield
  {journal} {\bibinfo  {journal} {Phys. Lett. B}\ }\textbf {\bibinfo {volume}
  {526}},\ \bibinfo {pages} {269} (\bibinfo {year} {2002})},\ \Eprint
  {https://arxiv.org/abs/hep-ex/0110073} {arXiv:hep-ex/0110073} \BibitemShut
  {NoStop}%
\end{thebibliography}%

\end{document}